\documentclass[%
twocolumn,
amsmath,amssymb,
 aps,
prb,
]{revtex4-2}

\usepackage{braket}
\usepackage{graphicx}% Include figure files
\usepackage{dcolumn}% Align table columns on decimal point
\usepackage{bm}% bold math
\usepackage{xcolor}
\usepackage{diagbox}
\usepackage{array, makecell}

\def\br{{\bf r}}
\def\bp{{\bf r}'}
\def\dd{{\rm d}}

\begin{document}

\preprint{APS/123-QED}

\title{Stochastic many-body calculations of moir\'{e} states in twisted bilayer graphene at high pressures}

\author{Mariya Romanova}
\author{Vojt\v{e}ch Vl\v{c}ek}%
 \email{vlcek@ucsb.edu}
\affiliation{Department of Chemistry and Biochemistry, University of California, Santa Barbara, CA 93106-9510, U.S.A.}%

\date{\today}

\begin{abstract}
 Atomistic first principles many-body computational studies were so far limited by the system size. In this work, we apply and expand the stochastic $GW$ method allowing calculations of quasiparticle energies of giant systems. We introduce three new developments to the stochastic many-body formalism: efficient evaluation of off-diagonal self-energy terms, construction of Dyson orbitals, and introduce stochastic cRPA, enabling efficient downfolding onto model Hamiltonians. We exemplify the new methodology by exploring twisted bilayer graphene (tBLG); its moir\'{e} periodicity is associated with giant unit-cells hosting correlated electrons in flat bands. The resulting behavior of tBLG is governed by the coupling between the weakly and strongly correlated electrons in individual monolayers, controlled by twist or applied pressure. Here we study the latter scenario. The calculations document the formation of charge carrier localization at the Fermi level while most weakly correlated states are merely shifted in energy. This contrasts with the mean-field results. We show how to efficiently downfold the correlated subspace on a model Hamiltonian, namely the Hubbard model with a screened frequency dependent on-site interaction, computed with newly developed stochastic cRPA. For the $6^{\circ}$ tBLG system, the onsite interactions range between 200 and 300 meV under compression. The Dyson orbitals exhibit spatial distribution similar to the mean-field single particle states. Under pressure, the electron-electron interactions in the localized states increase; the dynamical screening does not fully compensate the dominant bare Coulomb interaction. The correlated moir\'e states thus appear only moderately screened at high pressures suggesting the presence of insulating states in compressed tBLG.   
\end{abstract}

\maketitle

%\tableofcontents

\section{Background}
First-principles many-body theory provides invaluable insights for predicting and deciphering the behavior of electronic states without relying on empirical parameters. In particular, one encounters the largest need for predictive first principles theory in systems with emergent phenomena. For instance,  the coupling between nominally weakly correlated subsystems may lead to new states exhibiting all hallmarks of strong correlations. Describing such phenomena is often hindered by the large system sizes that pose an insurmountable challenge for conventional calculations. Here, by presenting new computational developments we expand our previous work that enabled the application of the ab-initio many-body theory to giant systems~\cite{brooks2020stochastic}. 

We exemplify the new methodology by studying twisted bilayer graphene (tBLG), which is a prototypical moir\'{e} superstructure in which the coupling of individual monolayers is controlled primarily by the twist angle, $\theta$.\cite{cao2018correlated} As $\theta$ approaches 1.1$^\circ$ ``magic angle'', tBLG transitions from a simple semimetal to a system hosting correlated electronic states\cite{cao2018correlated}. Under charge carrier doping tBLG at (or near) the magic angle exhibits superconducting, insulating, and magnetic properties~\cite{cao2018correlated,cao2018unconventional,yankowitz2019tuning,yankowitz2018dynamic,sharpe2019emergent,lu2019superconductors,saito2020independent,choi2019electronic,xie2019spectroscopic,kerelsky2019maximized,jiang2019charge}. These emergent states are associated with a shallow moir\'{e} potential, which localizes electrons in so-called AA stacking regions of the superstructure~\cite{trambly2010localization,koshino2018maximally,kang2018symmetry,calderon2020interactions,goodwin2019attractive,kerelsky2019maximized} and is responsible for the formation of flat (i.e., dispersionless) bands near the Fermi level~\cite{bistritzer2011moire,utama2020visualization}.

Fundamentally, the electron localization is governed by the strong interaction between the monolayers and the states' hybridization near the respective Dirac points. On the one hand, this is realized at small twists near magic angle, but equally well by the bilayer's in-plane strain or compression, as was shown by recent experiments and theoretical works~\cite{yankowitz2018dynamic,yankowitz2019tuning,bultnick2021}. In the latter case, the interlayer distance reduction drives electronic localization for angles substantially larger than 1.1$^\circ$~\cite{yankowitz2018dynamic,yankowitz2019tuning}. Unlike the twist angle, the degree of compression can be adjusted even after the deposition of the layers and hence represents a unique control mechanism for realizing correlated states. Yet, it has been so far studied to a lesser degree. 

At high pressures, graphene becomes thermodynamically unstable; a suitably chosen combination of pressure transmission media and encapsulation allows reaching pressures up to 37~GPa~\cite{tao2020raman,clark2013few} for tBLG. Even higher compression may be possible for up to 50~GPa before graphene inevitably transforms to diamond~\cite{clark2013few}. Nevertheless, even with the improved experimental setup, the tBLG electronic states cannot be probed in the same detail as at ambient conditions~\cite{tao2020raman,clark2013few}. Further, it is unknown whether the decreasing interlayer spacing affects the entire valence and conduction states or leads to selective hybridization of Dirac point states. Finally, weakly correlated states dynamically screen the many-body interactions within the flat bands and critically affect the emergent many-body phenomena~\cite{goodwin2019attractive,pizarro2019internal}; however, the effect of compression on screening is also unknown. 

Previous studies of the tBLG electronic structure were limited to tight-binding and continuum models~\cite{dosSantos2012,lian2019twisted,yuan2018model,yuan2018erratum,po2018origin,xu2018topological,roy2019unconventional,volovik2018graphite,padhi2018doped,dodaro2018phases,wu2018theory,isobe2018unconventional,huang2019antiferromagnetically,zhang2019nearly,kang2018symmetry,koshino2018maximally,kennes2018strong,zhang2018lowest,pizarro2019internal,guinea2018electrostatic,zou2018band,lima2017stacking,rademaker2018charge,angeli2018emergent,goodwin2020hartree} that demonstrated the magic angle-induced flat band formation at the Fermi level. The model parameters were usually determined from mean-field (DFT) calculations, which markedly deviate from quasiparticle energies.\cite{martin2016interacting} The ground state of tBLG at the magic angle was further investigated by atomistic Hartree and Hartree-Fock calculations based on the continuum model~\cite{xie2020nature,cea2020band,zhang2020correlated,bultinck2020ground,liu2021nematic,goodwin2020hartree,liu2021theories, gonzalez2020time, rademaker2019charge}. These calculations revealed that unscreened Coulomb interactions are responsible for stabilizing the insulating states in tBLG. Recently, exact diagonalization of downfolded many-body Hamiltonians (within a subspace of flat bands) was used to address the superconducting regime of tBLG.~\cite{potasz2021exact}  
Investigations of the high-pressure behavior remain scarce and limited to the MF or model Hamiltonian treatment,~\cite{munoz2016bilayer,chittari2018pressure,carr2018pressure,padhi2019pressure,lin2020pressure,green2020landau} which focused on describing the dispersion of states near the Fermi level. Further, the strength of the electron-electron interaction at high pressure was investigated neither. Thus, it remains unclear whether the flat bands' formation under compression is equivalent to that at (or near) the magic angle. While these questions can be answered by the first principles many-body approaches, they were not applied until now due to their enormous computational cost.

In this work, we \textit{overcome practical limitations} of \emph{ab initio} many-body method: we propose a series of new developments in the stochastic many-body perturbation theory (MBPT) techniques\cite{vlcek2017stochastic,Vlcek2018swift,neuhauser2014breaking,vlcek2019stochastic,vlvcek2018simple,vlvcek2018quasiparticle} which can readily elucidate how the electronic structure behaves in giant moir\'{e} systems. We investigate tBLG with a large twist angle of $\theta \approx6^{\circ}$ (with the supercell of size $4\times7$ nm containing 2184 atoms, i.e., 8736 valence electrons), which is weakly correlated at ambient conditions but develops flat bands at high compressions. The pressure-induced coupling of the two monolayers is substantially different and affects only states near the Fermi level.
Further, we develop a stochastic constrained random-phase approximation (s-cRPA), which efficiently (i.e., with minimal computational cost)  maps the correlated subspace on a Hubbard model with \textit{dynamical} on-site interactions, $U(\omega)$. We find that the electron-electron interactions in the flat bands are more screened under compression. However, the screening does not fully cancel out the bare Coulomb interaction. Thus, the effective interaction increases with pressure. As a result, the strong correlation is not only driven by vanishing band dispersion but also by increased on-site terms.  These results are the same for Dyson quasiparticle orbitals  and mean field canonical single-particle orbitals. We present our results first, then their significance and implications; the new theory is discussed in the Methods at the end of this work.

\section{Results}\label{sec:theory}
The simulations employ rectangular supercells of the graphene bilayer with twist angles $\theta=0^{\circ}$ ($24\times 12$ conventional unitcells  with 9216 valence electrons in total) and $\theta \approx6^{\circ}$ ($1\times3$ moir\'e conventional cells or $\sim 16.5\times 16.5$ in terms of conventional unitcells with 8736 valence electrons in total). Within our real-space methodology, the Brillouin zone of the supercell is sampled by the  $\Gamma$-point. The $1\times3$ moir\'e supercell makes our real-space grid commensurate with high-symmetry $\tilde K$ - point (coinciding with the Dirac point location), where the electronic localization principally occurs. We extract bandstructures with the projector-based energy-momentum analysis. \cite{brooks2020stochastic,popescu2012extracting,huang2014general,medeiros2014effects, PhysRevB.71.115215,boykin2007approximate} We study undoped tBLG, and thus, the Dirac point in our calculations is aligned with the Fermi level. Further, spin and valley symmetry breaking are not considered here. The ideal bilayer interlayer distance was first optimized with the first-principles DFT calculations with van der Waals corrections. For simplicity and to separate out the effects of electron-electron interactions treated by our novel methodology, we employed flat geometry, as lattice reconstructions are significant for small twist angles~\cite{yoo2019atomic,liang2020effect,cantele2020structural}. The equation of state is extracted from the total energy calculations for bilayers with variable interlayer distances (the details of the ground state calculations and the pressure estimation are provided in the Methods and SI). Our results are in excellent agreement with previous theoretical calculations employing weak interlayer interactions using the random phase approximation\cite{leconte2017}.

\subsection{Pressure-induced localization} \label{results1}

This section provides computational results obtained with the stochastic $GW$ approximation (see Methods) for both diagonal and off-diagonal parts and discusses the role of the orbital basis. We will first report results for the ideal Bernal-stacked graphene bilayer and then for the tBLG.

\begin{figure*}
    \centering
    \includegraphics[width=6.69in]{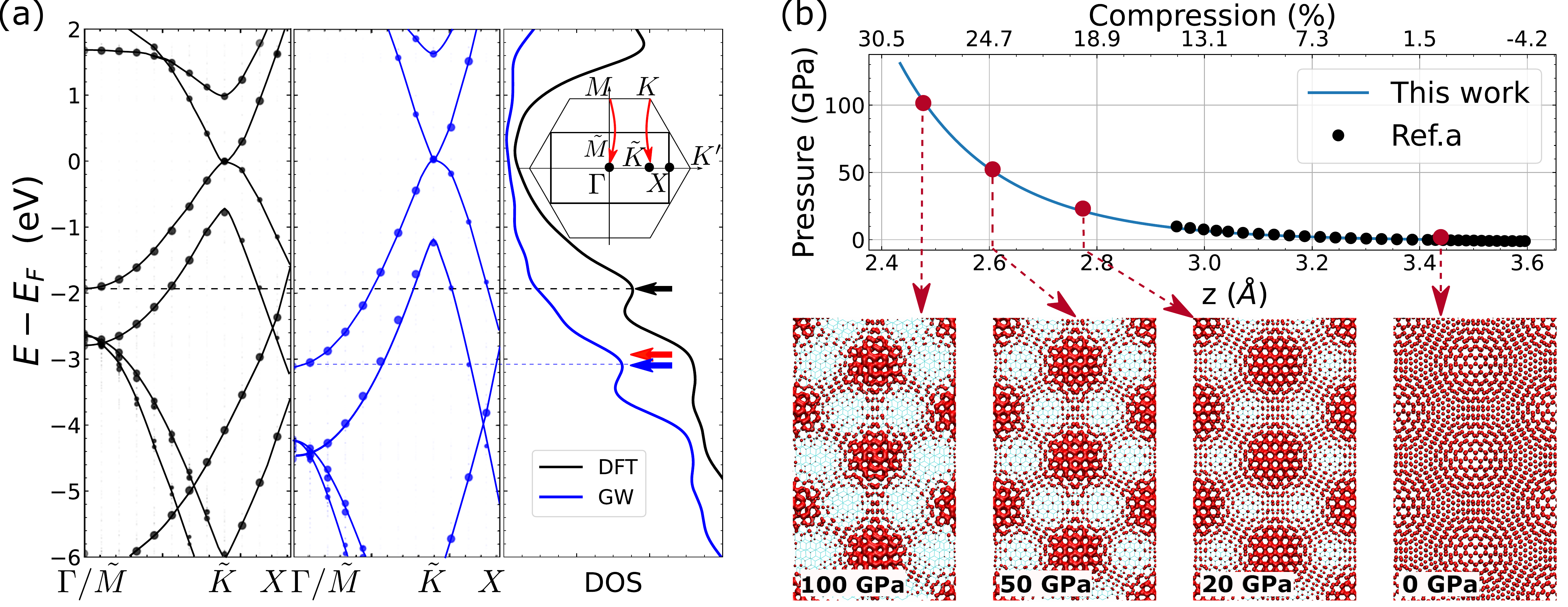}
    \caption{(a)Left: Black-DFT (blue-GW) bandstructure of ideal bilayer graphene. Due to the rectangular cell, corresponding bands are refolded onto $\Gamma$-point. Lines in band structure are guides for the eyes. Right: comparison of DFT and GW DOS. Red arrow indicates where should be an experimental peak corresponding to the flattening of the band in ARPES in Ref.~\cite{ohta2006controlling,ohta2007interlayer}. The onset figure schematically shows a comparison of the rectangular and hexagonal first Brillouin zones. (b) Top: Pressure- interlayer distance (compression) curve. Ref.~a\cite{leconte2017}. Bottom: Charge density of the Dirac point KS states at corresponding pressures. The isovalue is the same for all density plots. }
    \label{fig:ideal_dos}
\end{figure*}

As a first step, we investigate the role of non-local and dynamical correlations in an ideal graphene bilayer. We compare the mean-field DFT and QP energies computed using the diagonal approximation to the self-energy (i.e., $\Delta=0$ in Eq.~\eqref{eqp} of Methods). The corresponding bandstructures and the QP densities of states (DOS) are in Fig.~\ref{fig:ideal_dos}a. As expected,\cite{heske1999band,strocov2001photoemission,gruneis2008electron} MBPT significantly increases the bandwidth (compared to DFT), leading to an excellent agreement with the available experimental data.\cite{ohta2007interlayer,ohta2006controlling} 
Specifically, the experiments show a local minimum of a non-degenerate Dirac band at the M high-symmetry point, which is visible as a peak in the QP DOS at $-2.93$~eV (red arrow in Fig.~\ref{fig:ideal_dos}a). In the bandstructure for the rectangular cell in Fig.~\ref{fig:ideal_dos}a (obtained from momentum space projection~\cite{popescu2012extracting,huang2014general,medeiros2014effects,brooks2020stochastic}), the $\Gamma$ and M points coincide (due to the Brillouin zone reflection as depicted in the inset figure). The DFT calculation places the peak in DOS incorrectly at $-1.94$~eV (i.e., $1.0$~eV too close to the Fermi level, as shown by the black dashed line and an arrow). This agrees with previous DFT calculations that also underestimated the band dispersion by $10-20\%$ with respect to the experiments~\cite{heske1999band,zhou2005coexistence,ohta2007interlayer,gruneis2008electron,ohta2012evidence,kerelsky2019maximized,strocov2001photoemission}. In contrast, our $GW$ results predict a corresponding feature to appear at $-3.11$~eV (blue dashed line and an arrow), which is in excellent agreement with the experiment (the peak is placed only $0.18$~eV lower than the measured value).

Next, we explore tBLG systems under various pressures ranging from 0 up to 100 GPa (corresponding to maximal compression of $28\%$ of the interlayer distance -- see Figure~\ref{fig:ideal_dos}b and SI). The structure is characterized by a hexagonal symmetry with a periodicity of $23.4$~\AA\, between the AA stacking regions. Since the twisting angle is high, the coupling between the individual graphene layers is small at ambient conditions, and the system does not differ substantially from the ideal bilayer. First, the electronic states at the Dirac point are fairly delocalized (bottom of Figure~\ref{fig:ideal_dos}b), and the distinction between the increase of the orbital density in the AA region is hardly noticeable. Second, (due to the lack of localization) band dispersion is large, and there is no increase in the density of states at the Fermi level (Fig.~\ref{fig:qp_dos}). 

The situation changes with the compression (for the pressure of 20 GPa and higher). The electronic states become more localized in the AA stacking areas. Already at 20 GPa we clearly observe the spatial redistribution of the orbital density (see illustration in Fig.~\ref{fig:ideal_dos}b bottom). In our calculations, we further explore even higher pressures which may be difficult to realize experimentally (though the highest reported pressures achieved for tBLG was $P=$37~GPa~\cite{tao2020raman,clark2013few}). With increasing $P$, localization becomes even more pronounced, and at 100~GPa, roughly $75\%$ of the Dirac point states' orbital density is localized within $8$~\AA\, radius around the AA stacking point. This localization of the Dirac point states translates into the flat-band formation, corresponding to a peak around the Fermi level  DOS (Fig.~\ref{fig:qp_dos}). Note that the structure at the pressure of 100~GPa is thermodynamically unstable. Still, the observations are useful as an indicator of the electronic correlation in tBLG: based on experiments\cite{yankowitz2018dynamic}, the same type of localization is expected for lower $\theta$ angles at much lower pressures for which graphene is a stable polymorph. 

Note that the localization illustrated here lacks contributions beyond those included in the mean-field (DFT) Hamiltonian. In practice, the confinement of orbitals under pressure is driven by the external (ionic) potential and the degree of the localization is impacted by the delocalization error of semilocal DFT functionals. We address this question below and show that the mean field DFT orbitals, however, are remarkably close to the Dyson orbitals computed with MBPT.

\subsection{Mean-field vs MBPT energy spectrum}
\begin{figure}
    \centering
    \includegraphics[width=3.37in]{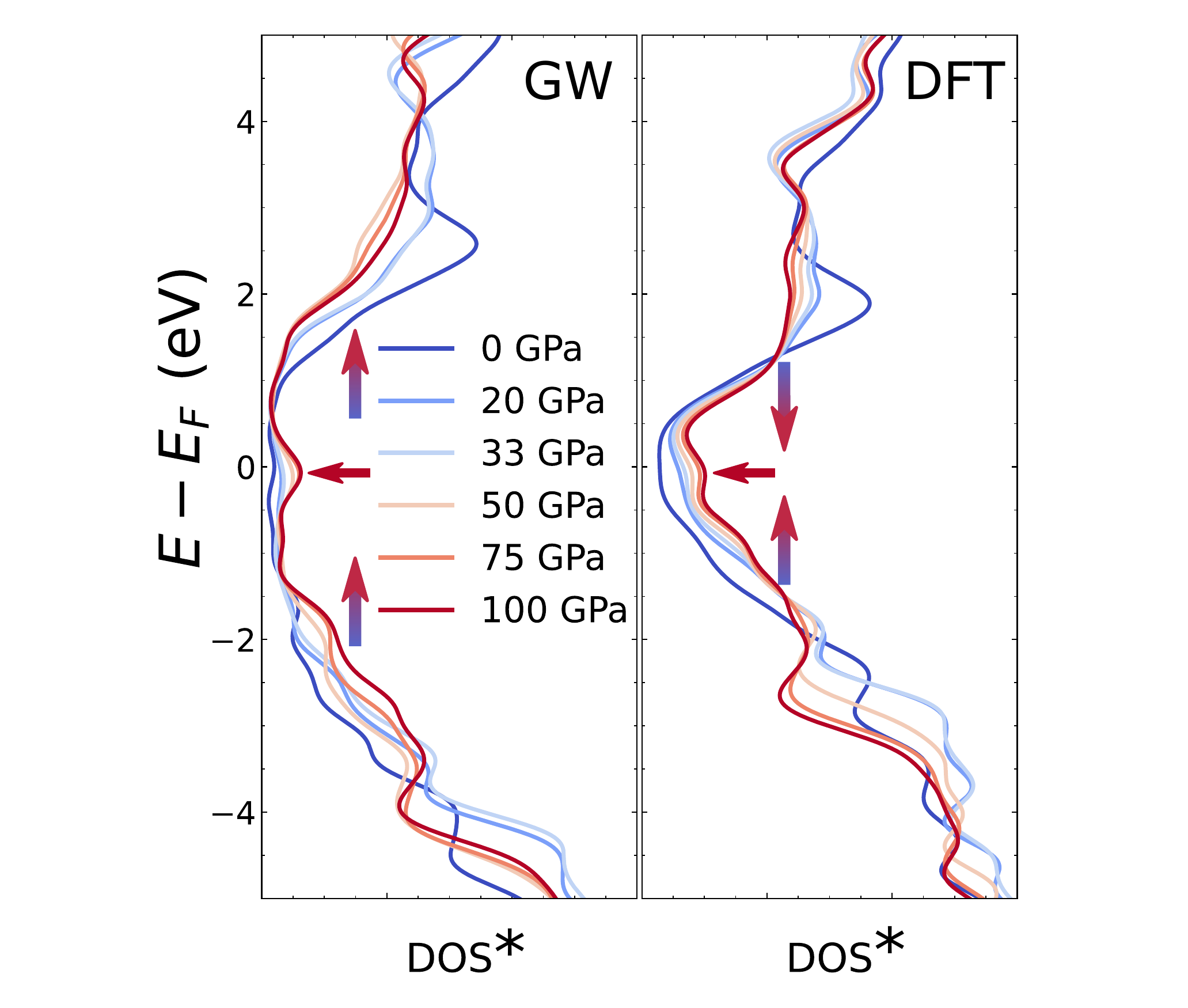}
    \caption{Left: qpDOS of a tBLG $\theta\approx 6^{\circ}$ as a function of pressure. The stochastic error on identifying the quasiparticle energy for the stochastic $GW$ calculation is $\sim20$~meV. Right: DFT DOS. Both DFT and qpDOS were constructed with Gaussian functions centered at each state (with broadening of 0.35~eV), for more details about qpDOS construction see Supplemental Information. (*) In SI we show that the DOS eigenenergies are well converged with the supercell size, we note, however, that some $k$-points are incommensurate with our real-space sampling grids, thus, some DOS features such as van Hove singularities~\cite{trambly2012numerical} are missing.}
    \label{fig:qp_dos}
\end{figure}

  At this stage, we compare the first-principles results obtained with the mean-field (DFT) and MBPT ($GW$) approaches (Fig.~2).
While the single particle energies are converged with respect to the supercell (see SI), the DOS curves however do not have sufficient resolution to capture certain details, such as the van Hove singularities near the Fermi level~\cite{brihuega2012unraveling}. We can extract their position, since it coincides with the energy of the M critical points of the Brillouin zone, which are refolded on the $\Gamma$-point. DFT places the singularity at $0.37$~eV away from the Fermi level. In contrast, GW positions them at 0.55 eV away from the Fermi level, in excellent agreement with the value of $0.56$~eV obtained experimentally~\cite{brihuega2012unraveling} (see SI). 
Overall, Fig~\ref{fig:qp_dos} shows that MBPT significantly increases the bandwidth and widens the DOS features at 0~GPa compared to the mean-field solution. However, the most apparent changes are for the high pressures at which the DFT DOS significantly contracts and predicts a strong reduction in the width of \textit{all} bands. While the flat-band formation leads to a peak at the Fermi level, its signature is suppressed by the proximity of the \textit{entire} set of top valence and bottom conduction bands, which become closer in energy. The DFT results show that occupied and unoccupied states' behavior is mostly symmetric around the chemical potential (moving up and down in energy, respectively). 

The many-body calculations show a different picture. 
Up to 50 GPa, the entire QP DOS shifts up in energy, i.e., the valence states are moving closer to the Fermi level while the conduction states away from it. The bandwidth of the states away from the Fermi level is mostly unaffected by the increased pressure. Simultaneously, we observe the flat band formation around the chemical potential (indicated by a red arrow in Figure~\ref{fig:qp_dos}), which does not overlap with the rest of the occupied and unoccupied states. The QP DOS peak comprises eight quasi-degenerate states (corresponding to the Dirac points at K and K' in the moir\'e hexagonal Brillouin zone). Increasing the pressure further (i.e., $P>50$~GPa) leads to more pronounced structures in the QP DOS, but the peaks' position remains roughly the same. This difference in the behavior can be understood from the compression curve shown in Figure~\ref{fig:ideal_dos}b: the change of the pressure between 50 and 100~GPa requires only a small decrease in the interlayer distance, i.e. small change of the coupling of the monolayers.  For 100 GPa, the flat band is clearly visible in between the conduction and valence bands. The key observation is that the compression-driven flat bands' formation leaves the rest of the states largely unaffected. Hence, despite the reduced interlayer spacing leads to the electron localization in the vicinity of the Dirac point, the increased coupling between the graphene monolayers is confined to a narrow energy range of the states near the chemical potential.

~\\
\subsection{Role of the off-diagonal self-energy elements}\label{sec:QPvsKS}

To explore the mutual coupling in the localized moir\'e states, we investigate the role of \textit{off-diagonal} self-energy terms $\Sigma_{j\neq k}=\langle \phi_j|\Sigma(\omega)|\phi_k\rangle$ in Eq.~\ref{eqp}, using our new development to the stochastic $GW$ described in the Methods section. In practice, we compute the contributions of $\Delta_{j}$ for the states up to $\pm0.5$~eV away from the highest occupied state. We note that the $\Sigma_{ij}$ is computed for a wide frequency range for all off-diagonal terms \textit{at once}. The cost of the $GW$ calculation is practically identical to the previously developed diagonal implementation (see Methods section). Next, we resort to a common procedure in the self-consistent $GW$\cite{faleev2004all,bruneval2006effect} and construct a symmetrized and self-adjoint quasiparticle Hamiltonian $H$ as:
\begin{equation}
    H_{jk} = H_{jk}^{KS} - v^{xc}_{jk} +\frac{1}{4}\bigg\{ \left[\Sigma_{jk}(\epsilon_{j}) + \Sigma_{kj}(\epsilon_{k})\right] + c.c.\bigg\}. 
\label{eq:Hqp}
\end{equation}
The QP energies of $H$, however, differ only negligibly from those computed with the diagonal approximation to Eq.~\ref{eqp}. We find that the difference for most of the states is less than $1\%$ and only for three states it is roughly $3\%$. The inclusion of the $\Delta_j$ terms can thus be safely neglected in the QP DOS. This is true also for the flat bands around the Fermi level. Here, the off-diagonal terms are not necessarily small, but they tend to ``average-out,'' and their net contribution is negligible. 

\begin{figure}
    \centering
    \includegraphics[width=3.37in]{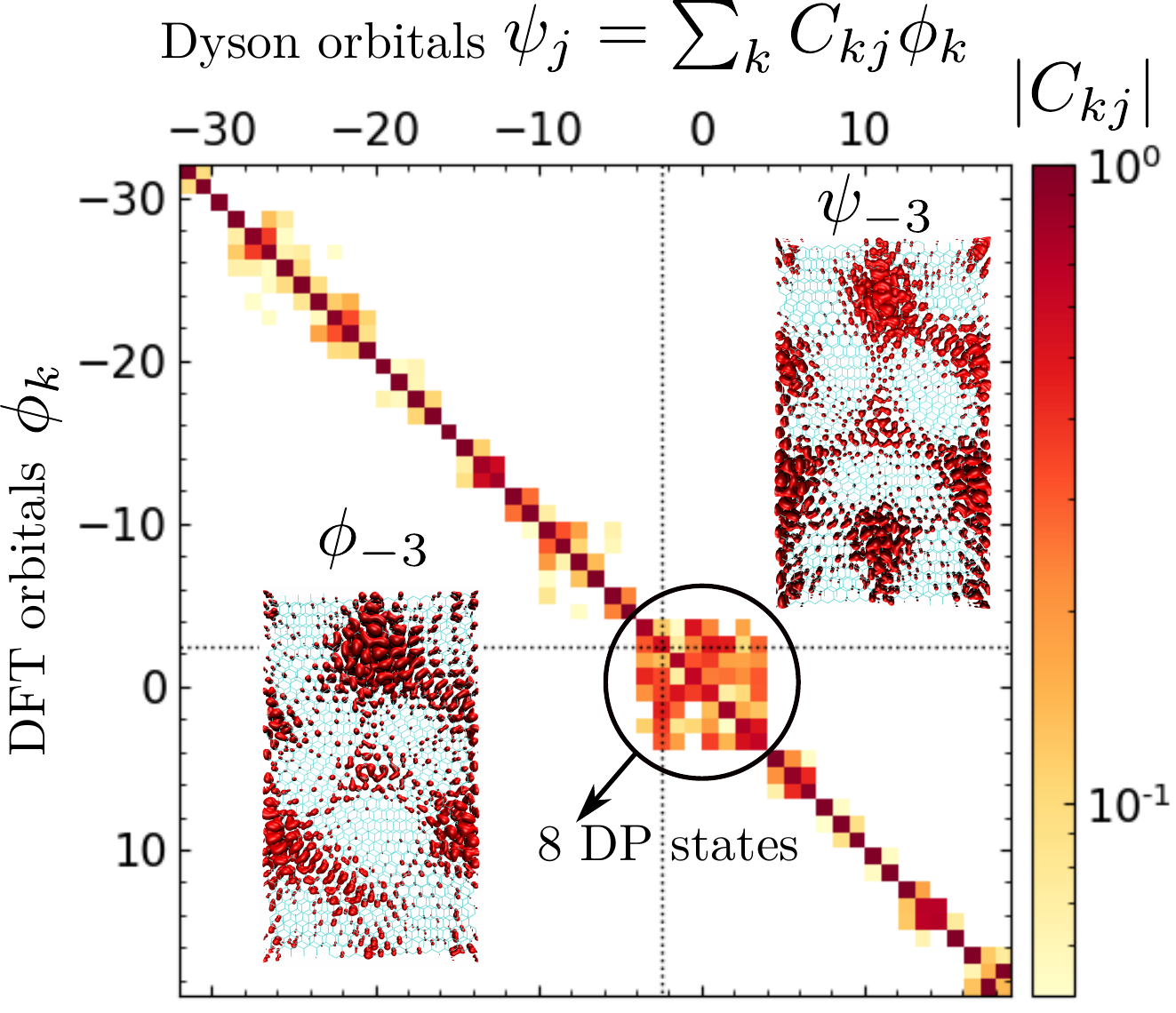}
    \caption{Coefficient matrix obtained from the exact diagonalization of the $51\times51$ $H^{QP}$ at 100 GPa pressure. The absolute values of the complex coefficients are plotted. A zero position is set between occupied (negative) and unoccupied (positive) states. With a black circle we indicate 8 Dirac point states that are quasi-degenerate in energy. Onset: We plot a density of state -3 to compare a KS orbital to the QP orbital. State -3 is highlighted with a dotted line in the figure.}
    \label{fig:Cqp}
\end{figure}

While the QP energies remain unaffected, this does not imply that the QP states, $\{\psi\}$, are close to the mean-field DFT orbitals $\{\phi\}$. To investigate this, we diagonalize $H_{jk}$ (Eq.~\eqref{eq:Hqp}) for the states in the $0.5$~eV vicinity of the Fermi level. Employing the DFT orbitals' basis, the new eigenstates are ${\psi_j} = \sum_k C_{kj} {\phi_k}$, with $\sum_k |C_{kj}|^2 = 1$. The expansion coefficients are illustrated graphically in Fig.~\ref{fig:Cqp}. For most states, the $C_{kj}$ matrix is practically diagonal. On the other hand, the nearly degenerate flat-band states, show substantial mixed character around the Fermi level and $\sum_{k\in \{\varphi \}} |C_{kj}|^2 = 0.995$, where $\{\varphi \}$ denotes a subspace of eight correlated states at the Fermi level. Nevertheless, each of them has a dominant contribution from a single DFT eigenvector. Indeed, the visual comparison (inset of Fig.~\ref{fig:Cqp}) shows that $\{\psi\}$ and $\{\phi\}$ are generally similar, e.g., both are localized in the AA stacking regions. 

We conclude that the stochastic approach efficiently computes both diagonal and off-diagonal terms (at the same cost). Further, our analysis shows that the off-diagonal terms have little impact on the QP density of states and that the pressure-induced coupling is limited to the subset of quasi-degenerate Dirac point states. The differences in the distribution of the single-particle orbitals are visually small, but a more quantitative analysis is provided in the next section.

~\\
\subsection{Pressure dependence of the dynamically screened on-site interaction}

In the remainder of the paper, we will demonstrate the stochastic methodology for extracting the dynamically screened on-site interaction $U(\omega)$, Eq.~\eqref{U}. With this approach we will explore the role of screening on the orbitals around the Fermi level at various pressures. 

We map the strongly correlated states (identified in the previous section and denoted $\{\varphi\}$) onto the Hubbard Hamiltonian (see Section \ref{sec:scRPA}), with the effective hopping, $t$, and on-site interaction, $U$, terms. The latter contains the information of all other electrons via the non-local and dynamical screening $\tilde W$ in Eq.~\eqref{U}. The electron correlation is commonly characterized by the interplay of the on-site interaction and the kinetic energy.\cite{kennes2018strong,goodwin2019twist,xian2019multiflat} In practice, the $U\gg t$ indicates the regime when the system is strongly correlated. For tBLG close to the magic angle, the strongly correlated regime was driven by the drastic reduction of the hopping (due to localization) to $t\le30$~meV.\cite{lin2020pressure} As a result, the $U/t$ ratio becomes very high,\cite{guo2018pairing,cao2018correlated,goodwin2019attractive, goodwin2019twist} although the screening was predicted to play an important role in reducing the value of the onsite Coulomb term $U$. Previous calculations suggested that the physics is dominated by the competition between low $t$ and dynamical screening: the dielectric constant was predicted to be 20 times larger at the magic angle than in the ideal bilayer\cite{goodwin2019attractive}. 

We estimate only the upper bound for the $t$ parameter from the localized states' bandwidth, extracted~\cite{goodwin2019twist,goodwin2019attractive,neto2009electronic} from the dispersion of the corresponding QP energies.\footnote{The hopping term is $1/6$ of the bandwidth associated with the flat bands.} For the system with the most pronounced localization, i.e., at 100 GPA, we find $t\approx40$~meV, which is in good agreement with the results for the correlated phase at much lower twisting angles. 
We note that, in contrast, the band dispersion at $0$~GPa is very large and $t \sim 600$~meV. Clearly, the pressure-induced localization is responsible for qualitative changes in the hopping (and the associated $t$ parameter decreases by order of magnitude).

Although the band flattening appears as the primary driver of electron-electron correlation, the on-site Coulomb interaction changes are equally important. Following the approach of Refs.~\cite{xian2019multiflat,miyake2009ab}, we provide the mean $U$ values in order to describe the correlation strength of the chosen subspace by a single number rather than comparing $U/t$ ratios for each state. However, we also discuss the variation of $U$ among distinct orbitals. In the text below, we first consider the total screened effective interaction $U(\omega)$. Next, we decompose $U(\omega)$ into the static bare contribution $U^b$ and dynamically screened counterpart $U^p(\omega)$. Note, while the $U$ is computed in the basis of correlated states, the idea of the scRPA is to include the dynamical renormalization due to screening from all electronic states in the orthogonal complement of the correlated subspace. The set of localized KS orbitals is a straightforward and convenient choice of basis for scRPA calculations since no additional localization or orthonormalization procedure is required~\cite{miyake2009ab}.  The approach to compute $U$ in the KS basis has been previously used in Refs.~\cite{xian2019multiflat,ma2021quantum}, but other options (e.g., Wannier basis~\cite{kang2018symmetry}) have been proposed. For simplicity, we resort to the first option and compute $U$ in the basis of KS orbitals $\{\phi\}$, next we compare the results to analogous calculations for QP orbital basis $\{\psi\}$.

\begin{figure}
    \centering
    \includegraphics[width=3.37in]{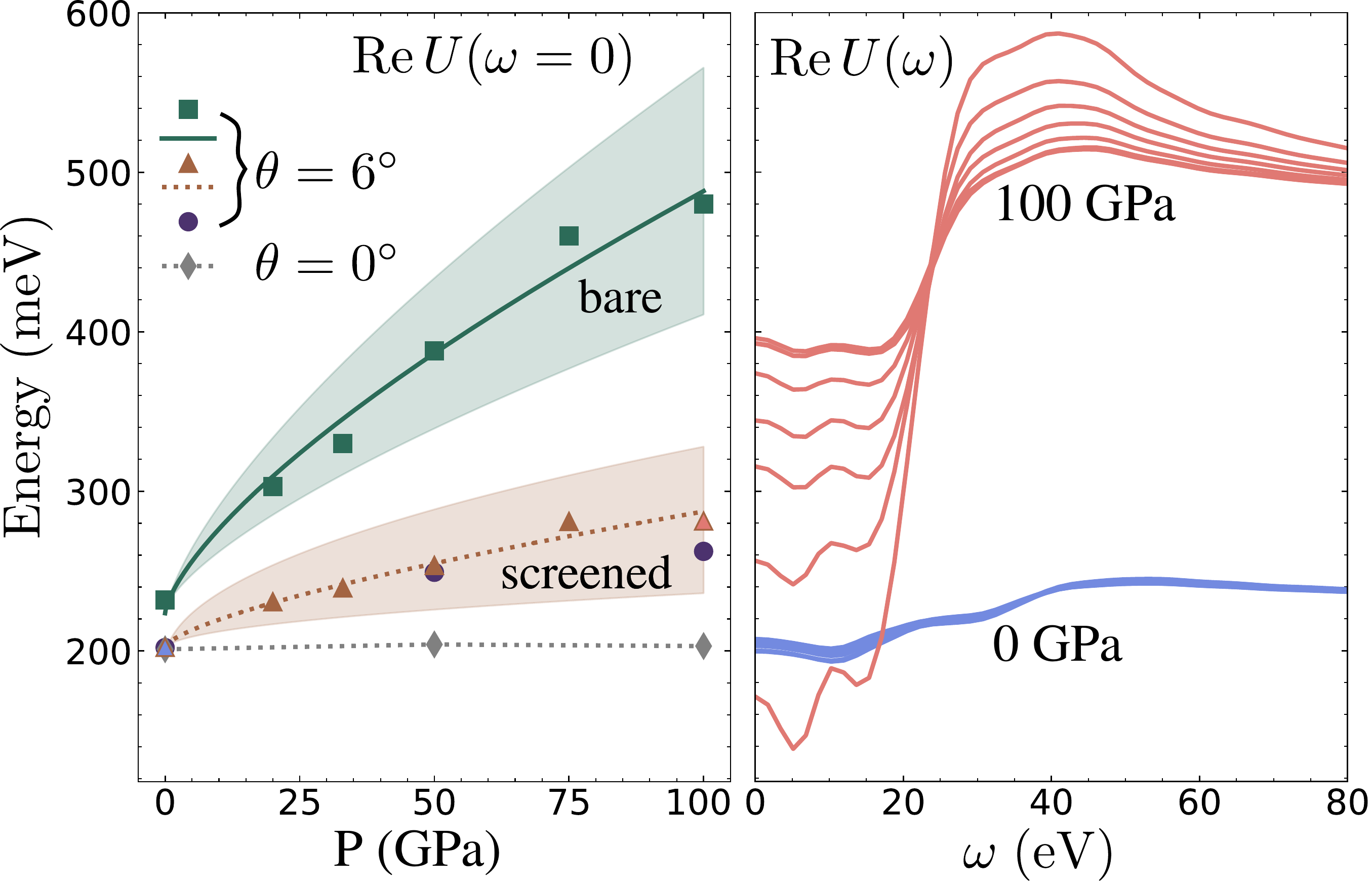}
    \caption{Left: Screened and bare $U(\omega=0)$ as a function of pressure for twisted bilayer graphene and screened for ideal bilayer graphene. Curve labeled with orange triangles is computed in the basis of DFT subspace orbitals, and the curves labeled with violet circles are in the QP basis. The lines connecting the mean values of bare and screened $U(\omega=0)$ are guide for the eyes. Stochastic error in determining the $U$ is smaller than the marker size. The shaded green and orange area provides the standard deviation from a mean value due to the difference of $U(\omega=0)$ for individual states within a correlated subspace.  Right: Frequency dependence of the screened $U(\omega)$ computed in the KS basis at 0 (red) and 100 (blue) GPa pressures. Different curves within one color represent individual 8 states within a correlated subspace. The colors of the curves match the triangles' colors from the left panel at 0 and 100 GPa. }
    \label{fig:U_pressure}
\end{figure}

Computed for the Dirac point states, see Eq.~\eqref{U}, the effective interaction $U(\omega)$, is screened by all the weakly interacting electrons confined to both monolayers. To account for the dynamical screening, we employ a set of random states which sample the dynamics of all weakly correlated states. The resulting $U(\omega)$ converges extremely fast (only 8 random vectors are necessary to yield a negligible stochastic error of $1$~meV for a wide range of frequencies -- see Fig~\ref{fig:U_pressure}) with minimal computational requirements ($<$120~CPU-hours; see section~\ref{sec:scRPA}). The frequency dependence of $U(\omega)$ is illustrated in Figure~\ref{fig:U_pressure}. In practice, similar to Refs.~\cite{miyake2009ab,ma2021quantum}, we will discuss the static limit ($\omega\to0$) since there is no elegant mathematical framework to solve the effective Hamiltonian with the frequency-dependent $U$.
For the tBLG at 0~GPa the total screened interaction $U(\omega\to0)=202$~meV. This is identical to the result for the uncompressed ideal bilayer at $0$~GPa where $U(\omega\to0)=201$~meV. Clearly, our initial assertion holds, and the $6^{\circ}$-tBLG indeed behaves like an ordinary bilayer at ambient conditions.

Under pressure, the screened on-site interaction in tBLG increases (as much as by $\approx 40\%$ for the highest compression ). At 100~GPa, the screened $U~=~282$~meV corresponds to a large ratio $U/t\approx7$ that suggests strongly correlated behavior. This is in striking contrast to the ideal bilayer, for which the $U$ parameter remains practically constant (see Figure~\ref{fig:U_pressure}). One can intuitively understand the difference in the on-site interaction behavior in tBLG and ideal bilayer from the charge density distributions of the DP states: in tBLG, the electrons in the DP states are trapped in the shallow moir\'{e} potential and become confined between the monolayers. The pressure-induced localization leads to increased on-site interaction; the weakly correlated states, on the other hand, remain spread over the entire system. In contrast, the DP electrons in the ideal bilayer experience no localizing potential.  Even under compression they remain fully delocalized in the in-plane direction, and their distribution is little affected along the normal to the bilayer. Further, the twist is a necessary prerequisite of coupling between layers since it allows to form energetically degenerate  states at K and K' points of the Brillouin zone. While in the ideal bilayer, K points of two Brillouin zones appear on top of each other forcing the energetic gap between corresponding states, and forbidding them from coupling. Hence, the on-site term in the ideal bilayer is insensitive to pressure. 

From the previous analysis (Section~\ref{sec:QPvsKS}) we saw that the Dyson orbitals $\{\psi\}$ are similar to the canonical KS states. Indeed, if we employ the approximate Dyson orbitals instead of the KS states, the $U$ parameter is only insignificantly smaller even at high pressures (cf.~Figure~\ref{fig:U_pressure}): at 100 GPa, the on-site term for the $\{\psi\}$ states is $260$~meV, and the $U/t$ is $\approx 6.5$. For this analysis, we use the mean values of the $U$ parameters. However, to emphasize that the individual states of the correlated subspace yield different values of $U$, we also provide the standard deviation with a shaded area (cf.~Figure~\ref{fig:U_pressure}, left panel). The spread of the individual $U$ values is particularly pronounced for higher pressures (cf.~Figure~\ref{fig:U_pressure}, right panel) and can be explained by lifting the degeneracy within the correlated subspace. In practice, QP and KS orbitals thus produce very similar $U$ interaction. This can be easily understood, given that $\{\psi\}$ and $\{ \phi\}$ are very similar.   

Finally, the right panel of Figure~\ref{fig:U_pressure} shows the frequency dependence of $U(\omega)$ for each DP state in tBLG at 0 and 100 GPa. Two main effects of the decreasing interlayer distance can be observed: a vertical shift of the entire $U(\omega)$ curve, and an eight-fold increase of the magnitude of the oscillations. 

To explore the $U(\omega)$ pressure behavior, we now turn to the analysis of the \textit{bare} and polarization terms,  $U^b$ and $U^p (\omega)$.  We find that the (average) values of the bare term are large: when the DFT states are used,  $U^b = 232$~meV at $0$GPa and it increases to $U^b = 480$~meV at 100 GPa (cf.~Figure~\ref{fig:U_pressure}). The approximate Dyson orbitals yield similar values of $230$~meV and $412$~meV at 0 and 100 GPa (not shown). Clearly, the difference between the bare terms partially accounts for the discrepancy between the average $U$ values for the $\{\phi\}$ and $\{\psi\}$ states.

The dynamical component of the on-site interaction, $U^p(\omega)$ contains the effect of the dynamical charge density fluctuations of all the weakly correlated states (outside of the correlated subspace); it is computed via s-cRPA (see the Methods sections). $U^p(\omega)$ increases under pressure and renormalizes the bare term: the low frequency limit $U^{p}(\omega\to0)$ at 0~GPa is $-30$~meV and at 100~GPa it is $-200$~meV. Apparently, this screening is not enough to entirely cancel out the $U^b$ contribution. Therefore, the total screened interaction $U(\omega\to0)$ is driven by the prevalent bare term and grows with compression. We also see that the magnitude of features in $U^p(\omega)$ curve increases as the (occupied) states shift closer to the Fermi level. Note that this result is independent of the $\{\phi\}$ or $\{\psi\}$ basis.\footnote{When computing the screening for the approximate Dyson orbitals, $ \psi$, the weakly correlated subspace remains identical, since $ \psi$ is composed of the combination of the quasi-degenerate states}

From the computational prospective we conclude that the stochastic approach is extremely efficient and yields $U(\omega)$ for even extremely large systems while requiring only minimal computational resources. This method enabled efficient downfolding and first principles calculations of the renormalized on-site interactions in undoped tBLG supercells.

\section{Discussion and Conclusions}
 
We presented and applied several new numerical developments within the stochastic many-body theory which efficiently treat even large systems. We illustrated the method on large scale many-body calculations for twisted bilayer graphene with nearly 9,000 valence electrons. We have expanded the stochastic computational toolkit to address the role of off-diagonal self-energy and the basis representation. Further, we have developed stochastic s-cRPA, enabling the downfolding of even giant systems onto model Hamiltonian problems. Our stochastic approaches are applicable to general systems and will find wide application in a wide variety of condensed matter problems.

Our $GW$ results show an excellent agreement with the experimental positions of the van Hove singularities. We also show the formation of electronic localization under compression of the tBLG, which is in agreement with available experimental data and indicate that the compression provides a unique path towards controlled coupling of monolayers and practical realization of moir\'e states. For systems that are weakly correlated at ambient conditions, the decreased interlayer spacing leads to the formation of flat bands associated with strong correlations. These localized states are found in the vicinity of the Fermi level. In contrast, the majority of the states (delocalized over the individual monolayers) are weakly correlated and remain practically unaffected by the compression.

We compare then effects of pressure to those by varying twist angle on the interplay between screening and electron-electron interaction. The previous theoretical investigations showed that the screening has a crucial role in reducing the on-site interactions, $U$, as the system approaches the magic twist angle. Thus, the correlations (characterized by large $U/t$ ratios) are primarily driven by the vanishing dispersion (i.e., $t\to0$) of the states near the Fermi level.\cite{lu2016local,stauber2016quasi,liu2020tuning,pizarro2019internal,goodwin2019attractive,zhu2101dynamical,vanhala2020constrained,calderon2020interactions} In contrast, our stochastic calculations reveal a different scenario: the dynamical screening of the on-site term, $U$, increases relatively slowly with pressure and the effect of the screening does not fully compensate the bare term. Due to the interlayer coupling, the localized states are strongly affected by compression, and the electronic correlation stems from both small $t$ and \textit{large} $U$.

Our calculations indicate that dynamical electronic correlations lead to only small changes in the single-particle orbitals. Consequently, the corresponding $U/t$ ratios in KS and QP basis are practically the same. 

By neglecting the graphene layer reconstruction (i.e., in the absence of corrugation) we overestimate the screening effect: structural relaxations in tBLG tend to separate the flat bands from the rest of the spectrum ($\sim20$~meV for small angles) \cite{nam2017lattice,angeli2018emergent,leconte2019relaxation,lin2020pressure, cantele2020structural} and increased gap would translate to a reduced static limit of the screening. At the same time, the effect of structural relaxations on screening in this system is likely to be small since their effect on bandstructure was shown to be prominent only for twist angles $<2^{\circ}$~\cite{yoo2019atomic,liang2020effect}. In addition, tBLG is usually encapsulated with hBN in high pressure experiments, effectively suppressing the out-of-plane relaxation. While further investigation of the role of structural reconstruction is needed, our methodology will likely play a critical role in performing such calculations.

Since the electron-electron interactions in the flat bands appear only mildly screened even at large compression, the electronic structure at high pressures is likely associated with robust insulating states.\cite{liu2020tuning} However, the absence of internal screening can be efficiently modified, e.g., by encapsulation and by extrinsic adjustable screening~\cite{pizarro2019internal,liu2020tuning}.

Our methodology is critical in providing the \emph{ab initio} information about internal screening effects. It informs future works combining dielectric materials and high pressures. Further, it opens a route to the theoretical understanding of precise control of quasiparticle states.

\section{Methods}
\subsection{Stochastic many-body theory}
To compute the QP energies and analyze the MB interactions, we employ a combination of MBPT and mapping of the selected (strongly correlated) subspace on the Hubbard model. Within MBPT, the central quantity is the self-energy, $\Sigma(\omega)$, which is a dynamical and non-local potential acting on a single QP state and incorporates all many-body effects. The QP energies correspond to the poles of the Green's function, $G$, representing a QP propagator that is expressed in terms of the Dyson series: $G^{-1}(\omega)=G_0^{-1}(\omega)-\Sigma(\omega)$, where $G_0$ is the reference (non-interacting) Green's function (GF). Here, the reference GF is taken from DFT calculations with PBE exchange-correlation functional\cite{perdew1996generalized}; $\Sigma$ is found using a perturbation expansion on top of $G_0$ and is responsible for capturing the dynamical correlation effects. 

Here, we employ the basis of single particle states, $\{\phi_j\}$, obtained from the ground state DFT calculations and the self-energy thus becomes a matrix composed of elements $\Sigma_{j,k}(\omega) \equiv \braket{\phi_j|\Sigma(\omega)|\phi_k}$. The QP energies, $\varepsilon$, are:
 \begin{equation}
    \epsilon_j = \epsilon_j^0 - v^{xc}_j+  \operatorname{Re}\left[\Sigma_{j,j}(\omega = \epsilon_j)\right] + \operatorname{Re}\left[\Delta_{j}(\omega = \epsilon_j)\right].
     \label{eqp}
 \end{equation}
where  $\varepsilon_j^0$ is the Kohn-Sham (KS) eigenvalue, $v^{xc}_j$ is the exchange-correlation potential, and $\Delta_{j}(\omega)$ comprises the coupling due to the off-diagonal elements of $\Sigma_{j,k}(\omega)$. The frequency dependent self-energy, $\Sigma(\omega)$, is evaluated at $\varepsilon$. We resort to the $GW$ approximation to the self-energy, which contains exchange and correlation parts; the latter is approximated by the dynamical effects of the charge density fluctuations due to the addition (removal) of an electron to (from) $\left| \phi_j\right\rangle$.\cite{Hedin1965,hybertsen1986electron,Aryasetiawan1998,martin2016interacting} 
In general, the off-diagonal contributions to the self-energy $\Delta_j$ capture the deviation of the $\{\phi\}$-states from the QP (Dyson) orbitals $\left|\psi_j(x)\right\rangle$\footnote{The Dyson orbital is defined by an overlap of a many-body wavefunction for the ground state of $N$ particles, $\Psi_0^{N}$, and $N\pm1$ particles state, $\Psi_j^{N}$ for the $j^{\rm th}$ excited state. For a hole in the $j^{\rm th}$ state, the Dyson orbital is $\psi_j(x)\equiv \sqrt{N} \Psi_j^{N-1}(\bar x_1,\bar x_2,\dots,\bar x_N)\Psi_0^{N}(\bar x_1,\bar x_2,\dots,\bar x_N,x)$, where $x_k$ is the spin-space coordinate of electron $k$ and one integrates over all coordinates with bar on top}. For common weakly correlated systems, the diagonal contributions, $\Sigma_{jj}$, strongly dominate while the off-diagonal terms are orders of magnitude smaller and can be neglected (i.e., $\Delta_j =0$).\cite{faleev2004all,kaplan2015off}  

Our first development efficiently expands the \textit{stochastic} methodology\cite{neuhauser2014breaking,Vlcek2018swift} and computes \textit{both types of contributions} using a single-step correction. The expectation values of $\Sigma_{jk}$ are sampled via decomposition of the Green's function into random vectors $\zeta$ spanning the occupied and unoccupied subspace and propagated backward and forward in time, hence representing particle and hole components of the time-ordered Green's function.
The resulting expression is, e.g. for $t<0$, $iG(\br,\bp,t) \equiv \{\zeta(\bp,t)\zeta(\br)\} $, where $\{\cdots\}$ denotes stochastic averaging and $\ket{\zeta(t)}\equiv e^{-i \hat H t}\ket{\zeta}$, and $\hat H$ is the system Hamiltonian. For non-interacting Green's function, $G_0$ ,(i.e., in the one shot correction scheme), the time evolution is governed by the underlying DFT Hamiltonian $\hat H_0$. 
The real-time sampling of the induced densities is performed by another set of random vectors $\eta$ representing the charge density fluctuations, i.e., $\delta n(\br,t) \approx |{{\eta}}(\br,t)|^2$. For nanoscale systems with thousands of atoms, $\sim 100$ samples suffice to represent the GF, with only $\sim10$ needed to represent $\delta n(\br,t)$.~\cite{vlcek2017stochastic,Vlcek2018swift,vlcek2019stochastic} 

The stochastic methodology capitalizes on the fact that the key quantities ($G$ and $W$) are determined by collective properties, which are inherently low-rank and captured by the dynamics of a few (random) states within the Hilbert space of single-particle states. This approach leads to a linear scaling algorithm that can treat thousands of atoms.\cite{neuhauser2014breaking,Vlcek2018swift} The new implementation expands this methodology and efficiently yields also $\Delta_j$ terms (the details are provided in section~\ref{offdiag}). Further, using the QP hamiltonian matrix (represented in the $\{\phi\}$-state basis in the  Eq.~\eqref{eq:Hqp}), we compute the QP orbitals $\psi$, corresponding to the first step of the self-consistent renormalization loop. 

\subsection{Off-diagonal self-energy\label{offdiag}}
The off-diagonal terms in the polarization self-energy have been implemented in our development version of the stochastic $GW$ code~\cite{Vlcek2018swift}. In the stochastic $GW$ formalism, the non-interacting Green's function $G_0$ and the screened Coulomb potential $W$ are sampled with two independent sets of random functions $\{\zeta\}$ and $\{\eta\}$ respectively. Additional set of random vectors is used for the  sparse  stochastic  compression in the time-ordering procedure. As a result, the expectation value for the polarization self-energy is a statistical estimator with a statistic error decreasing with number of random vectors as $1/\sqrt{N}$. A specific offdiagonal term of the self-energy has the following expression:
\begin{equation}
    \langle \phi_j |\Sigma_P| \phi_k \rangle  \simeq \frac{1}{N_{ \bar{\zeta}}} \sum_{\bar{\zeta}} \int \phi_j(\br)\zeta(\br,t)u_{\zeta,k}(\br,t) d^3\br
\label{eq:offdiag_sigma}    
\end{equation}
where $\simeq$ denotes that the expression is exact in the limit of $N_{\bar \zeta} \to \infty$. The function $\zeta$ at time $t$ is defined with a help of the time evolution operator $U_{0}(t) \equiv e^{-i H_0 t}$ and the projector $P_\mu(t)$ that selects the states above or below the chemical potential, $\mu$, depending on the sign of $t$.:
\begin{equation}\label{tevolve}
|\zeta(t)\rangle~\equiv~U_{0}(t) P_\mu(t)|{\zeta}\rangle,    
\end{equation}
The $\zeta$ vectors in the occupied and unoccupied subspace are propagated backward or forward in time and contribute selectively to the hole and particle non-interacting Green's functions.

In Eq.\eqref{eq:offdiag_sigma} the overlap with $\phi_k$ is hidden within $u_{\zeta,k}(\br,t)$~--~an induced charge density potential:
\begin{equation}
u_{\zeta,k}(\br,t) = \int W_P(\br,\br',t) \bar{\zeta}(\br')\phi_k(\br')d^3\br',
\label{eff_potential}
\end{equation}
$u_{\zeta,k}(\br,t)$ represents the  time-ordered potential of the response to the charge addition or removal. $u_{\zeta,k}$ is calculated from the retarded response potential, which is $\tilde u_{\zeta,k} = \int \tilde W_P (\br,\br',t) \bar{\zeta}(\br')\phi_k(\br')d^3\br'$ with a subsequent time-ordering procedure.\cite{FetterWalecka,vlcek2017stochastic,Vlcek2018swift}

Further, the retarded response is related to the time-evolved charge density 
$
\delta n (\br,t) \equiv \frac{1}{\lambda} \left[n(\br,t) - n(\br, 0)  \right]  $
induced by a scaled perturbing potential $\delta v = \lambda [ \nu(\br,\br'){\bar\zeta}(\br')\phi_k(\br')]$. Here $\lambda$ is selected to be small, i.e., inducing a linear response; in our case we chose $\lambda=10^{-4}$~a.u.. The retarded response becomes:
\begin{align}\label{eqn:u}
&\tilde u_{\zeta,k}(\br,t) = \nonumber\\
\iiint \nu(\br,\br'')& \chi(\br'',\br''',t) \delta v(\br''',\br')\dd\br'\dd \br''\dd\br''' \nonumber \\
&\equiv \int \nu(\br,\bp) \delta n(\bp,t) {\dd}\bp
\end{align}
Instead of computing $\delta n(\br,t)$ by a sum over single-particle states,  we employ the set of random vectors $\left\{ \eta \right\}$ confined to the occupied subspace. This reduces tremendously a cost of the computation $\tilde u_{\zeta,k}$. Time-dependent density $n(\br,t)$  is thus\cite{neuhauser2014breaking,vlcek2017stochastic,Vlcek2018swift,gao2015sublinear,Rabani2015,Neuhauser_2016} 
\begin{equation}
n(\br,t)=\lim_{N_\eta \to \infty} \frac{1}{N_{\eta}}\sum_{l}^{N_{\eta}}|\eta_l(\br,t)|^2,
\label{TDdensity}
\end{equation}
where $\eta_l$ is propagated in time using $U_{0,t}$
\begin{equation}\label{tpropeta_Un}
    \left| \eta(t)\right\rangle = U_{0,t} [n(t)] \left| \eta\right\rangle.
\end{equation} 

Note, the perturbing potential $\delta v$ depends explicitly on the state $\phi_k$ and the $\bar\zeta$ vector, which samples the whole Hilbert space. 

We employ RPA, i.e., performing evolution within the time-dependent Hartree approximation\cite{Baroni2001,BaerNeuhauser2004,Neuhauser2005}, to calculate $\tilde u_{\zeta,k}$.

\subsection{Stochastic Hamiltonian downfolding}\label{sec:scRPA}
The second new development presented in this work enables efficient Hamiltonian downfolding and extracting effective parameters for model approaches using first principles. We identify correlated states using the QP orbitals analysis (discussed in the main text) and map the corresponding subspace on a dynamically screened Hubbard model\cite{hubbard1963electron,springer1998frequency,kotani2000ab}:
\begin{equation}\label{eq:hub_ham}
     \hat{H}= - \sum_{i,j, \sigma}t_{ij}\hat{c}^{\dagger}_{i\sigma}\hat{c}_{j\sigma}
 + \sum_{i\sigma}^{} U\hat{n}^{\dagger}_{i\uparrow}\hat{n}_{i\downarrow},
\end{equation}
where $\hat{c}^{\dagger}_{i\sigma}$, $\hat{c}_{i,\sigma}$ are creation and annihilation operators, $\hat{n}_{i\sigma}=\hat{c}^{\dagger}_{i\sigma}\hat{c}_{i\sigma}$ is a particle number operator.

In practice, we extract the hopping and on-site Coulomb terms, $t$ and $U$ from the first-principles calculations. The latter is :\cite{miyake2009ab}  
\begin{align}\label{U}
     U(\omega) = 
     \frac{1}{N}\sum_{i=1}^{N}\iint  d\br d\br' |\varphi_i(\br)|^2\tilde W(\br,\br',\omega)|\varphi_i(\br')|^2.
\end{align}
Here, $\{\varphi_i\}$ is a set of KS or QP states spanning the correlated subspace (represented by $\{\phi\}$ and $\{\psi\}$ sets respectively). They are subject to Coulomb interaction $\tilde W$ that contains both bare (instantaneous) and screened (i.e., dynamical) terms. Symbolically, the interaction is $\tilde W = \nu + \nu \tilde\chi \nu, $ where $\nu$ is the bare Coulomb kernel and $\tilde\chi$ is the polarizability due to electronic states orthogonal to the $\{\varphi_i\}$-subspace that contains the DP states. 

The cost of the conventional calculation is huge, as it needs to be evaluated by considering all possible transitions between occupied and unoccupied states (see below). Calculations for large systems (such as those studied here) were thus out of reach. In contrast, we propose an efficient approach in which  Eq.~\eqref{U} is evaluated \textit{stochastically} within the constrained random-phase approximation (cRPA): the real-time formalism samples the dynamics of all occupied states using a new set of random vectors confined to the occupied subspace and orthogonal to $\{\varphi_i\}$. The separation of the Hilbert space employs our recently developed decomposition technique\cite{romanova2020decomposition}. This technique is computationally \textit{inexpensive}: $U(\omega)$ screened by 4364 valence bands require \textit{merely} $<$120 CPU$\cdot$hrs on a 2.5~GHz processor.\footnote{The testing configuration was AMD EPYC 7502 with 2.5~GHz frequency using 10 out of 32 physical cores. The total computational time was 9.6~hrs.} This methodology thus enables Hamiltonian downfolding even for extremely large systems. The details of the implementation are provided in the next sections on the bare Coulomb interaction (\ref{sec:Ub}) and its dynamical screening evaluated by s-cRPA (\ref{cRPA}).

~\\
\subsection{The effective bare Coulomb interaction\label{sec:Ub}}

We calculate the bare effective interaction parameter, the Hubbard $U^b$, both in a basis of KS or QP wavefunctions (see discussion on the orbital construction in Sec.~\ref{results1}) of a chosen subspace of $N=8$ states:
\begin{equation}\label{eq:Ub}
U^{b} =    \frac{1}{N}\sum_{i=1}^{N}\iint d\br d\br' |\varphi_i(\br)|^2\nu(\br,\br')|\varphi_i(\br')|^2, 
\end{equation}
where $\nu(\br,\br')$ is a bare Coulomb kernel.
The full Hubbard is $U(\omega)$ is given by Eq.~\eqref{U} and contains, besides $U^b$, also the dynamical screened \textit{polarization} term. The latter part is computed stochastically as detailed below, in Sec.~\ref{cRPA}.

~\\
\subsection{Stochastic constrained RPA (s-cRPA)\label{cRPA}}
Here we discuss the implementation of the dynamical Hubbard term, $U(\omega) = U^b + U^p(\omega)$, where the latter is: \begin{align}
     U^p(\omega) = 
     \frac{1}{N}\sum_{i=1}^{N}\iint  d\br d\br' |\varphi_i(\br)|^2\tilde W_P(\br,\br',\omega)|\varphi_i(\br')|^2.
\end{align}
The polarization operator $\tilde W_P =\nu \tilde\chi \nu$ is computed by the stochastic constrained random-phase approximation (s-cRPA). The key idea is to capture the effect of the entire system on the correlated electrons in states $\{\varphi\}$ described by the (downfolded) Hubbard Hamiltonian Eq.\eqref{eq:hub_ham}. In practice, one accounts for the screening through a projection on the subspace, which \textit{excludes} all correlated states $\{\varphi\}$. In cRPA, $\tilde W_P$ thus contains contributions of the induced density fluctuations in the weakly correlated portion of the system.

Conventional techniques evaluate $\tilde W_P=\nu \tilde\chi \nu$ in frequency domain by the sum over all single-particle transitions outside the correlated subspace ($i,j \notin \{ \varphi \}$), requiring operation on both the entire occupied and unoccupied space~\cite{miyake2009ab}: 

\begin{align}
    \tilde\chi(\br,\br',\omega) = \sum_i^{occ} \sum_j^{unocc} \phi_i(\br)\phi_i(\br')^*\phi_j(\br)^*\phi_j(\br') \times \nonumber\\
    \times \left( \frac{1}{\omega - \varepsilon_j+\varepsilon_i+i\lambda}-\frac{1}{\omega+\varepsilon_j-\varepsilon_i-i\lambda}  \right). 
\end{align} 
Hence, these calculations become expensive for large systems.
 In contrast, we compute $\tilde W_P$ term stochastically in real-time domain:
  \begin{equation}
    \langle \varphi_j \varphi_j|\tilde W_P| \varphi_j \varphi_j\rangle  \simeq  \int |\varphi_j(\br)|^2 \tilde u(\br,t) d^3\br.
\label{eq:wp}    
\end{equation}
This expression is computed by time-ordering from the retarded charge density potential:
\begin{equation}
    \tilde u^r(\br,t) =\int \nu(\br,\bp) \delta \tilde n(\bp,t) {\dd}\bp.
\end{equation}
where $\delta \tilde n$ is the induced charge density in the weakly correlated subspace perturbed by a potential due to $|\varphi_j(\br)|^2$. This is formally equivalent to Eq.~\ref{eff_potential} (representing the action of the self-energy in the $GW$ approximation). In practice, the density is constructed from random vectors $\{\tilde \eta\}$:
\begin{equation}
    \ket{\tilde\eta} = \left(1-P_\varphi\right)\ket{\eta}
\end{equation}
where the $\{\eta\}$-states are described in Sec.~\ref{offdiag} and $P_\varphi$ is the projection operator on the $\{ \varphi\}$-subspace:
\begin{equation}\label{projectorphi_occ}
    P_\varphi = \sum_{k\in \left\{\varphi\right\} } f_k \left| k\middle\rangle \middle \langle k \right|.
\end{equation}
where $f_k$ is the occupation of state $\ket k$. Note that the time evolution of $\{\tilde \eta\}$ vectors follows Eq.~\ref{tpropeta_Un}, which depends on the \textit{total} density. For details of the time-evolution of subspaces, see Ref.~\onlinecite{romanova2020decomposition}.

The method is implemented alongside the stochastic $GW$ formalism and both can be evaluated at once. However, in practice, the statistical error in $U^p(\omega)$ is orders of magnitude smaller since: (i) it stems from \textit{one} random sampling of $W$; in contrast, the $GW$ self-energy suffers from larger statistical errors due to the additional random vectors sampling the Green's function. (ii) it contains only contributions of states orthogonal to those which it is acting on; as a result, the dynamics is ``well-behaved'' and characterized by only a few dominant (resonant) frequencies which can be efficiently sampled by a small number of random vectors.

\subsection{Equilibrium geometry and equation of state}\label{sec:eos}
The tBLG cells at a specific out-of-plane pressure have been approximated using the interlayer distance of the ideal bilayer graphene in the Bernal stacking at a corresponding pressure. All the calculations for the ideal bilayer graphene have been performed using hexagonal unit cell in QuantumESPRESSO code\cite{QE2017} and Tkatchenko-Scheffler's total energy Van der Waals corrections\cite{TS_2009} and Effective Screening Medium Method\cite{Otani_2006}. Troullier-Martins pseudopotentials\cite{TroullierMartins1991}, and the PBE\cite{PerdewWang} functional have been employed.
To calculate the pressure-distance curves in the ideal bilayer, we have fitted the total energy $E$ as a function of the volume $V$  with the Murnaghan equation of state\cite{murnaghan1944}.
\begin{equation}
E(V) = E(V_0) + \frac{B_0V}{B'_0}\left[\frac{(V_0/V)^{B'_0}}{B'_0-1}+1\right] - \frac{V_0B_0}{B'_0-1} ,
\label{eq:eos}    
\end{equation}
where $V=S \cdot z$ is the volume confined by two graphene layers, $S=a_{lat}^2$ is a surface of the layer, and $z$ is the interlayer distance. $S$ was kept constant using the equilibrium lattice parameter $a_{lat}=2.464$~\AA, while $z$ was varied. The neglect of the pressure-induced in plane expansion is, in part, justified by the large anisotropy of the bulk modulus in the in- and out-of-plane directions.\cite{yankowitz2018dynamic} $B_0$ and $B'_0$ are the bulk modulus and its pressure derivative at the equilibrium volume $V_0$. The resulting fit and fitter parameters are provided in the Supplemental Information. Using fitted parameters $P(z)$ pressure - distance curves were calculated as a derivative $P = dE/dV$:

\begin{equation}
    P = \frac{B_0}{B'_0}\left[1 - \left(\frac{V0}{S\cdot z}\right)^{B'_0}\right]
\end{equation}

\subsection{Starting point DFT calculations}
The starting-point calculations are performed with the density functional theory (DFT) in a real-space implementation, employing regular grids, Troullier-Martins pseudopotentials,\cite{TroullierMartins1991} and the PBE\cite{PerdewWang} functional for exchange and correlation. We investigate tBLG infinite systems using modified periodic boundary conditions with Coulomb interaction cutoffs.\cite{Rozzi_2006} To converge the occupied $H_0$ eigenvalues to $< 5$~meV, we use a kinetic energy cutoff of 26~$E_h$ and $192\times 132 \times 66$ real-space grid with the step of $0.4 \times 0.4 \times 0.5$~$a_0$, where the $z$-direction is aligned with the normal of the bilayer plane.

\subsection{$G_0W_0$ calculations}
The $GW$ calculations were performed using a development version of the StochasticGW code.\cite{neuhauser2014breaking, Vlcek2018swift, vlcek2017stochastic}
The calculations employ an additional set of 20,000 random vectors for the sparse stochastic compression used for time-ordering of $\tilde u_{\zeta}$.\cite{Vlcek2018swift} The sampling of the Green's function $G$ was performed using $N_{\zeta}$=500 random vectors. $N_{\eta}$=8 was used to sample the induced charge density.\cite{Vlcek2018swift} The final stochastic error on quasiparticle energies is $\le 20$~meV.
The time propagation of the induced charge density was performed for a maximum propagation time of 50~a.u., with the time-step of 0.05~a.u.
\section{Data availability}
All the data supporting the results of this study are available upon reasonable request to the corresponding author
\section{Code availability}
The public version of the stochastic $GW$ code is available at www.stochasticGW.com. We used a development version of the stochastic $GW$ to perform calculations, which will be released soon and is available upon reasonable request.

\begin{acknowledgments}
The development of the off-diagonal self-energy and s-cRPA (VV) was supported by the NSF through NSF CAREER award Grant No. DMR-1945098. The development of the downfolding and the implementation (M.R.)~were supported by the Materials
Research Science and Engineering Centers (MRSEC)
Program through Grant No. DMR-1720256 (Seed
Program). M.R.'s work was also supported by the NSF Quantum Foundry through Q-AMASE-i
program Award No. DMR-1906325. The calculations were performed as part of the XSEDE\cite{Towns_2014} computational Project No.~TG-CHE180051. Use was made of computational facilities purchased with
funds from the National Science Foundation (CNS-1725797)
and administered by the Center for Scientific Computing
(CSC). The CSC is supported by the California NanoSystems
Institute and the Materials Research Science and Engineering
Center (MRSEC; NSF DMR-1720256) at UC Santa Barbara.
\end{acknowledgments}

\subsection{Contributions}
M.R. conducted the research work under the guidance of V.V. All authors contributed and reviewed the manuscript.
\subsection{Competing interests}
The authors declare no competing interests.
%\bibliography{biblio}

\begin{thebibliography}{128}%
\makeatletter
\providecommand \@ifxundefined [1]{%
 \@ifx{#1\undefined}
}%
\providecommand \@ifnum [1]{%
 \ifnum #1\expandafter \@firstoftwo
 \else \expandafter \@secondoftwo
 \fi
}%
\providecommand \@ifx [1]{%
 \ifx #1\expandafter \@firstoftwo
 \else \expandafter \@secondoftwo
 \fi
}%
\providecommand \natexlab [1]{#1}%
\providecommand \enquote  [1]{``#1''}%
\providecommand \bibnamefont  [1]{#1}%
\providecommand \bibfnamefont [1]{#1}%
\providecommand \citenamefont [1]{#1}%
\providecommand \href@noop [0]{\@secondoftwo}%
\providecommand \href [0]{\begingroup \@sanitize@url \@href}%
\providecommand \@href[1]{\@@startlink{#1}\@@href}%
\providecommand \@@href[1]{\endgroup#1\@@endlink}%
\providecommand \@sanitize@url [0]{\catcode `\\12\catcode `\$12\catcode
  `\&12\catcode `\#12\catcode `\^12\catcode `\_12\catcode `\%12\relax}%
\providecommand \@@startlink[1]{}%
\providecommand \@@endlink[0]{}%
\providecommand \url  [0]{\begingroup\@sanitize@url \@url }%
\providecommand \@url [1]{\endgroup\@href {#1}{\urlprefix }}%
\providecommand \urlprefix  [0]{URL }%
\providecommand \Eprint [0]{\href }%
\providecommand \doibase [0]{https://doi.org/}%
\providecommand \selectlanguage [0]{\@gobble}%
\providecommand \bibinfo  [0]{\@secondoftwo}%
\providecommand \bibfield  [0]{\@secondoftwo}%
\providecommand \translation [1]{[#1]}%
\providecommand \BibitemOpen [0]{}%
\providecommand \bibitemStop [0]{}%
\providecommand \bibitemNoStop [0]{.\EOS\space}%
\providecommand \EOS [0]{\spacefactor3000\relax}%
\providecommand \BibitemShut  [1]{\csname bibitem#1\endcsname}%
\let\auto@bib@innerbib\@empty
%</preamble>
\bibitem [{\citenamefont {Brooks}\ \emph {et~al.}(2020)\citenamefont {Brooks},
  \citenamefont {Weng}, \citenamefont {Taylor},\ and\ \citenamefont
  {Vlcek}}]{brooks2020stochastic}%
  \BibitemOpen
  \bibfield  {author} {\bibinfo {author} {\bibfnamefont {J.}~\bibnamefont
  {Brooks}}, \bibinfo {author} {\bibfnamefont {G.}~\bibnamefont {Weng}},
  \bibinfo {author} {\bibfnamefont {S.}~\bibnamefont {Taylor}},\ and\ \bibinfo
  {author} {\bibfnamefont {V.}~\bibnamefont {Vlcek}},\ }\bibfield  {title}
  {\bibinfo {title} {Stochastic many-body perturbation theory for moir{\'e}
  states in twisted bilayer phosphorene},\ }\href@noop {} {\bibfield  {journal}
  {\bibinfo  {journal} {Journal of Physics: Condensed Matter}\ }\textbf
  {\bibinfo {volume} {32}},\ \bibinfo {pages} {234001} (\bibinfo {year}
  {2020})}\BibitemShut {NoStop}%
\bibitem [{\citenamefont {Cao}\ \emph {et~al.}(2018{\natexlab{a}})\citenamefont
  {Cao}, \citenamefont {Fatemi}, \citenamefont {Demir}, \citenamefont {Fang},
  \citenamefont {Tomarken}, \citenamefont {Luo}, \citenamefont
  {Sanchez-Yamagishi}, \citenamefont {Watanabe}, \citenamefont {Taniguchi},
  \citenamefont {Kaxiras} \emph {et~al.}}]{cao2018correlated}%
  \BibitemOpen
  \bibfield  {author} {\bibinfo {author} {\bibfnamefont {Y.}~\bibnamefont
  {Cao}}, \bibinfo {author} {\bibfnamefont {V.}~\bibnamefont {Fatemi}},
  \bibinfo {author} {\bibfnamefont {A.}~\bibnamefont {Demir}}, \bibinfo
  {author} {\bibfnamefont {S.}~\bibnamefont {Fang}}, \bibinfo {author}
  {\bibfnamefont {S.~L.}\ \bibnamefont {Tomarken}}, \bibinfo {author}
  {\bibfnamefont {J.~Y.}\ \bibnamefont {Luo}}, \bibinfo {author} {\bibfnamefont
  {J.~D.}\ \bibnamefont {Sanchez-Yamagishi}}, \bibinfo {author} {\bibfnamefont
  {K.}~\bibnamefont {Watanabe}}, \bibinfo {author} {\bibfnamefont
  {T.}~\bibnamefont {Taniguchi}}, \bibinfo {author} {\bibfnamefont
  {E.}~\bibnamefont {Kaxiras}}, \emph {et~al.},\ }\bibfield  {title} {\bibinfo
  {title} {Correlated insulator behaviour at half-filling in magic-angle
  graphene superlattices},\ }\href@noop {} {\bibfield  {journal} {\bibinfo
  {journal} {Nature}\ }\textbf {\bibinfo {volume} {556}},\ \bibinfo {pages}
  {80} (\bibinfo {year} {2018}{\natexlab{a}})}\BibitemShut {NoStop}%
\bibitem [{\citenamefont {Cao}\ \emph {et~al.}(2018{\natexlab{b}})\citenamefont
  {Cao}, \citenamefont {Fatemi}, \citenamefont {Fang}, \citenamefont
  {Watanabe}, \citenamefont {Taniguchi}, \citenamefont {Kaxiras},\ and\
  \citenamefont {Jarillo-Herrero}}]{cao2018unconventional}%
  \BibitemOpen
  \bibfield  {author} {\bibinfo {author} {\bibfnamefont {Y.}~\bibnamefont
  {Cao}}, \bibinfo {author} {\bibfnamefont {V.}~\bibnamefont {Fatemi}},
  \bibinfo {author} {\bibfnamefont {S.}~\bibnamefont {Fang}}, \bibinfo {author}
  {\bibfnamefont {K.}~\bibnamefont {Watanabe}}, \bibinfo {author}
  {\bibfnamefont {T.}~\bibnamefont {Taniguchi}}, \bibinfo {author}
  {\bibfnamefont {E.}~\bibnamefont {Kaxiras}},\ and\ \bibinfo {author}
  {\bibfnamefont {P.}~\bibnamefont {Jarillo-Herrero}},\ }\bibfield  {title}
  {\bibinfo {title} {Unconventional superconductivity in magic-angle graphene
  superlattices},\ }\href@noop {} {\bibfield  {journal} {\bibinfo  {journal}
  {Nature}\ }\textbf {\bibinfo {volume} {556}},\ \bibinfo {pages} {43}
  (\bibinfo {year} {2018}{\natexlab{b}})}\BibitemShut {NoStop}%
\bibitem [{\citenamefont {Yankowitz}\ \emph {et~al.}(2019)\citenamefont
  {Yankowitz}, \citenamefont {Chen}, \citenamefont {Polshyn}, \citenamefont
  {Zhang}, \citenamefont {Watanabe}, \citenamefont {Taniguchi}, \citenamefont
  {Graf}, \citenamefont {Young},\ and\ \citenamefont
  {Dean}}]{yankowitz2019tuning}%
  \BibitemOpen
  \bibfield  {author} {\bibinfo {author} {\bibfnamefont {M.}~\bibnamefont
  {Yankowitz}}, \bibinfo {author} {\bibfnamefont {S.}~\bibnamefont {Chen}},
  \bibinfo {author} {\bibfnamefont {H.}~\bibnamefont {Polshyn}}, \bibinfo
  {author} {\bibfnamefont {Y.}~\bibnamefont {Zhang}}, \bibinfo {author}
  {\bibfnamefont {K.}~\bibnamefont {Watanabe}}, \bibinfo {author}
  {\bibfnamefont {T.}~\bibnamefont {Taniguchi}}, \bibinfo {author}
  {\bibfnamefont {D.}~\bibnamefont {Graf}}, \bibinfo {author} {\bibfnamefont
  {A.~F.}\ \bibnamefont {Young}},\ and\ \bibinfo {author} {\bibfnamefont
  {C.~R.}\ \bibnamefont {Dean}},\ }\bibfield  {title} {\bibinfo {title} {Tuning
  superconductivity in twisted bilayer graphene},\ }\href@noop {} {\bibfield
  {journal} {\bibinfo  {journal} {Science}\ }\textbf {\bibinfo {volume}
  {363}},\ \bibinfo {pages} {1059} (\bibinfo {year} {2019})}\BibitemShut
  {NoStop}%
\bibitem [{\citenamefont {Yankowitz}\ \emph {et~al.}(2018)\citenamefont
  {Yankowitz}, \citenamefont {Jung}, \citenamefont {Laksono}, \citenamefont
  {Leconte}, \citenamefont {Chittari}, \citenamefont {Watanabe}, \citenamefont
  {Taniguchi}, \citenamefont {Adam}, \citenamefont {Graf},\ and\ \citenamefont
  {Dean}}]{yankowitz2018dynamic}%
  \BibitemOpen
  \bibfield  {author} {\bibinfo {author} {\bibfnamefont {M.}~\bibnamefont
  {Yankowitz}}, \bibinfo {author} {\bibfnamefont {J.}~\bibnamefont {Jung}},
  \bibinfo {author} {\bibfnamefont {E.}~\bibnamefont {Laksono}}, \bibinfo
  {author} {\bibfnamefont {N.}~\bibnamefont {Leconte}}, \bibinfo {author}
  {\bibfnamefont {B.~L.}\ \bibnamefont {Chittari}}, \bibinfo {author}
  {\bibfnamefont {K.}~\bibnamefont {Watanabe}}, \bibinfo {author}
  {\bibfnamefont {T.}~\bibnamefont {Taniguchi}}, \bibinfo {author}
  {\bibfnamefont {S.}~\bibnamefont {Adam}}, \bibinfo {author} {\bibfnamefont
  {D.}~\bibnamefont {Graf}},\ and\ \bibinfo {author} {\bibfnamefont {C.~R.}\
  \bibnamefont {Dean}},\ }\bibfield  {title} {\bibinfo {title} {Dynamic
  band-structure tuning of graphene moir{\'e} superlattices with pressure},\
  }\href@noop {} {\bibfield  {journal} {\bibinfo  {journal} {Nature}\ }\textbf
  {\bibinfo {volume} {557}},\ \bibinfo {pages} {404} (\bibinfo {year}
  {2018})}\BibitemShut {NoStop}%
\bibitem [{\citenamefont {Sharpe}\ \emph {et~al.}(2019)\citenamefont {Sharpe},
  \citenamefont {Fox}, \citenamefont {Barnard}, \citenamefont {Finney},
  \citenamefont {Watanabe}, \citenamefont {Taniguchi}, \citenamefont
  {Kastner},\ and\ \citenamefont {Goldhaber-Gordon}}]{sharpe2019emergent}%
  \BibitemOpen
  \bibfield  {author} {\bibinfo {author} {\bibfnamefont {A.~L.}\ \bibnamefont
  {Sharpe}}, \bibinfo {author} {\bibfnamefont {E.~J.}\ \bibnamefont {Fox}},
  \bibinfo {author} {\bibfnamefont {A.~W.}\ \bibnamefont {Barnard}}, \bibinfo
  {author} {\bibfnamefont {J.}~\bibnamefont {Finney}}, \bibinfo {author}
  {\bibfnamefont {K.}~\bibnamefont {Watanabe}}, \bibinfo {author}
  {\bibfnamefont {T.}~\bibnamefont {Taniguchi}}, \bibinfo {author}
  {\bibfnamefont {M.}~\bibnamefont {Kastner}},\ and\ \bibinfo {author}
  {\bibfnamefont {D.}~\bibnamefont {Goldhaber-Gordon}},\ }\bibfield  {title}
  {\bibinfo {title} {Emergent ferromagnetism near three-quarters filling in
  twisted bilayer graphene},\ }\href@noop {} {\bibfield  {journal} {\bibinfo
  {journal} {Science}\ }\textbf {\bibinfo {volume} {365}},\ \bibinfo {pages}
  {605} (\bibinfo {year} {2019})}\BibitemShut {NoStop}%
\bibitem [{\citenamefont {Lu}\ \emph {et~al.}(2019)\citenamefont {Lu},
  \citenamefont {Stepanov}, \citenamefont {Yang}, \citenamefont {Xie},
  \citenamefont {Aamir}, \citenamefont {Das}, \citenamefont {Urgell},
  \citenamefont {Watanabe}, \citenamefont {Taniguchi}, \citenamefont {Zhang}
  \emph {et~al.}}]{lu2019superconductors}%
  \BibitemOpen
  \bibfield  {author} {\bibinfo {author} {\bibfnamefont {X.}~\bibnamefont
  {Lu}}, \bibinfo {author} {\bibfnamefont {P.}~\bibnamefont {Stepanov}},
  \bibinfo {author} {\bibfnamefont {W.}~\bibnamefont {Yang}}, \bibinfo {author}
  {\bibfnamefont {M.}~\bibnamefont {Xie}}, \bibinfo {author} {\bibfnamefont
  {M.~A.}\ \bibnamefont {Aamir}}, \bibinfo {author} {\bibfnamefont
  {I.}~\bibnamefont {Das}}, \bibinfo {author} {\bibfnamefont {C.}~\bibnamefont
  {Urgell}}, \bibinfo {author} {\bibfnamefont {K.}~\bibnamefont {Watanabe}},
  \bibinfo {author} {\bibfnamefont {T.}~\bibnamefont {Taniguchi}}, \bibinfo
  {author} {\bibfnamefont {G.}~\bibnamefont {Zhang}}, \emph {et~al.},\
  }\bibfield  {title} {\bibinfo {title} {Superconductors, orbital magnets and
  correlated states in magic-angle bilayer graphene},\ }\href@noop {}
  {\bibfield  {journal} {\bibinfo  {journal} {Nature}\ }\textbf {\bibinfo
  {volume} {574}},\ \bibinfo {pages} {653} (\bibinfo {year}
  {2019})}\BibitemShut {NoStop}%
\bibitem [{\citenamefont {Saito}\ \emph {et~al.}(2020)\citenamefont {Saito},
  \citenamefont {Ge}, \citenamefont {Watanabe}, \citenamefont {Taniguchi},\
  and\ \citenamefont {Young}}]{saito2020independent}%
  \BibitemOpen
  \bibfield  {author} {\bibinfo {author} {\bibfnamefont {Y.}~\bibnamefont
  {Saito}}, \bibinfo {author} {\bibfnamefont {J.}~\bibnamefont {Ge}}, \bibinfo
  {author} {\bibfnamefont {K.}~\bibnamefont {Watanabe}}, \bibinfo {author}
  {\bibfnamefont {T.}~\bibnamefont {Taniguchi}},\ and\ \bibinfo {author}
  {\bibfnamefont {A.~F.}\ \bibnamefont {Young}},\ }\bibfield  {title} {\bibinfo
  {title} {Independent superconductors and correlated insulators in twisted
  bilayer graphene},\ }\href@noop {} {\bibfield  {journal} {\bibinfo  {journal}
  {Nature Physics}\ }\textbf {\bibinfo {volume} {16}},\ \bibinfo {pages} {926}
  (\bibinfo {year} {2020})}\BibitemShut {NoStop}%
\bibitem [{\citenamefont {Choi}\ \emph {et~al.}(2019)\citenamefont {Choi},
  \citenamefont {Kemmer}, \citenamefont {Peng}, \citenamefont {Thomson},
  \citenamefont {Arora}, \citenamefont {Polski}, \citenamefont {Zhang},
  \citenamefont {Ren}, \citenamefont {Alicea}, \citenamefont {Refael} \emph
  {et~al.}}]{choi2019electronic}%
  \BibitemOpen
  \bibfield  {author} {\bibinfo {author} {\bibfnamefont {Y.}~\bibnamefont
  {Choi}}, \bibinfo {author} {\bibfnamefont {J.}~\bibnamefont {Kemmer}},
  \bibinfo {author} {\bibfnamefont {Y.}~\bibnamefont {Peng}}, \bibinfo {author}
  {\bibfnamefont {A.}~\bibnamefont {Thomson}}, \bibinfo {author} {\bibfnamefont
  {H.}~\bibnamefont {Arora}}, \bibinfo {author} {\bibfnamefont
  {R.}~\bibnamefont {Polski}}, \bibinfo {author} {\bibfnamefont
  {Y.}~\bibnamefont {Zhang}}, \bibinfo {author} {\bibfnamefont
  {H.}~\bibnamefont {Ren}}, \bibinfo {author} {\bibfnamefont {J.}~\bibnamefont
  {Alicea}}, \bibinfo {author} {\bibfnamefont {G.}~\bibnamefont {Refael}},
  \emph {et~al.},\ }\bibfield  {title} {\bibinfo {title} {Electronic
  correlations in twisted bilayer graphene near the magic angle},\ }\href@noop
  {} {\bibfield  {journal} {\bibinfo  {journal} {Nature Physics}\ }\textbf
  {\bibinfo {volume} {15}},\ \bibinfo {pages} {1174} (\bibinfo {year}
  {2019})}\BibitemShut {NoStop}%
\bibitem [{\citenamefont {Xie}\ \emph {et~al.}(2019)\citenamefont {Xie},
  \citenamefont {Lian}, \citenamefont {J{\"a}ck}, \citenamefont {Liu},
  \citenamefont {Chiu}, \citenamefont {Watanabe}, \citenamefont {Taniguchi},
  \citenamefont {Bernevig},\ and\ \citenamefont
  {Yazdani}}]{xie2019spectroscopic}%
  \BibitemOpen
  \bibfield  {author} {\bibinfo {author} {\bibfnamefont {Y.}~\bibnamefont
  {Xie}}, \bibinfo {author} {\bibfnamefont {B.}~\bibnamefont {Lian}}, \bibinfo
  {author} {\bibfnamefont {B.}~\bibnamefont {J{\"a}ck}}, \bibinfo {author}
  {\bibfnamefont {X.}~\bibnamefont {Liu}}, \bibinfo {author} {\bibfnamefont
  {C.-L.}\ \bibnamefont {Chiu}}, \bibinfo {author} {\bibfnamefont
  {K.}~\bibnamefont {Watanabe}}, \bibinfo {author} {\bibfnamefont
  {T.}~\bibnamefont {Taniguchi}}, \bibinfo {author} {\bibfnamefont {B.~A.}\
  \bibnamefont {Bernevig}},\ and\ \bibinfo {author} {\bibfnamefont
  {A.}~\bibnamefont {Yazdani}},\ }\bibfield  {title} {\bibinfo {title}
  {Spectroscopic signatures of many-body correlations in magic-angle twisted
  bilayer graphene},\ }\href@noop {} {\bibfield  {journal} {\bibinfo  {journal}
  {Nature}\ }\textbf {\bibinfo {volume} {572}},\ \bibinfo {pages} {101}
  (\bibinfo {year} {2019})}\BibitemShut {NoStop}%
\bibitem [{\citenamefont {Kerelsky}\ \emph {et~al.}(2019)\citenamefont
  {Kerelsky}, \citenamefont {McGilly}, \citenamefont {Kennes}, \citenamefont
  {Xian}, \citenamefont {Yankowitz}, \citenamefont {Chen}, \citenamefont
  {Watanabe}, \citenamefont {Taniguchi}, \citenamefont {Hone}, \citenamefont
  {Dean} \emph {et~al.}}]{kerelsky2019maximized}%
  \BibitemOpen
  \bibfield  {author} {\bibinfo {author} {\bibfnamefont {A.}~\bibnamefont
  {Kerelsky}}, \bibinfo {author} {\bibfnamefont {L.~J.}\ \bibnamefont
  {McGilly}}, \bibinfo {author} {\bibfnamefont {D.~M.}\ \bibnamefont {Kennes}},
  \bibinfo {author} {\bibfnamefont {L.}~\bibnamefont {Xian}}, \bibinfo {author}
  {\bibfnamefont {M.}~\bibnamefont {Yankowitz}}, \bibinfo {author}
  {\bibfnamefont {S.}~\bibnamefont {Chen}}, \bibinfo {author} {\bibfnamefont
  {K.}~\bibnamefont {Watanabe}}, \bibinfo {author} {\bibfnamefont
  {T.}~\bibnamefont {Taniguchi}}, \bibinfo {author} {\bibfnamefont
  {J.}~\bibnamefont {Hone}}, \bibinfo {author} {\bibfnamefont {C.}~\bibnamefont
  {Dean}}, \emph {et~al.},\ }\bibfield  {title} {\bibinfo {title} {Maximized
  electron interactions at the magic angle in twisted bilayer graphene},\
  }\href@noop {} {\bibfield  {journal} {\bibinfo  {journal} {Nature}\ }\textbf
  {\bibinfo {volume} {572}},\ \bibinfo {pages} {95} (\bibinfo {year}
  {2019})}\BibitemShut {NoStop}%
\bibitem [{\citenamefont {Jiang}\ \emph {et~al.}(2019)\citenamefont {Jiang},
  \citenamefont {Lai}, \citenamefont {Watanabe}, \citenamefont {Taniguchi},
  \citenamefont {Haule}, \citenamefont {Mao},\ and\ \citenamefont
  {Andrei}}]{jiang2019charge}%
  \BibitemOpen
  \bibfield  {author} {\bibinfo {author} {\bibfnamefont {Y.}~\bibnamefont
  {Jiang}}, \bibinfo {author} {\bibfnamefont {X.}~\bibnamefont {Lai}}, \bibinfo
  {author} {\bibfnamefont {K.}~\bibnamefont {Watanabe}}, \bibinfo {author}
  {\bibfnamefont {T.}~\bibnamefont {Taniguchi}}, \bibinfo {author}
  {\bibfnamefont {K.}~\bibnamefont {Haule}}, \bibinfo {author} {\bibfnamefont
  {J.}~\bibnamefont {Mao}},\ and\ \bibinfo {author} {\bibfnamefont {E.~Y.}\
  \bibnamefont {Andrei}},\ }\bibfield  {title} {\bibinfo {title} {Charge order
  and broken rotational symmetry in magic-angle twisted bilayer graphene},\
  }\href@noop {} {\bibfield  {journal} {\bibinfo  {journal} {Nature}\ }\textbf
  {\bibinfo {volume} {573}},\ \bibinfo {pages} {91} (\bibinfo {year}
  {2019})}\BibitemShut {NoStop}%
\bibitem [{\citenamefont {Trambly~de Laissardi{\`e}re}\ \emph
  {et~al.}(2010)\citenamefont {Trambly~de Laissardi{\`e}re}, \citenamefont
  {Mayou},\ and\ \citenamefont {Magaud}}]{trambly2010localization}%
  \BibitemOpen
  \bibfield  {author} {\bibinfo {author} {\bibfnamefont {G.}~\bibnamefont
  {Trambly~de Laissardi{\`e}re}}, \bibinfo {author} {\bibfnamefont
  {D.}~\bibnamefont {Mayou}},\ and\ \bibinfo {author} {\bibfnamefont
  {L.}~\bibnamefont {Magaud}},\ }\bibfield  {title} {\bibinfo {title}
  {Localization of dirac electrons in rotated graphene bilayers},\ }\href@noop
  {} {\bibfield  {journal} {\bibinfo  {journal} {Nano letters}\ }\textbf
  {\bibinfo {volume} {10}},\ \bibinfo {pages} {804} (\bibinfo {year}
  {2010})}\BibitemShut {NoStop}%
\bibitem [{\citenamefont {Koshino}\ \emph {et~al.}(2018)\citenamefont
  {Koshino}, \citenamefont {Yuan}, \citenamefont {Koretsune}, \citenamefont
  {Ochi}, \citenamefont {Kuroki},\ and\ \citenamefont
  {Fu}}]{koshino2018maximally}%
  \BibitemOpen
  \bibfield  {author} {\bibinfo {author} {\bibfnamefont {M.}~\bibnamefont
  {Koshino}}, \bibinfo {author} {\bibfnamefont {N.~F.}\ \bibnamefont {Yuan}},
  \bibinfo {author} {\bibfnamefont {T.}~\bibnamefont {Koretsune}}, \bibinfo
  {author} {\bibfnamefont {M.}~\bibnamefont {Ochi}}, \bibinfo {author}
  {\bibfnamefont {K.}~\bibnamefont {Kuroki}},\ and\ \bibinfo {author}
  {\bibfnamefont {L.}~\bibnamefont {Fu}},\ }\bibfield  {title} {\bibinfo
  {title} {Maximally localized wannier orbitals and the extended hubbard model
  for twisted bilayer graphene},\ }\href@noop {} {\bibfield  {journal}
  {\bibinfo  {journal} {Physical Review X}\ }\textbf {\bibinfo {volume} {8}},\
  \bibinfo {pages} {031087} (\bibinfo {year} {2018})}\BibitemShut {NoStop}%
\bibitem [{\citenamefont {Kang}\ and\ \citenamefont
  {Vafek}(2018)}]{kang2018symmetry}%
  \BibitemOpen
  \bibfield  {author} {\bibinfo {author} {\bibfnamefont {J.}~\bibnamefont
  {Kang}}\ and\ \bibinfo {author} {\bibfnamefont {O.}~\bibnamefont {Vafek}},\
  }\bibfield  {title} {\bibinfo {title} {Symmetry, maximally localized wannier
  states, and a low-energy model for twisted bilayer graphene narrow bands},\
  }\href@noop {} {\bibfield  {journal} {\bibinfo  {journal} {Physical Review
  X}\ }\textbf {\bibinfo {volume} {8}},\ \bibinfo {pages} {031088} (\bibinfo
  {year} {2018})}\BibitemShut {NoStop}%
\bibitem [{\citenamefont {Calder{\'o}n}\ and\ \citenamefont
  {Bascones}(2020)}]{calderon2020interactions}%
  \BibitemOpen
  \bibfield  {author} {\bibinfo {author} {\bibfnamefont {M.}~\bibnamefont
  {Calder{\'o}n}}\ and\ \bibinfo {author} {\bibfnamefont {E.}~\bibnamefont
  {Bascones}},\ }\bibfield  {title} {\bibinfo {title} {Interactions in the
  8-orbital model for twisted bilayer graphene},\ }\href@noop {} {\bibfield
  {journal} {\bibinfo  {journal} {Physical Review B}\ }\textbf {\bibinfo
  {volume} {102}},\ \bibinfo {pages} {155149} (\bibinfo {year}
  {2020})}\BibitemShut {NoStop}%
\bibitem [{\citenamefont {Goodwin}\ \emph
  {et~al.}(2019{\natexlab{a}})\citenamefont {Goodwin}, \citenamefont
  {Corsetti}, \citenamefont {Mostofi},\ and\ \citenamefont
  {Lischner}}]{goodwin2019attractive}%
  \BibitemOpen
  \bibfield  {author} {\bibinfo {author} {\bibfnamefont {Z.~A.}\ \bibnamefont
  {Goodwin}}, \bibinfo {author} {\bibfnamefont {F.}~\bibnamefont {Corsetti}},
  \bibinfo {author} {\bibfnamefont {A.~A.}\ \bibnamefont {Mostofi}},\ and\
  \bibinfo {author} {\bibfnamefont {J.}~\bibnamefont {Lischner}},\ }\bibfield
  {title} {\bibinfo {title} {Attractive electron-electron interactions from
  internal screening in magic-angle twisted bilayer graphene},\ }\href@noop {}
  {\bibfield  {journal} {\bibinfo  {journal} {Physical Review B}\ }\textbf
  {\bibinfo {volume} {100}},\ \bibinfo {pages} {235424} (\bibinfo {year}
  {2019}{\natexlab{a}})}\BibitemShut {NoStop}%
\bibitem [{\citenamefont {Bistritzer}\ and\ \citenamefont
  {MacDonald}(2011)}]{bistritzer2011moire}%
  \BibitemOpen
  \bibfield  {author} {\bibinfo {author} {\bibfnamefont {R.}~\bibnamefont
  {Bistritzer}}\ and\ \bibinfo {author} {\bibfnamefont {A.~H.}\ \bibnamefont
  {MacDonald}},\ }\bibfield  {title} {\bibinfo {title} {Moir{\'e} bands in
  twisted double-layer graphene},\ }\href@noop {} {\bibfield  {journal}
  {\bibinfo  {journal} {Proceedings of the National Academy of Sciences}\
  }\textbf {\bibinfo {volume} {108}},\ \bibinfo {pages} {12233} (\bibinfo
  {year} {2011})}\BibitemShut {NoStop}%
\bibitem [{\citenamefont {Utama}\ \emph {et~al.}(2020)\citenamefont {Utama},
  \citenamefont {Koch}, \citenamefont {Lee}, \citenamefont {Leconte},
  \citenamefont {Li}, \citenamefont {Zhao}, \citenamefont {Jiang},
  \citenamefont {Zhu}, \citenamefont {Watanabe}, \citenamefont {Taniguchi}
  \emph {et~al.}}]{utama2020visualization}%
  \BibitemOpen
  \bibfield  {author} {\bibinfo {author} {\bibfnamefont {M.~I.~B.}\
  \bibnamefont {Utama}}, \bibinfo {author} {\bibfnamefont {R.~J.}\ \bibnamefont
  {Koch}}, \bibinfo {author} {\bibfnamefont {K.}~\bibnamefont {Lee}}, \bibinfo
  {author} {\bibfnamefont {N.}~\bibnamefont {Leconte}}, \bibinfo {author}
  {\bibfnamefont {H.}~\bibnamefont {Li}}, \bibinfo {author} {\bibfnamefont
  {S.}~\bibnamefont {Zhao}}, \bibinfo {author} {\bibfnamefont {L.}~\bibnamefont
  {Jiang}}, \bibinfo {author} {\bibfnamefont {J.}~\bibnamefont {Zhu}}, \bibinfo
  {author} {\bibfnamefont {K.}~\bibnamefont {Watanabe}}, \bibinfo {author}
  {\bibfnamefont {T.}~\bibnamefont {Taniguchi}}, \emph {et~al.},\ }\bibfield
  {title} {\bibinfo {title} {Visualization of the flat electronic band in
  twisted bilayer graphene near the magic angle twist},\ }\href@noop {}
  {\bibfield  {journal} {\bibinfo  {journal} {Nature Physics}\ ,\ \bibinfo
  {pages} {1}} (\bibinfo {year} {2020})}\BibitemShut {NoStop}%
\bibitem [{\citenamefont {Parker}\ \emph {et~al.}(2021)\citenamefont {Parker},
  \citenamefont {Soejima}, \citenamefont {Hauschild}, \citenamefont {Zaletel},\
  and\ \citenamefont {Bultinck}}]{bultnick2021}%
  \BibitemOpen
  \bibfield  {author} {\bibinfo {author} {\bibfnamefont {D.~E.}\ \bibnamefont
  {Parker}}, \bibinfo {author} {\bibfnamefont {T.}~\bibnamefont {Soejima}},
  \bibinfo {author} {\bibfnamefont {J.}~\bibnamefont {Hauschild}}, \bibinfo
  {author} {\bibfnamefont {M.~P.}\ \bibnamefont {Zaletel}},\ and\ \bibinfo
  {author} {\bibfnamefont {N.}~\bibnamefont {Bultinck}},\ }\bibfield  {title}
  {\bibinfo {title} {Strain-induced quantum phase transitions in magic-angle
  graphene},\ }\href {https://doi.org/10.1103/PhysRevLett.127.027601}
  {\bibfield  {journal} {\bibinfo  {journal} {Phys. Rev. Lett.}\ }\textbf
  {\bibinfo {volume} {127}},\ \bibinfo {pages} {027601} (\bibinfo {year}
  {2021})}\BibitemShut {NoStop}%
\bibitem [{\citenamefont {Tao}\ \emph {et~al.}(2020)\citenamefont {Tao},
  \citenamefont {Du}, \citenamefont {Qi}, \citenamefont {Ni}, \citenamefont
  {Jiang},\ and\ \citenamefont {Zhu}}]{tao2020raman}%
  \BibitemOpen
  \bibfield  {author} {\bibinfo {author} {\bibfnamefont {Z.}~\bibnamefont
  {Tao}}, \bibinfo {author} {\bibfnamefont {J.}~\bibnamefont {Du}}, \bibinfo
  {author} {\bibfnamefont {Z.}~\bibnamefont {Qi}}, \bibinfo {author}
  {\bibfnamefont {K.}~\bibnamefont {Ni}}, \bibinfo {author} {\bibfnamefont
  {S.}~\bibnamefont {Jiang}},\ and\ \bibinfo {author} {\bibfnamefont
  {Y.}~\bibnamefont {Zhu}},\ }\bibfield  {title} {\bibinfo {title} {Raman
  spectroscopy study of sp2 to sp3 transition in bilayer graphene under high
  pressures},\ }\href@noop {} {\bibfield  {journal} {\bibinfo  {journal}
  {Applied Physics Letters}\ }\textbf {\bibinfo {volume} {116}},\ \bibinfo
  {pages} {133101} (\bibinfo {year} {2020})}\BibitemShut {NoStop}%
\bibitem [{\citenamefont {Clark}\ \emph {et~al.}(2013)\citenamefont {Clark},
  \citenamefont {Jeon}, \citenamefont {Chen},\ and\ \citenamefont
  {Yoo}}]{clark2013few}%
  \BibitemOpen
  \bibfield  {author} {\bibinfo {author} {\bibfnamefont {S.}~\bibnamefont
  {Clark}}, \bibinfo {author} {\bibfnamefont {K.-J.}\ \bibnamefont {Jeon}},
  \bibinfo {author} {\bibfnamefont {J.-Y.}\ \bibnamefont {Chen}},\ and\
  \bibinfo {author} {\bibfnamefont {C.-S.}\ \bibnamefont {Yoo}},\ }\bibfield
  {title} {\bibinfo {title} {Few-layer graphene under high pressure: Raman and
  x-ray diffraction studies},\ }\href@noop {} {\bibfield  {journal} {\bibinfo
  {journal} {Solid State Communications}\ }\textbf {\bibinfo {volume} {154}},\
  \bibinfo {pages} {15} (\bibinfo {year} {2013})}\BibitemShut {NoStop}%
\bibitem [{\citenamefont {Pizarro}\ \emph {et~al.}(2019)\citenamefont
  {Pizarro}, \citenamefont {R{\"o}sner}, \citenamefont {Thomale}, \citenamefont
  {Valent{\'\i}},\ and\ \citenamefont {Wehling}}]{pizarro2019internal}%
  \BibitemOpen
  \bibfield  {author} {\bibinfo {author} {\bibfnamefont {J.}~\bibnamefont
  {Pizarro}}, \bibinfo {author} {\bibfnamefont {M.}~\bibnamefont {R{\"o}sner}},
  \bibinfo {author} {\bibfnamefont {R.}~\bibnamefont {Thomale}}, \bibinfo
  {author} {\bibfnamefont {R.}~\bibnamefont {Valent{\'\i}}},\ and\ \bibinfo
  {author} {\bibfnamefont {T.}~\bibnamefont {Wehling}},\ }\bibfield  {title}
  {\bibinfo {title} {Internal screening and dielectric engineering in
  magic-angle twisted bilayer graphene},\ }\href@noop {} {\bibfield  {journal}
  {\bibinfo  {journal} {Physical Review B}\ }\textbf {\bibinfo {volume}
  {100}},\ \bibinfo {pages} {161102} (\bibinfo {year} {2019})}\BibitemShut
  {NoStop}%
\bibitem [{\citenamefont {Dos~Santos}\ \emph {et~al.}(2012)\citenamefont
  {Dos~Santos}, \citenamefont {Peres},\ and\ \citenamefont
  {Neto}}]{dosSantos2012}%
  \BibitemOpen
  \bibfield  {author} {\bibinfo {author} {\bibfnamefont {J.~L.}\ \bibnamefont
  {Dos~Santos}}, \bibinfo {author} {\bibfnamefont {N.}~\bibnamefont {Peres}},\
  and\ \bibinfo {author} {\bibfnamefont {A.~C.}\ \bibnamefont {Neto}},\
  }\bibfield  {title} {\bibinfo {title} {Continuum model of the twisted
  graphene bilayer},\ }\href@noop {} {\bibfield  {journal} {\bibinfo  {journal}
  {Physical Review B}\ }\textbf {\bibinfo {volume} {86}},\ \bibinfo {pages}
  {155449} (\bibinfo {year} {2012})}\BibitemShut {NoStop}%
\bibitem [{\citenamefont {Lian}\ \emph {et~al.}(2019)\citenamefont {Lian},
  \citenamefont {Wang},\ and\ \citenamefont {Bernevig}}]{lian2019twisted}%
  \BibitemOpen
  \bibfield  {author} {\bibinfo {author} {\bibfnamefont {B.}~\bibnamefont
  {Lian}}, \bibinfo {author} {\bibfnamefont {Z.}~\bibnamefont {Wang}},\ and\
  \bibinfo {author} {\bibfnamefont {B.~A.}\ \bibnamefont {Bernevig}},\
  }\bibfield  {title} {\bibinfo {title} {Twisted bilayer graphene: a
  phonon-driven superconductor},\ }\href@noop {} {\bibfield  {journal}
  {\bibinfo  {journal} {Physical review letters}\ }\textbf {\bibinfo {volume}
  {122}},\ \bibinfo {pages} {257002} (\bibinfo {year} {2019})}\BibitemShut
  {NoStop}%
\bibitem [{\citenamefont {Yuan}\ and\ \citenamefont
  {Fu}(2018{\natexlab{a}})}]{yuan2018model}%
  \BibitemOpen
  \bibfield  {author} {\bibinfo {author} {\bibfnamefont {N.~F.}\ \bibnamefont
  {Yuan}}\ and\ \bibinfo {author} {\bibfnamefont {L.}~\bibnamefont {Fu}},\
  }\bibfield  {title} {\bibinfo {title} {Model for the metal-insulator
  transition in graphene superlattices and beyond},\ }\href@noop {} {\bibfield
  {journal} {\bibinfo  {journal} {Physical Review B}\ }\textbf {\bibinfo
  {volume} {98}},\ \bibinfo {pages} {045103} (\bibinfo {year}
  {2018}{\natexlab{a}})}\BibitemShut {NoStop}%
\bibitem [{\citenamefont {Yuan}\ and\ \citenamefont
  {Fu}(2018{\natexlab{b}})}]{yuan2018erratum}%
  \BibitemOpen
  \bibfield  {author} {\bibinfo {author} {\bibfnamefont {N.~F.}\ \bibnamefont
  {Yuan}}\ and\ \bibinfo {author} {\bibfnamefont {L.}~\bibnamefont {Fu}},\
  }\bibfield  {title} {\bibinfo {title} {Erratum: Model for the metal-insulator
  transition in graphene superlattices and beyond [phys. rev. b 98, 045103
  (2018)]},\ }\href@noop {} {\bibfield  {journal} {\bibinfo  {journal}
  {Physical Review B}\ }\textbf {\bibinfo {volume} {98}},\ \bibinfo {pages}
  {079901} (\bibinfo {year} {2018}{\natexlab{b}})}\BibitemShut {NoStop}%
\bibitem [{\citenamefont {Po}\ \emph {et~al.}(2018)\citenamefont {Po},
  \citenamefont {Zou}, \citenamefont {Vishwanath},\ and\ \citenamefont
  {Senthil}}]{po2018origin}%
  \BibitemOpen
  \bibfield  {author} {\bibinfo {author} {\bibfnamefont {H.~C.}\ \bibnamefont
  {Po}}, \bibinfo {author} {\bibfnamefont {L.}~\bibnamefont {Zou}}, \bibinfo
  {author} {\bibfnamefont {A.}~\bibnamefont {Vishwanath}},\ and\ \bibinfo
  {author} {\bibfnamefont {T.}~\bibnamefont {Senthil}},\ }\bibfield  {title}
  {\bibinfo {title} {Origin of mott insulating behavior and superconductivity
  in twisted bilayer graphene},\ }\href@noop {} {\bibfield  {journal} {\bibinfo
   {journal} {Physical Review X}\ }\textbf {\bibinfo {volume} {8}},\ \bibinfo
  {pages} {031089} (\bibinfo {year} {2018})}\BibitemShut {NoStop}%
\bibitem [{\citenamefont {Xu}\ and\ \citenamefont
  {Balents}(2018)}]{xu2018topological}%
  \BibitemOpen
  \bibfield  {author} {\bibinfo {author} {\bibfnamefont {C.}~\bibnamefont
  {Xu}}\ and\ \bibinfo {author} {\bibfnamefont {L.}~\bibnamefont {Balents}},\
  }\bibfield  {title} {\bibinfo {title} {Topological superconductivity in
  twisted multilayer graphene},\ }\href@noop {} {\bibfield  {journal} {\bibinfo
   {journal} {Physical review letters}\ }\textbf {\bibinfo {volume} {121}},\
  \bibinfo {pages} {087001} (\bibinfo {year} {2018})}\BibitemShut {NoStop}%
\bibitem [{\citenamefont {Roy}\ and\ \citenamefont
  {Juri{\v{c}}i{\'c}}(2019)}]{roy2019unconventional}%
  \BibitemOpen
  \bibfield  {author} {\bibinfo {author} {\bibfnamefont {B.}~\bibnamefont
  {Roy}}\ and\ \bibinfo {author} {\bibfnamefont {V.}~\bibnamefont
  {Juri{\v{c}}i{\'c}}},\ }\bibfield  {title} {\bibinfo {title} {Unconventional
  superconductivity in nearly flat bands in twisted bilayer graphene},\
  }\href@noop {} {\bibfield  {journal} {\bibinfo  {journal} {Physical Review
  B}\ }\textbf {\bibinfo {volume} {99}},\ \bibinfo {pages} {121407} (\bibinfo
  {year} {2019})}\BibitemShut {NoStop}%
\bibitem [{\citenamefont {Volovik}(2018)}]{volovik2018graphite}%
  \BibitemOpen
  \bibfield  {author} {\bibinfo {author} {\bibfnamefont {G.~E.}\ \bibnamefont
  {Volovik}},\ }\bibfield  {title} {\bibinfo {title} {Graphite, graphene, and
  the flat band superconductivity},\ }\href@noop {} {\bibfield  {journal}
  {\bibinfo  {journal} {JETP Letters}\ }\textbf {\bibinfo {volume} {107}},\
  \bibinfo {pages} {516} (\bibinfo {year} {2018})}\BibitemShut {NoStop}%
\bibitem [{\citenamefont {Padhi}\ \emph {et~al.}(2018)\citenamefont {Padhi},
  \citenamefont {Setty},\ and\ \citenamefont {Phillips}}]{padhi2018doped}%
  \BibitemOpen
  \bibfield  {author} {\bibinfo {author} {\bibfnamefont {B.}~\bibnamefont
  {Padhi}}, \bibinfo {author} {\bibfnamefont {C.}~\bibnamefont {Setty}},\ and\
  \bibinfo {author} {\bibfnamefont {P.~W.}\ \bibnamefont {Phillips}},\
  }\bibfield  {title} {\bibinfo {title} {Doped twisted bilayer graphene near
  magic angles: proximity to wigner crystallization, not mott insulation},\
  }\href@noop {} {\bibfield  {journal} {\bibinfo  {journal} {Nano letters}\
  }\textbf {\bibinfo {volume} {18}},\ \bibinfo {pages} {6175} (\bibinfo {year}
  {2018})}\BibitemShut {NoStop}%
\bibitem [{\citenamefont {Dodaro}\ \emph {et~al.}(2018)\citenamefont {Dodaro},
  \citenamefont {Kivelson}, \citenamefont {Schattner}, \citenamefont {Sun},\
  and\ \citenamefont {Wang}}]{dodaro2018phases}%
  \BibitemOpen
  \bibfield  {author} {\bibinfo {author} {\bibfnamefont {J.~F.}\ \bibnamefont
  {Dodaro}}, \bibinfo {author} {\bibfnamefont {S.~A.}\ \bibnamefont
  {Kivelson}}, \bibinfo {author} {\bibfnamefont {Y.}~\bibnamefont {Schattner}},
  \bibinfo {author} {\bibfnamefont {X.-Q.}\ \bibnamefont {Sun}},\ and\ \bibinfo
  {author} {\bibfnamefont {C.}~\bibnamefont {Wang}},\ }\bibfield  {title}
  {\bibinfo {title} {Phases of a phenomenological model of twisted bilayer
  graphene},\ }\href@noop {} {\bibfield  {journal} {\bibinfo  {journal}
  {Physical Review B}\ }\textbf {\bibinfo {volume} {98}},\ \bibinfo {pages}
  {075154} (\bibinfo {year} {2018})}\BibitemShut {NoStop}%
\bibitem [{\citenamefont {Wu}\ \emph {et~al.}(2018)\citenamefont {Wu},
  \citenamefont {MacDonald},\ and\ \citenamefont {Martin}}]{wu2018theory}%
  \BibitemOpen
  \bibfield  {author} {\bibinfo {author} {\bibfnamefont {F.}~\bibnamefont
  {Wu}}, \bibinfo {author} {\bibfnamefont {A.}~\bibnamefont {MacDonald}},\ and\
  \bibinfo {author} {\bibfnamefont {I.}~\bibnamefont {Martin}},\ }\bibfield
  {title} {\bibinfo {title} {Theory of phonon-mediated superconductivity in
  twisted bilayer graphene},\ }\href@noop {} {\bibfield  {journal} {\bibinfo
  {journal} {Physical review letters}\ }\textbf {\bibinfo {volume} {121}},\
  \bibinfo {pages} {257001} (\bibinfo {year} {2018})}\BibitemShut {NoStop}%
\bibitem [{\citenamefont {Isobe}\ \emph {et~al.}(2018)\citenamefont {Isobe},
  \citenamefont {Yuan},\ and\ \citenamefont {Fu}}]{isobe2018unconventional}%
  \BibitemOpen
  \bibfield  {author} {\bibinfo {author} {\bibfnamefont {H.}~\bibnamefont
  {Isobe}}, \bibinfo {author} {\bibfnamefont {N.~F.}\ \bibnamefont {Yuan}},\
  and\ \bibinfo {author} {\bibfnamefont {L.}~\bibnamefont {Fu}},\ }\bibfield
  {title} {\bibinfo {title} {Unconventional superconductivity and density waves
  in twisted bilayer graphene},\ }\href@noop {} {\bibfield  {journal} {\bibinfo
   {journal} {Physical Review X}\ }\textbf {\bibinfo {volume} {8}},\ \bibinfo
  {pages} {041041} (\bibinfo {year} {2018})}\BibitemShut {NoStop}%
\bibitem [{\citenamefont {Huang}\ \emph {et~al.}(2019)\citenamefont {Huang},
  \citenamefont {Zhang},\ and\ \citenamefont
  {Ma}}]{huang2019antiferromagnetically}%
  \BibitemOpen
  \bibfield  {author} {\bibinfo {author} {\bibfnamefont {T.}~\bibnamefont
  {Huang}}, \bibinfo {author} {\bibfnamefont {L.}~\bibnamefont {Zhang}},\ and\
  \bibinfo {author} {\bibfnamefont {T.}~\bibnamefont {Ma}},\ }\bibfield
  {title} {\bibinfo {title} {Antiferromagnetically ordered mott insulator and
  d+ id superconductivity in twisted bilayer graphene: A quantum monte carlo
  study},\ }\href@noop {} {\bibfield  {journal} {\bibinfo  {journal} {Science
  Bulletin}\ }\textbf {\bibinfo {volume} {64}},\ \bibinfo {pages} {310}
  (\bibinfo {year} {2019})}\BibitemShut {NoStop}%
\bibitem [{\citenamefont {Zhang}\ \emph {et~al.}(2019)\citenamefont {Zhang},
  \citenamefont {Mao}, \citenamefont {Cao}, \citenamefont {Jarillo-Herrero},\
  and\ \citenamefont {Senthil}}]{zhang2019nearly}%
  \BibitemOpen
  \bibfield  {author} {\bibinfo {author} {\bibfnamefont {Y.-H.}\ \bibnamefont
  {Zhang}}, \bibinfo {author} {\bibfnamefont {D.}~\bibnamefont {Mao}}, \bibinfo
  {author} {\bibfnamefont {Y.}~\bibnamefont {Cao}}, \bibinfo {author}
  {\bibfnamefont {P.}~\bibnamefont {Jarillo-Herrero}},\ and\ \bibinfo {author}
  {\bibfnamefont {T.}~\bibnamefont {Senthil}},\ }\bibfield  {title} {\bibinfo
  {title} {Nearly flat chern bands in moir{\'e} superlattices},\ }\href@noop {}
  {\bibfield  {journal} {\bibinfo  {journal} {Physical Review B}\ }\textbf
  {\bibinfo {volume} {99}},\ \bibinfo {pages} {075127} (\bibinfo {year}
  {2019})}\BibitemShut {NoStop}%
\bibitem [{\citenamefont {Kennes}\ \emph {et~al.}(2018)\citenamefont {Kennes},
  \citenamefont {Lischner},\ and\ \citenamefont {Karrasch}}]{kennes2018strong}%
  \BibitemOpen
  \bibfield  {author} {\bibinfo {author} {\bibfnamefont {D.~M.}\ \bibnamefont
  {Kennes}}, \bibinfo {author} {\bibfnamefont {J.}~\bibnamefont {Lischner}},\
  and\ \bibinfo {author} {\bibfnamefont {C.}~\bibnamefont {Karrasch}},\
  }\bibfield  {title} {\bibinfo {title} {Strong correlations and d+ id
  superconductivity in twisted bilayer graphene},\ }\href@noop {} {\bibfield
  {journal} {\bibinfo  {journal} {Physical Review B}\ }\textbf {\bibinfo
  {volume} {98}},\ \bibinfo {pages} {241407} (\bibinfo {year}
  {2018})}\BibitemShut {NoStop}%
\bibitem [{\citenamefont {Zhang}(2018)}]{zhang2018lowest}%
  \BibitemOpen
  \bibfield  {author} {\bibinfo {author} {\bibfnamefont {L.}~\bibnamefont
  {Zhang}},\ }\bibfield  {title} {\bibinfo {title} {Lowest-energy
  moir$\backslash$'e band formed by dirac zero modes in twisted bilayer
  graphene},\ }\href@noop {} {\bibfield  {journal} {\bibinfo  {journal} {arXiv
  preprint arXiv:1804.09047}\ } (\bibinfo {year} {2018})}\BibitemShut {NoStop}%
\bibitem [{\citenamefont {Guinea}\ and\ \citenamefont
  {Walet}(2018)}]{guinea2018electrostatic}%
  \BibitemOpen
  \bibfield  {author} {\bibinfo {author} {\bibfnamefont {F.}~\bibnamefont
  {Guinea}}\ and\ \bibinfo {author} {\bibfnamefont {N.~R.}\ \bibnamefont
  {Walet}},\ }\bibfield  {title} {\bibinfo {title} {Electrostatic effects, band
  distortions, and superconductivity in twisted graphene bilayers},\
  }\href@noop {} {\bibfield  {journal} {\bibinfo  {journal} {Proceedings of the
  National Academy of Sciences}\ }\textbf {\bibinfo {volume} {115}},\ \bibinfo
  {pages} {13174} (\bibinfo {year} {2018})}\BibitemShut {NoStop}%
\bibitem [{\citenamefont {Zou}\ \emph {et~al.}(2018)\citenamefont {Zou},
  \citenamefont {Po}, \citenamefont {Vishwanath},\ and\ \citenamefont
  {Senthil}}]{zou2018band}%
  \BibitemOpen
  \bibfield  {author} {\bibinfo {author} {\bibfnamefont {L.}~\bibnamefont
  {Zou}}, \bibinfo {author} {\bibfnamefont {H.~C.}\ \bibnamefont {Po}},
  \bibinfo {author} {\bibfnamefont {A.}~\bibnamefont {Vishwanath}},\ and\
  \bibinfo {author} {\bibfnamefont {T.}~\bibnamefont {Senthil}},\ }\bibfield
  {title} {\bibinfo {title} {Band structure of twisted bilayer graphene:
  Emergent symmetries, commensurate approximants, and wannier obstructions},\
  }\href@noop {} {\bibfield  {journal} {\bibinfo  {journal} {Physical Review
  B}\ }\textbf {\bibinfo {volume} {98}},\ \bibinfo {pages} {085435} (\bibinfo
  {year} {2018})}\BibitemShut {NoStop}%
\bibitem [{\citenamefont {Lima}\ \emph {et~al.}(2017)\citenamefont {Lima},
  \citenamefont {Padilha}, \citenamefont {Pontes}, \citenamefont {Fazzio},\
  and\ \citenamefont {da~Silva}}]{lima2017stacking}%
  \BibitemOpen
  \bibfield  {author} {\bibinfo {author} {\bibfnamefont {M.~P.}\ \bibnamefont
  {Lima}}, \bibinfo {author} {\bibfnamefont {J.~E.}\ \bibnamefont {Padilha}},
  \bibinfo {author} {\bibfnamefont {R.~B.}\ \bibnamefont {Pontes}}, \bibinfo
  {author} {\bibfnamefont {A.}~\bibnamefont {Fazzio}},\ and\ \bibinfo {author}
  {\bibfnamefont {A.~J.~R.}\ \bibnamefont {da~Silva}},\ }\bibfield  {title}
  {\bibinfo {title} {Stacking-dependent transport properties in few-layers
  graphene},\ }\href@noop {} {\bibfield  {journal} {\bibinfo  {journal} {Solid
  State Communications}\ }\textbf {\bibinfo {volume} {250}},\ \bibinfo {pages}
  {70} (\bibinfo {year} {2017})}\BibitemShut {NoStop}%
\bibitem [{\citenamefont {Rademaker}\ and\ \citenamefont
  {Mellado}(2018)}]{rademaker2018charge}%
  \BibitemOpen
  \bibfield  {author} {\bibinfo {author} {\bibfnamefont {L.}~\bibnamefont
  {Rademaker}}\ and\ \bibinfo {author} {\bibfnamefont {P.}~\bibnamefont
  {Mellado}},\ }\bibfield  {title} {\bibinfo {title} {Charge-transfer
  insulation in twisted bilayer graphene},\ }\href@noop {} {\bibfield
  {journal} {\bibinfo  {journal} {Physical Review B}\ }\textbf {\bibinfo
  {volume} {98}},\ \bibinfo {pages} {235158} (\bibinfo {year}
  {2018})}\BibitemShut {NoStop}%
\bibitem [{\citenamefont {Angeli}\ \emph {et~al.}(2018)\citenamefont {Angeli},
  \citenamefont {Mandelli}, \citenamefont {Valli}, \citenamefont {Amaricci},
  \citenamefont {Capone}, \citenamefont {Tosatti},\ and\ \citenamefont
  {Fabrizio}}]{angeli2018emergent}%
  \BibitemOpen
  \bibfield  {author} {\bibinfo {author} {\bibfnamefont {M.}~\bibnamefont
  {Angeli}}, \bibinfo {author} {\bibfnamefont {D.}~\bibnamefont {Mandelli}},
  \bibinfo {author} {\bibfnamefont {A.}~\bibnamefont {Valli}}, \bibinfo
  {author} {\bibfnamefont {A.}~\bibnamefont {Amaricci}}, \bibinfo {author}
  {\bibfnamefont {M.}~\bibnamefont {Capone}}, \bibinfo {author} {\bibfnamefont
  {E.}~\bibnamefont {Tosatti}},\ and\ \bibinfo {author} {\bibfnamefont
  {M.}~\bibnamefont {Fabrizio}},\ }\bibfield  {title} {\bibinfo {title}
  {Emergent d 6 symmetry in fully relaxed magic-angle twisted bilayer
  graphene},\ }\href@noop {} {\bibfield  {journal} {\bibinfo  {journal}
  {Physical Review B}\ }\textbf {\bibinfo {volume} {98}},\ \bibinfo {pages}
  {235137} (\bibinfo {year} {2018})}\BibitemShut {NoStop}%
\bibitem [{\citenamefont {Goodwin}\ \emph {et~al.}(2020)\citenamefont
  {Goodwin}, \citenamefont {Vitale}, \citenamefont {Liang}, \citenamefont
  {Mostofi},\ and\ \citenamefont {Lischner}}]{goodwin2020hartree}%
  \BibitemOpen
  \bibfield  {author} {\bibinfo {author} {\bibfnamefont {Z.~A.}\ \bibnamefont
  {Goodwin}}, \bibinfo {author} {\bibfnamefont {V.}~\bibnamefont {Vitale}},
  \bibinfo {author} {\bibfnamefont {X.}~\bibnamefont {Liang}}, \bibinfo
  {author} {\bibfnamefont {A.~A.}\ \bibnamefont {Mostofi}},\ and\ \bibinfo
  {author} {\bibfnamefont {J.}~\bibnamefont {Lischner}},\ }\bibfield  {title}
  {\bibinfo {title} {Hartree theory calculations of quasiparticle properties in
  twisted bilayer graphene},\ }\href@noop {} {\bibfield  {journal} {\bibinfo
  {journal} {Electronic Structure}\ }\textbf {\bibinfo {volume} {2}},\ \bibinfo
  {pages} {034001} (\bibinfo {year} {2020})}\BibitemShut {NoStop}%
\bibitem [{\citenamefont {Martin}\ \emph {et~al.}(2016)\citenamefont {Martin},
  \citenamefont {Reining},\ and\ \citenamefont
  {Ceperley}}]{martin2016interacting}%
  \BibitemOpen
  \bibfield  {author} {\bibinfo {author} {\bibfnamefont {R.~M.}\ \bibnamefont
  {Martin}}, \bibinfo {author} {\bibfnamefont {L.}~\bibnamefont {Reining}},\
  and\ \bibinfo {author} {\bibfnamefont {D.~M.}\ \bibnamefont {Ceperley}},\
  }\href@noop {} {\emph {\bibinfo {title} {Interacting Electrons}}}\ (\bibinfo
  {publisher} {Cambridge University Press},\ \bibinfo {year}
  {2016})\BibitemShut {NoStop}%
\bibitem [{\citenamefont {Xie}\ and\ \citenamefont
  {MacDonald}(2020)}]{xie2020nature}%
  \BibitemOpen
  \bibfield  {author} {\bibinfo {author} {\bibfnamefont {M.}~\bibnamefont
  {Xie}}\ and\ \bibinfo {author} {\bibfnamefont {A.~H.}\ \bibnamefont
  {MacDonald}},\ }\bibfield  {title} {\bibinfo {title} {Nature of the
  correlated insulator states in twisted bilayer graphene},\ }\href@noop {}
  {\bibfield  {journal} {\bibinfo  {journal} {Physical review letters}\
  }\textbf {\bibinfo {volume} {124}},\ \bibinfo {pages} {097601} (\bibinfo
  {year} {2020})}\BibitemShut {NoStop}%
\bibitem [{\citenamefont {Cea}\ and\ \citenamefont
  {Guinea}(2020)}]{cea2020band}%
  \BibitemOpen
  \bibfield  {author} {\bibinfo {author} {\bibfnamefont {T.}~\bibnamefont
  {Cea}}\ and\ \bibinfo {author} {\bibfnamefont {F.}~\bibnamefont {Guinea}},\
  }\bibfield  {title} {\bibinfo {title} {Band structure and insulating states
  driven by coulomb interaction in twisted bilayer graphene},\ }\href@noop {}
  {\bibfield  {journal} {\bibinfo  {journal} {Physical Review B}\ }\textbf
  {\bibinfo {volume} {102}},\ \bibinfo {pages} {045107} (\bibinfo {year}
  {2020})}\BibitemShut {NoStop}%
\bibitem [{\citenamefont {Zhang}\ \emph {et~al.}(2020)\citenamefont {Zhang},
  \citenamefont {Jiang}, \citenamefont {Wang},\ and\ \citenamefont
  {Zhang}}]{zhang2020correlated}%
  \BibitemOpen
  \bibfield  {author} {\bibinfo {author} {\bibfnamefont {Y.}~\bibnamefont
  {Zhang}}, \bibinfo {author} {\bibfnamefont {K.}~\bibnamefont {Jiang}},
  \bibinfo {author} {\bibfnamefont {Z.}~\bibnamefont {Wang}},\ and\ \bibinfo
  {author} {\bibfnamefont {F.}~\bibnamefont {Zhang}},\ }\bibfield  {title}
  {\bibinfo {title} {Correlated insulating phases of twisted bilayer graphene
  at commensurate filling fractions: a hartree-fock study},\ }\href@noop {}
  {\bibfield  {journal} {\bibinfo  {journal} {Physical Review B}\ }\textbf
  {\bibinfo {volume} {102}},\ \bibinfo {pages} {035136} (\bibinfo {year}
  {2020})}\BibitemShut {NoStop}%
\bibitem [{\citenamefont {Bultinck}\ \emph {et~al.}(2020)\citenamefont
  {Bultinck}, \citenamefont {Khalaf}, \citenamefont {Liu}, \citenamefont
  {Chatterjee}, \citenamefont {Vishwanath},\ and\ \citenamefont
  {Zaletel}}]{bultinck2020ground}%
  \BibitemOpen
  \bibfield  {author} {\bibinfo {author} {\bibfnamefont {N.}~\bibnamefont
  {Bultinck}}, \bibinfo {author} {\bibfnamefont {E.}~\bibnamefont {Khalaf}},
  \bibinfo {author} {\bibfnamefont {S.}~\bibnamefont {Liu}}, \bibinfo {author}
  {\bibfnamefont {S.}~\bibnamefont {Chatterjee}}, \bibinfo {author}
  {\bibfnamefont {A.}~\bibnamefont {Vishwanath}},\ and\ \bibinfo {author}
  {\bibfnamefont {M.~P.}\ \bibnamefont {Zaletel}},\ }\bibfield  {title}
  {\bibinfo {title} {Ground state and hidden symmetry of magic-angle graphene
  at even integer filling},\ }\href@noop {} {\bibfield  {journal} {\bibinfo
  {journal} {Physical Review X}\ }\textbf {\bibinfo {volume} {10}},\ \bibinfo
  {pages} {031034} (\bibinfo {year} {2020})}\BibitemShut {NoStop}%
\bibitem [{\citenamefont {Liu}\ \emph {et~al.}(2021{\natexlab{a}})\citenamefont
  {Liu}, \citenamefont {Khalaf}, \citenamefont {Lee},\ and\ \citenamefont
  {Vishwanath}}]{liu2021nematic}%
  \BibitemOpen
  \bibfield  {author} {\bibinfo {author} {\bibfnamefont {S.}~\bibnamefont
  {Liu}}, \bibinfo {author} {\bibfnamefont {E.}~\bibnamefont {Khalaf}},
  \bibinfo {author} {\bibfnamefont {J.~Y.}\ \bibnamefont {Lee}},\ and\ \bibinfo
  {author} {\bibfnamefont {A.}~\bibnamefont {Vishwanath}},\ }\bibfield  {title}
  {\bibinfo {title} {Nematic topological semimetal and insulator in magic-angle
  bilayer graphene at charge neutrality},\ }\href@noop {} {\bibfield  {journal}
  {\bibinfo  {journal} {Physical Review Research}\ }\textbf {\bibinfo {volume}
  {3}},\ \bibinfo {pages} {013033} (\bibinfo {year}
  {2021}{\natexlab{a}})}\BibitemShut {NoStop}%
\bibitem [{\citenamefont {Liu}\ and\ \citenamefont
  {Dai}(2021)}]{liu2021theories}%
  \BibitemOpen
  \bibfield  {author} {\bibinfo {author} {\bibfnamefont {J.}~\bibnamefont
  {Liu}}\ and\ \bibinfo {author} {\bibfnamefont {X.}~\bibnamefont {Dai}},\
  }\bibfield  {title} {\bibinfo {title} {Theories for the correlated insulating
  states and quantum anomalous hall effect phenomena in twisted bilayer
  graphene},\ }\href@noop {} {\bibfield  {journal} {\bibinfo  {journal}
  {Physical Review B}\ }\textbf {\bibinfo {volume} {103}},\ \bibinfo {pages}
  {035427} (\bibinfo {year} {2021})}\BibitemShut {NoStop}%
\bibitem [{\citenamefont {Gonz{\'a}lez}\ and\ \citenamefont
  {Stauber}(2020)}]{gonzalez2020time}%
  \BibitemOpen
  \bibfield  {author} {\bibinfo {author} {\bibfnamefont {J.}~\bibnamefont
  {Gonz{\'a}lez}}\ and\ \bibinfo {author} {\bibfnamefont {T.}~\bibnamefont
  {Stauber}},\ }\bibfield  {title} {\bibinfo {title} {Time-reversal symmetry
  breaking versus chiral symmetry breaking in twisted bilayer graphene},\
  }\href@noop {} {\bibfield  {journal} {\bibinfo  {journal} {Physical Review
  B}\ }\textbf {\bibinfo {volume} {102}},\ \bibinfo {pages} {081118} (\bibinfo
  {year} {2020})}\BibitemShut {NoStop}%
\bibitem [{\citenamefont {Rademaker}\ \emph {et~al.}(2019)\citenamefont
  {Rademaker}, \citenamefont {Abanin},\ and\ \citenamefont
  {Mellado}}]{rademaker2019charge}%
  \BibitemOpen
  \bibfield  {author} {\bibinfo {author} {\bibfnamefont {L.}~\bibnamefont
  {Rademaker}}, \bibinfo {author} {\bibfnamefont {D.~A.}\ \bibnamefont
  {Abanin}},\ and\ \bibinfo {author} {\bibfnamefont {P.}~\bibnamefont
  {Mellado}},\ }\bibfield  {title} {\bibinfo {title} {Charge smoothening and
  band flattening due to hartree corrections in twisted bilayer graphene},\
  }\href@noop {} {\bibfield  {journal} {\bibinfo  {journal} {Physical Review
  B}\ }\textbf {\bibinfo {volume} {100}},\ \bibinfo {pages} {205114} (\bibinfo
  {year} {2019})}\BibitemShut {NoStop}%
\bibitem [{\citenamefont {Potasz}\ \emph {et~al.}(2021)\citenamefont {Potasz},
  \citenamefont {Xie},\ and\ \citenamefont {MacDonald}}]{potasz2021exact}%
  \BibitemOpen
  \bibfield  {author} {\bibinfo {author} {\bibfnamefont {P.}~\bibnamefont
  {Potasz}}, \bibinfo {author} {\bibfnamefont {M.}~\bibnamefont {Xie}},\ and\
  \bibinfo {author} {\bibfnamefont {A.~H.}\ \bibnamefont {MacDonald}},\
  }\bibfield  {title} {\bibinfo {title} {Exact diagonalization for magic-angle
  twisted bilayer graphene},\ }\href@noop {} {\bibfield  {journal} {\bibinfo
  {journal} {arXiv preprint arXiv:2102.02256}\ } (\bibinfo {year}
  {2021})}\BibitemShut {NoStop}%
\bibitem [{\citenamefont {Munoz}\ \emph {et~al.}(2016)\citenamefont {Munoz},
  \citenamefont {Collado}, \citenamefont {Usaj}, \citenamefont {Sofo},\ and\
  \citenamefont {Balseiro}}]{munoz2016bilayer}%
  \BibitemOpen
  \bibfield  {author} {\bibinfo {author} {\bibfnamefont {F.}~\bibnamefont
  {Munoz}}, \bibinfo {author} {\bibfnamefont {H.~O.}\ \bibnamefont {Collado}},
  \bibinfo {author} {\bibfnamefont {G.}~\bibnamefont {Usaj}}, \bibinfo {author}
  {\bibfnamefont {J.~O.}\ \bibnamefont {Sofo}},\ and\ \bibinfo {author}
  {\bibfnamefont {C.}~\bibnamefont {Balseiro}},\ }\bibfield  {title} {\bibinfo
  {title} {Bilayer graphene under pressure: Electron-hole symmetry breaking,
  valley hall effect, and landau levels},\ }\href@noop {} {\bibfield  {journal}
  {\bibinfo  {journal} {Physical Review B}\ }\textbf {\bibinfo {volume} {93}},\
  \bibinfo {pages} {235443} (\bibinfo {year} {2016})}\BibitemShut {NoStop}%
\bibitem [{\citenamefont {Chittari}\ \emph {et~al.}(2018)\citenamefont
  {Chittari}, \citenamefont {Leconte}, \citenamefont {Javvaji},\ and\
  \citenamefont {Jung}}]{chittari2018pressure}%
  \BibitemOpen
  \bibfield  {author} {\bibinfo {author} {\bibfnamefont {B.~L.}\ \bibnamefont
  {Chittari}}, \bibinfo {author} {\bibfnamefont {N.}~\bibnamefont {Leconte}},
  \bibinfo {author} {\bibfnamefont {S.}~\bibnamefont {Javvaji}},\ and\ \bibinfo
  {author} {\bibfnamefont {J.}~\bibnamefont {Jung}},\ }\bibfield  {title}
  {\bibinfo {title} {Pressure induced compression of flatbands in twisted
  bilayer graphene},\ }\href@noop {} {\bibfield  {journal} {\bibinfo  {journal}
  {Electronic Structure}\ }\textbf {\bibinfo {volume} {1}},\ \bibinfo {pages}
  {015001} (\bibinfo {year} {2018})}\BibitemShut {NoStop}%
\bibitem [{\citenamefont {Carr}\ \emph {et~al.}(2018)\citenamefont {Carr},
  \citenamefont {Fang}, \citenamefont {Jarillo-Herrero},\ and\ \citenamefont
  {Kaxiras}}]{carr2018pressure}%
  \BibitemOpen
  \bibfield  {author} {\bibinfo {author} {\bibfnamefont {S.}~\bibnamefont
  {Carr}}, \bibinfo {author} {\bibfnamefont {S.}~\bibnamefont {Fang}}, \bibinfo
  {author} {\bibfnamefont {P.}~\bibnamefont {Jarillo-Herrero}},\ and\ \bibinfo
  {author} {\bibfnamefont {E.}~\bibnamefont {Kaxiras}},\ }\bibfield  {title}
  {\bibinfo {title} {Pressure dependence of the magic twist angle in graphene
  superlattices},\ }\href@noop {} {\bibfield  {journal} {\bibinfo  {journal}
  {Physical Review B}\ }\textbf {\bibinfo {volume} {98}},\ \bibinfo {pages}
  {085144} (\bibinfo {year} {2018})}\BibitemShut {NoStop}%
\bibitem [{\citenamefont {Padhi}\ and\ \citenamefont
  {Phillips}(2019)}]{padhi2019pressure}%
  \BibitemOpen
  \bibfield  {author} {\bibinfo {author} {\bibfnamefont {B.}~\bibnamefont
  {Padhi}}\ and\ \bibinfo {author} {\bibfnamefont {P.~W.}\ \bibnamefont
  {Phillips}},\ }\bibfield  {title} {\bibinfo {title} {Pressure-induced
  metal-insulator transition in twisted bilayer graphene},\ }\href@noop {}
  {\bibfield  {journal} {\bibinfo  {journal} {Physical Review B}\ }\textbf
  {\bibinfo {volume} {99}},\ \bibinfo {pages} {205141} (\bibinfo {year}
  {2019})}\BibitemShut {NoStop}%
\bibitem [{\citenamefont {Lin}\ \emph {et~al.}(2020)\citenamefont {Lin},
  \citenamefont {Zhu},\ and\ \citenamefont {Ni}}]{lin2020pressure}%
  \BibitemOpen
  \bibfield  {author} {\bibinfo {author} {\bibfnamefont {X.}~\bibnamefont
  {Lin}}, \bibinfo {author} {\bibfnamefont {H.}~\bibnamefont {Zhu}},\ and\
  \bibinfo {author} {\bibfnamefont {J.}~\bibnamefont {Ni}},\ }\bibfield
  {title} {\bibinfo {title} {Pressure-induced gap modulation and topological
  transitions in twisted bilayer and twisted double bilayer graphene},\
  }\href@noop {} {\bibfield  {journal} {\bibinfo  {journal} {Physical Review
  B}\ }\textbf {\bibinfo {volume} {101}},\ \bibinfo {pages} {155405} (\bibinfo
  {year} {2020})}\BibitemShut {NoStop}%
\bibitem [{\citenamefont {Green}\ and\ \citenamefont
  {Sofo}(2020)}]{green2020landau}%
  \BibitemOpen
  \bibfield  {author} {\bibinfo {author} {\bibfnamefont {B.~R.}\ \bibnamefont
  {Green}}\ and\ \bibinfo {author} {\bibfnamefont {J.~O.}\ \bibnamefont
  {Sofo}},\ }\bibfield  {title} {\bibinfo {title} {Landau level phases in
  bilayer graphene under pressure at charge neutrality},\ }\href@noop {}
  {\bibfield  {journal} {\bibinfo  {journal} {Physical Review B}\ }\textbf
  {\bibinfo {volume} {101}},\ \bibinfo {pages} {195432} (\bibinfo {year}
  {2020})}\BibitemShut {NoStop}%
\bibitem [{\citenamefont {Vl\v{c}ek}\ \emph {et~al.}(2017)\citenamefont
  {Vl\v{c}ek}, \citenamefont {Rabani}, \citenamefont {Neuhauser},\ and\
  \citenamefont {Baer}}]{vlcek2017stochastic}%
  \BibitemOpen
  \bibfield  {author} {\bibinfo {author} {\bibfnamefont {V.}~\bibnamefont
  {Vl\v{c}ek}}, \bibinfo {author} {\bibfnamefont {E.}~\bibnamefont {Rabani}},
  \bibinfo {author} {\bibfnamefont {D.}~\bibnamefont {Neuhauser}},\ and\
  \bibinfo {author} {\bibfnamefont {R.}~\bibnamefont {Baer}},\ }\bibfield
  {title} {\bibinfo {title} {Stochastic gw calculations for molecules},\
  }\href@noop {} {\bibfield  {journal} {\bibinfo  {journal} {J. Chem. Theory
  Comput.}\ }\textbf {\bibinfo {volume} {13}},\ \bibinfo {pages} {4997}
  (\bibinfo {year} {2017})}\BibitemShut {NoStop}%
\bibitem [{\citenamefont {Vl\v{c}ek}\ \emph {et~al.}(2018)\citenamefont
  {Vl\v{c}ek}, \citenamefont {Li}, \citenamefont {Baer}, \citenamefont
  {Rabani},\ and\ \citenamefont {Neuhauser}}]{Vlcek2018swift}%
  \BibitemOpen
  \bibfield  {author} {\bibinfo {author} {\bibfnamefont {V.}~\bibnamefont
  {Vl\v{c}ek}}, \bibinfo {author} {\bibfnamefont {W.}~\bibnamefont {Li}},
  \bibinfo {author} {\bibfnamefont {R.}~\bibnamefont {Baer}}, \bibinfo {author}
  {\bibfnamefont {E.}~\bibnamefont {Rabani}},\ and\ \bibinfo {author}
  {\bibfnamefont {D.}~\bibnamefont {Neuhauser}},\ }\bibfield  {title} {\bibinfo
  {title} {Swift $gw$ beyond 10,000 electrons using sparse stochastic
  compression},\ }\href {https://doi.org/10.1103/PhysRevB.98.075107} {\bibfield
   {journal} {\bibinfo  {journal} {Phys. Rev. B}\ }\textbf {\bibinfo {volume}
  {98}},\ \bibinfo {pages} {075107} (\bibinfo {year} {2018})}\BibitemShut
  {NoStop}%
\bibitem [{\citenamefont {Neuhauser}\ \emph {et~al.}(2014)\citenamefont
  {Neuhauser}, \citenamefont {Gao}, \citenamefont {Arntsen}, \citenamefont
  {Karshenas}, \citenamefont {Rabani},\ and\ \citenamefont
  {Baer}}]{neuhauser2014breaking}%
  \BibitemOpen
  \bibfield  {author} {\bibinfo {author} {\bibfnamefont {D.}~\bibnamefont
  {Neuhauser}}, \bibinfo {author} {\bibfnamefont {Y.}~\bibnamefont {Gao}},
  \bibinfo {author} {\bibfnamefont {C.}~\bibnamefont {Arntsen}}, \bibinfo
  {author} {\bibfnamefont {C.}~\bibnamefont {Karshenas}}, \bibinfo {author}
  {\bibfnamefont {E.}~\bibnamefont {Rabani}},\ and\ \bibinfo {author}
  {\bibfnamefont {R.}~\bibnamefont {Baer}},\ }\bibfield  {title} {\bibinfo
  {title} {{Breaking the Theoretical Scaling Limit for Predicting Quasiparticle
  Energies: The Stochastic G W Approach}},\ }\href@noop {} {\bibfield
  {journal} {\bibinfo  {journal} {Phys. Rev. Lett.}\ }\textbf {\bibinfo
  {volume} {113}},\ \bibinfo {pages} {076402} (\bibinfo {year}
  {2014})}\BibitemShut {NoStop}%
\bibitem [{\citenamefont {Vlcek}(2019)}]{vlcek2019stochastic}%
  \BibitemOpen
  \bibfield  {author} {\bibinfo {author} {\bibfnamefont {V.}~\bibnamefont
  {Vlcek}},\ }\bibfield  {title} {\bibinfo {title} {Stochastic vertex
  corrections: Linear scaling methods for accurate quasiparticle energies},\
  }\href@noop {} {\bibfield  {journal} {\bibinfo  {journal} {Journal of
  chemical theory and computation}\ }\textbf {\bibinfo {volume} {15}},\
  \bibinfo {pages} {6254} (\bibinfo {year} {2019})}\BibitemShut {NoStop}%
\bibitem [{\citenamefont {Vl{\v{c}}ek}\ \emph
  {et~al.}(2018{\natexlab{a}})\citenamefont {Vl{\v{c}}ek}, \citenamefont
  {Baer}, \citenamefont {Rabani},\ and\ \citenamefont
  {Neuhauser}}]{vlvcek2018simple}%
  \BibitemOpen
  \bibfield  {author} {\bibinfo {author} {\bibfnamefont {V.}~\bibnamefont
  {Vl{\v{c}}ek}}, \bibinfo {author} {\bibfnamefont {R.}~\bibnamefont {Baer}},
  \bibinfo {author} {\bibfnamefont {E.}~\bibnamefont {Rabani}},\ and\ \bibinfo
  {author} {\bibfnamefont {D.}~\bibnamefont {Neuhauser}},\ }\bibfield  {title}
  {\bibinfo {title} {Simple eigenvalue-self-consistent $\delta${\={}} gw 0},\
  }\href@noop {} {\bibfield  {journal} {\bibinfo  {journal} {The Journal of
  chemical physics}\ }\textbf {\bibinfo {volume} {149}},\ \bibinfo {pages}
  {174107} (\bibinfo {year} {2018}{\natexlab{a}})}\BibitemShut {NoStop}%
\bibitem [{\citenamefont {Vl{\v{c}}ek}\ \emph
  {et~al.}(2018{\natexlab{b}})\citenamefont {Vl{\v{c}}ek}, \citenamefont
  {Rabani},\ and\ \citenamefont {Neuhauser}}]{vlvcek2018quasiparticle}%
  \BibitemOpen
  \bibfield  {author} {\bibinfo {author} {\bibfnamefont {V.}~\bibnamefont
  {Vl{\v{c}}ek}}, \bibinfo {author} {\bibfnamefont {E.}~\bibnamefont
  {Rabani}},\ and\ \bibinfo {author} {\bibfnamefont {D.}~\bibnamefont
  {Neuhauser}},\ }\bibfield  {title} {\bibinfo {title} {Quasiparticle spectra
  from molecules to bulk},\ }\href@noop {} {\bibfield  {journal} {\bibinfo
  {journal} {Physical Review Materials}\ }\textbf {\bibinfo {volume} {2}},\
  \bibinfo {pages} {030801} (\bibinfo {year} {2018}{\natexlab{b}})}\BibitemShut
  {NoStop}%
\bibitem [{\citenamefont {Popescu}\ and\ \citenamefont
  {Zunger}(2012)}]{popescu2012extracting}%
  \BibitemOpen
  \bibfield  {author} {\bibinfo {author} {\bibfnamefont {V.}~\bibnamefont
  {Popescu}}\ and\ \bibinfo {author} {\bibfnamefont {A.}~\bibnamefont
  {Zunger}},\ }\bibfield  {title} {\bibinfo {title} {Extracting e versus k
  effective band structure from supercell calculations on alloys and
  impurities},\ }\href@noop {} {\bibfield  {journal} {\bibinfo  {journal}
  {Physical Review B}\ }\textbf {\bibinfo {volume} {85}},\ \bibinfo {pages}
  {085201} (\bibinfo {year} {2012})}\BibitemShut {NoStop}%
\bibitem [{\citenamefont {Huang}\ \emph {et~al.}(2014)\citenamefont {Huang},
  \citenamefont {Zheng}, \citenamefont {Zhang}, \citenamefont {Wu},
  \citenamefont {Gu},\ and\ \citenamefont {Duan}}]{huang2014general}%
  \BibitemOpen
  \bibfield  {author} {\bibinfo {author} {\bibfnamefont {H.}~\bibnamefont
  {Huang}}, \bibinfo {author} {\bibfnamefont {F.}~\bibnamefont {Zheng}},
  \bibinfo {author} {\bibfnamefont {P.}~\bibnamefont {Zhang}}, \bibinfo
  {author} {\bibfnamefont {J.}~\bibnamefont {Wu}}, \bibinfo {author}
  {\bibfnamefont {B.-L.}\ \bibnamefont {Gu}},\ and\ \bibinfo {author}
  {\bibfnamefont {W.}~\bibnamefont {Duan}},\ }\bibfield  {title} {\bibinfo
  {title} {A general group theoretical method to unfold band structures and its
  application},\ }\href@noop {} {\bibfield  {journal} {\bibinfo  {journal} {New
  Journal of Physics}\ }\textbf {\bibinfo {volume} {16}},\ \bibinfo {pages}
  {033034} (\bibinfo {year} {2014})}\BibitemShut {NoStop}%
\bibitem [{\citenamefont {Medeiros}\ \emph {et~al.}(2014)\citenamefont
  {Medeiros}, \citenamefont {Stafstr{\"o}m},\ and\ \citenamefont
  {Bj{\"o}rk}}]{medeiros2014effects}%
  \BibitemOpen
  \bibfield  {author} {\bibinfo {author} {\bibfnamefont {P.~V.}\ \bibnamefont
  {Medeiros}}, \bibinfo {author} {\bibfnamefont {S.}~\bibnamefont
  {Stafstr{\"o}m}},\ and\ \bibinfo {author} {\bibfnamefont {J.}~\bibnamefont
  {Bj{\"o}rk}},\ }\bibfield  {title} {\bibinfo {title} {Effects of extrinsic
  and intrinsic perturbations on the electronic structure of graphene:
  Retaining an effective primitive cell band structure by band unfolding},\
  }\href@noop {} {\bibfield  {journal} {\bibinfo  {journal} {Physical Review
  B}\ }\textbf {\bibinfo {volume} {89}},\ \bibinfo {pages} {041407} (\bibinfo
  {year} {2014})}\BibitemShut {NoStop}%
\bibitem [{\citenamefont {Boykin}\ and\ \citenamefont
  {Klimeck}(2005)}]{PhysRevB.71.115215}%
  \BibitemOpen
  \bibfield  {author} {\bibinfo {author} {\bibfnamefont {T.~B.}\ \bibnamefont
  {Boykin}}\ and\ \bibinfo {author} {\bibfnamefont {G.}~\bibnamefont
  {Klimeck}},\ }\bibfield  {title} {\bibinfo {title} {Practical application of
  zone-folding concepts in tight-binding calculations},\ }\href
  {https://doi.org/10.1103/PhysRevB.71.115215} {\bibfield  {journal} {\bibinfo
  {journal} {Phys. Rev. B}\ }\textbf {\bibinfo {volume} {71}},\ \bibinfo
  {pages} {115215} (\bibinfo {year} {2005})}\BibitemShut {NoStop}%
\bibitem [{\citenamefont {Boykin}\ \emph {et~al.}(2007)\citenamefont {Boykin},
  \citenamefont {Kharche}, \citenamefont {Klimeck},\ and\ \citenamefont
  {Korkusinski}}]{boykin2007approximate}%
  \BibitemOpen
  \bibfield  {author} {\bibinfo {author} {\bibfnamefont {T.~B.}\ \bibnamefont
  {Boykin}}, \bibinfo {author} {\bibfnamefont {N.}~\bibnamefont {Kharche}},
  \bibinfo {author} {\bibfnamefont {G.}~\bibnamefont {Klimeck}},\ and\ \bibinfo
  {author} {\bibfnamefont {M.}~\bibnamefont {Korkusinski}},\ }\bibfield
  {title} {\bibinfo {title} {Approximate bandstructures of semiconductor alloys
  from tight-binding supercell calculations},\ }\href@noop {} {\bibfield
  {journal} {\bibinfo  {journal} {Journal of Physics: Condensed Matter}\
  }\textbf {\bibinfo {volume} {19}},\ \bibinfo {pages} {036203} (\bibinfo
  {year} {2007})}\BibitemShut {NoStop}%
\bibitem [{\citenamefont {Yoo}\ \emph {et~al.}(2019)\citenamefont {Yoo},
  \citenamefont {Engelke}, \citenamefont {Carr}, \citenamefont {Fang},
  \citenamefont {Zhang}, \citenamefont {Cazeaux}, \citenamefont {Sung},
  \citenamefont {Hovden}, \citenamefont {Tsen}, \citenamefont {Taniguchi} \emph
  {et~al.}}]{yoo2019atomic}%
  \BibitemOpen
  \bibfield  {author} {\bibinfo {author} {\bibfnamefont {H.}~\bibnamefont
  {Yoo}}, \bibinfo {author} {\bibfnamefont {R.}~\bibnamefont {Engelke}},
  \bibinfo {author} {\bibfnamefont {S.}~\bibnamefont {Carr}}, \bibinfo {author}
  {\bibfnamefont {S.}~\bibnamefont {Fang}}, \bibinfo {author} {\bibfnamefont
  {K.}~\bibnamefont {Zhang}}, \bibinfo {author} {\bibfnamefont
  {P.}~\bibnamefont {Cazeaux}}, \bibinfo {author} {\bibfnamefont {S.~H.}\
  \bibnamefont {Sung}}, \bibinfo {author} {\bibfnamefont {R.}~\bibnamefont
  {Hovden}}, \bibinfo {author} {\bibfnamefont {A.~W.}\ \bibnamefont {Tsen}},
  \bibinfo {author} {\bibfnamefont {T.}~\bibnamefont {Taniguchi}}, \emph
  {et~al.},\ }\bibfield  {title} {\bibinfo {title} {Atomic and electronic
  reconstruction at the van der waals interface in twisted bilayer graphene},\
  }\href@noop {} {\bibfield  {journal} {\bibinfo  {journal} {Nature materials}\
  }\textbf {\bibinfo {volume} {18}},\ \bibinfo {pages} {448} (\bibinfo {year}
  {2019})}\BibitemShut {NoStop}%
\bibitem [{\citenamefont {Liang}\ \emph {et~al.}(2020)\citenamefont {Liang},
  \citenamefont {Goodwin}, \citenamefont {Vitale}, \citenamefont {Corsetti},
  \citenamefont {Mostofi},\ and\ \citenamefont {Lischner}}]{liang2020effect}%
  \BibitemOpen
  \bibfield  {author} {\bibinfo {author} {\bibfnamefont {X.}~\bibnamefont
  {Liang}}, \bibinfo {author} {\bibfnamefont {Z.~A.}\ \bibnamefont {Goodwin}},
  \bibinfo {author} {\bibfnamefont {V.}~\bibnamefont {Vitale}}, \bibinfo
  {author} {\bibfnamefont {F.}~\bibnamefont {Corsetti}}, \bibinfo {author}
  {\bibfnamefont {A.~A.}\ \bibnamefont {Mostofi}},\ and\ \bibinfo {author}
  {\bibfnamefont {J.}~\bibnamefont {Lischner}},\ }\bibfield  {title} {\bibinfo
  {title} {Effect of bilayer stacking on the atomic and electronic structure of
  twisted double bilayer graphene},\ }\href@noop {} {\bibfield  {journal}
  {\bibinfo  {journal} {Physical Review B}\ }\textbf {\bibinfo {volume}
  {102}},\ \bibinfo {pages} {155146} (\bibinfo {year} {2020})}\BibitemShut
  {NoStop}%
\bibitem [{\citenamefont {Cantele}\ \emph {et~al.}(2020)\citenamefont
  {Cantele}, \citenamefont {Alf{\`e}}, \citenamefont {Conte}, \citenamefont
  {Cataudella}, \citenamefont {Ninno},\ and\ \citenamefont
  {Lucignano}}]{cantele2020structural}%
  \BibitemOpen
  \bibfield  {author} {\bibinfo {author} {\bibfnamefont {G.}~\bibnamefont
  {Cantele}}, \bibinfo {author} {\bibfnamefont {D.}~\bibnamefont {Alf{\`e}}},
  \bibinfo {author} {\bibfnamefont {F.}~\bibnamefont {Conte}}, \bibinfo
  {author} {\bibfnamefont {V.}~\bibnamefont {Cataudella}}, \bibinfo {author}
  {\bibfnamefont {D.}~\bibnamefont {Ninno}},\ and\ \bibinfo {author}
  {\bibfnamefont {P.}~\bibnamefont {Lucignano}},\ }\bibfield  {title} {\bibinfo
  {title} {Structural relaxation and low-energy properties of twisted bilayer
  graphene},\ }\href@noop {} {\bibfield  {journal} {\bibinfo  {journal}
  {Physical Review Research}\ }\textbf {\bibinfo {volume} {2}},\ \bibinfo
  {pages} {043127} (\bibinfo {year} {2020})}\BibitemShut {NoStop}%
\bibitem [{\citenamefont {Leconte}\ \emph {et~al.}(2017)\citenamefont
  {Leconte}, \citenamefont {Jung}, \citenamefont {Leb{\`e}gue},\ and\
  \citenamefont {Gould}}]{leconte2017}%
  \BibitemOpen
  \bibfield  {author} {\bibinfo {author} {\bibfnamefont {N.}~\bibnamefont
  {Leconte}}, \bibinfo {author} {\bibfnamefont {J.}~\bibnamefont {Jung}},
  \bibinfo {author} {\bibfnamefont {S.}~\bibnamefont {Leb{\`e}gue}},\ and\
  \bibinfo {author} {\bibfnamefont {T.}~\bibnamefont {Gould}},\ }\bibfield
  {title} {\bibinfo {title} {Moir{\'e}-pattern interlayer potentials in van der
  waals materials in the random-phase approximation},\ }\href@noop {}
  {\bibfield  {journal} {\bibinfo  {journal} {Physical Review B}\ }\textbf
  {\bibinfo {volume} {96}},\ \bibinfo {pages} {195431} (\bibinfo {year}
  {2017})}\BibitemShut {NoStop}%
\bibitem [{\citenamefont {Heske}\ \emph {et~al.}(1999)\citenamefont {Heske},
  \citenamefont {Treusch}, \citenamefont {Himpsel}, \citenamefont {Kakar},
  \citenamefont {Terminello}, \citenamefont {Weyer},\ and\ \citenamefont
  {Shirley}}]{heske1999band}%
  \BibitemOpen
  \bibfield  {author} {\bibinfo {author} {\bibfnamefont {C.}~\bibnamefont
  {Heske}}, \bibinfo {author} {\bibfnamefont {R.}~\bibnamefont {Treusch}},
  \bibinfo {author} {\bibfnamefont {F.}~\bibnamefont {Himpsel}}, \bibinfo
  {author} {\bibfnamefont {S.}~\bibnamefont {Kakar}}, \bibinfo {author}
  {\bibfnamefont {L.}~\bibnamefont {Terminello}}, \bibinfo {author}
  {\bibfnamefont {H.}~\bibnamefont {Weyer}},\ and\ \bibinfo {author}
  {\bibfnamefont {E.~L.}\ \bibnamefont {Shirley}},\ }\bibfield  {title}
  {\bibinfo {title} {Band widening in graphite},\ }\href@noop {} {\bibfield
  {journal} {\bibinfo  {journal} {Physical Review B}\ }\textbf {\bibinfo
  {volume} {59}},\ \bibinfo {pages} {4680} (\bibinfo {year}
  {1999})}\BibitemShut {NoStop}%
\bibitem [{\citenamefont {Strocov}\ \emph {et~al.}(2001)\citenamefont
  {Strocov}, \citenamefont {Charrier}, \citenamefont {Themlin}, \citenamefont
  {Rohlfing}, \citenamefont {Claessen}, \citenamefont {Barrett}, \citenamefont
  {Avila}, \citenamefont {Sanchez},\ and\ \citenamefont
  {Asensio}}]{strocov2001photoemission}%
  \BibitemOpen
  \bibfield  {author} {\bibinfo {author} {\bibfnamefont {V.}~\bibnamefont
  {Strocov}}, \bibinfo {author} {\bibfnamefont {A.}~\bibnamefont {Charrier}},
  \bibinfo {author} {\bibfnamefont {J.-M.}\ \bibnamefont {Themlin}}, \bibinfo
  {author} {\bibfnamefont {M.}~\bibnamefont {Rohlfing}}, \bibinfo {author}
  {\bibfnamefont {R.}~\bibnamefont {Claessen}}, \bibinfo {author}
  {\bibfnamefont {N.}~\bibnamefont {Barrett}}, \bibinfo {author} {\bibfnamefont
  {J.}~\bibnamefont {Avila}}, \bibinfo {author} {\bibfnamefont
  {J.}~\bibnamefont {Sanchez}},\ and\ \bibinfo {author} {\bibfnamefont {M.-C.}\
  \bibnamefont {Asensio}},\ }\bibfield  {title} {\bibinfo {title}
  {Photoemission from graphite: Intrinsic and self-energy effects},\
  }\href@noop {} {\bibfield  {journal} {\bibinfo  {journal} {Physical Review
  B}\ }\textbf {\bibinfo {volume} {64}},\ \bibinfo {pages} {075105} (\bibinfo
  {year} {2001})}\BibitemShut {NoStop}%
\bibitem [{\citenamefont {Gr{\"u}neis}\ \emph {et~al.}(2008)\citenamefont
  {Gr{\"u}neis}, \citenamefont {Attaccalite}, \citenamefont {Pichler},
  \citenamefont {Zabolotnyy}, \citenamefont {Shiozawa}, \citenamefont
  {Molodtsov}, \citenamefont {Inosov}, \citenamefont {Koitzsch}, \citenamefont
  {Knupfer}, \citenamefont {Schiessling} \emph {et~al.}}]{gruneis2008electron}%
  \BibitemOpen
  \bibfield  {author} {\bibinfo {author} {\bibfnamefont {A.}~\bibnamefont
  {Gr{\"u}neis}}, \bibinfo {author} {\bibfnamefont {C.}~\bibnamefont
  {Attaccalite}}, \bibinfo {author} {\bibfnamefont {T.}~\bibnamefont
  {Pichler}}, \bibinfo {author} {\bibfnamefont {V.}~\bibnamefont {Zabolotnyy}},
  \bibinfo {author} {\bibfnamefont {H.}~\bibnamefont {Shiozawa}}, \bibinfo
  {author} {\bibfnamefont {S.}~\bibnamefont {Molodtsov}}, \bibinfo {author}
  {\bibfnamefont {D.}~\bibnamefont {Inosov}}, \bibinfo {author} {\bibfnamefont
  {A.}~\bibnamefont {Koitzsch}}, \bibinfo {author} {\bibfnamefont
  {M.}~\bibnamefont {Knupfer}}, \bibinfo {author} {\bibfnamefont
  {J.}~\bibnamefont {Schiessling}}, \emph {et~al.},\ }\bibfield  {title}
  {\bibinfo {title} {Electron-electron correlation in graphite: a combined
  angle-resolved photoemission and first-principles study},\ }\href@noop {}
  {\bibfield  {journal} {\bibinfo  {journal} {Physical review letters}\
  }\textbf {\bibinfo {volume} {100}},\ \bibinfo {pages} {037601} (\bibinfo
  {year} {2008})}\BibitemShut {NoStop}%
\bibitem [{\citenamefont {Ohta}\ \emph {et~al.}(2007)\citenamefont {Ohta},
  \citenamefont {Bostwick}, \citenamefont {McChesney}, \citenamefont {Seyller},
  \citenamefont {Horn},\ and\ \citenamefont {Rotenberg}}]{ohta2007interlayer}%
  \BibitemOpen
  \bibfield  {author} {\bibinfo {author} {\bibfnamefont {T.}~\bibnamefont
  {Ohta}}, \bibinfo {author} {\bibfnamefont {A.}~\bibnamefont {Bostwick}},
  \bibinfo {author} {\bibfnamefont {J.~L.}\ \bibnamefont {McChesney}}, \bibinfo
  {author} {\bibfnamefont {T.}~\bibnamefont {Seyller}}, \bibinfo {author}
  {\bibfnamefont {K.}~\bibnamefont {Horn}},\ and\ \bibinfo {author}
  {\bibfnamefont {E.}~\bibnamefont {Rotenberg}},\ }\bibfield  {title} {\bibinfo
  {title} {Interlayer interaction and electronic screening in multilayer
  graphene investigated with angle-resolved photoemission spectroscopy},\
  }\href@noop {} {\bibfield  {journal} {\bibinfo  {journal} {Physical Review
  Letters}\ }\textbf {\bibinfo {volume} {98}},\ \bibinfo {pages} {206802}
  (\bibinfo {year} {2007})}\BibitemShut {NoStop}%
\bibitem [{\citenamefont {Ohta}\ \emph {et~al.}(2006)\citenamefont {Ohta},
  \citenamefont {Bostwick}, \citenamefont {Seyller}, \citenamefont {Horn},\
  and\ \citenamefont {Rotenberg}}]{ohta2006controlling}%
  \BibitemOpen
  \bibfield  {author} {\bibinfo {author} {\bibfnamefont {T.}~\bibnamefont
  {Ohta}}, \bibinfo {author} {\bibfnamefont {A.}~\bibnamefont {Bostwick}},
  \bibinfo {author} {\bibfnamefont {T.}~\bibnamefont {Seyller}}, \bibinfo
  {author} {\bibfnamefont {K.}~\bibnamefont {Horn}},\ and\ \bibinfo {author}
  {\bibfnamefont {E.}~\bibnamefont {Rotenberg}},\ }\bibfield  {title} {\bibinfo
  {title} {Controlling the electronic structure of bilayer graphene},\
  }\href@noop {} {\bibfield  {journal} {\bibinfo  {journal} {Science}\ }\textbf
  {\bibinfo {volume} {313}},\ \bibinfo {pages} {951} (\bibinfo {year}
  {2006})}\BibitemShut {NoStop}%
\bibitem [{\citenamefont {Zhou}\ \emph {et~al.}(2005)\citenamefont {Zhou},
  \citenamefont {Gweon}, \citenamefont {Spataru}, \citenamefont {Graf},
  \citenamefont {Lee}, \citenamefont {Louie},\ and\ \citenamefont
  {Lanzara}}]{zhou2005coexistence}%
  \BibitemOpen
  \bibfield  {author} {\bibinfo {author} {\bibfnamefont {S.}~\bibnamefont
  {Zhou}}, \bibinfo {author} {\bibfnamefont {G.-H.}\ \bibnamefont {Gweon}},
  \bibinfo {author} {\bibfnamefont {C.}~\bibnamefont {Spataru}}, \bibinfo
  {author} {\bibfnamefont {J.}~\bibnamefont {Graf}}, \bibinfo {author}
  {\bibfnamefont {D.-H.}\ \bibnamefont {Lee}}, \bibinfo {author} {\bibfnamefont
  {S.~G.}\ \bibnamefont {Louie}},\ and\ \bibinfo {author} {\bibfnamefont
  {A.}~\bibnamefont {Lanzara}},\ }\bibfield  {title} {\bibinfo {title}
  {Coexistence of sharp quasiparticle dispersions and disorder features in
  graphite},\ }\href@noop {} {\bibfield  {journal} {\bibinfo  {journal}
  {Physical Review B}\ }\textbf {\bibinfo {volume} {71}},\ \bibinfo {pages}
  {161403} (\bibinfo {year} {2005})}\BibitemShut {NoStop}%
\bibitem [{\citenamefont {Ohta}\ \emph {et~al.}(2012)\citenamefont {Ohta},
  \citenamefont {Robinson}, \citenamefont {Feibelman}, \citenamefont
  {Bostwick}, \citenamefont {Rotenberg},\ and\ \citenamefont
  {Beechem}}]{ohta2012evidence}%
  \BibitemOpen
  \bibfield  {author} {\bibinfo {author} {\bibfnamefont {T.}~\bibnamefont
  {Ohta}}, \bibinfo {author} {\bibfnamefont {J.~T.}\ \bibnamefont {Robinson}},
  \bibinfo {author} {\bibfnamefont {P.~J.}\ \bibnamefont {Feibelman}}, \bibinfo
  {author} {\bibfnamefont {A.}~\bibnamefont {Bostwick}}, \bibinfo {author}
  {\bibfnamefont {E.}~\bibnamefont {Rotenberg}},\ and\ \bibinfo {author}
  {\bibfnamefont {T.~E.}\ \bibnamefont {Beechem}},\ }\bibfield  {title}
  {\bibinfo {title} {Evidence for interlayer coupling and moir{\'e} periodic
  potentials in twisted bilayer graphene},\ }\href@noop {} {\bibfield
  {journal} {\bibinfo  {journal} {Physical review letters}\ }\textbf {\bibinfo
  {volume} {109}},\ \bibinfo {pages} {186807} (\bibinfo {year}
  {2012})}\BibitemShut {NoStop}%
\bibitem [{\citenamefont {Brihuega}\ \emph {et~al.}(2012)\citenamefont
  {Brihuega}, \citenamefont {Mallet}, \citenamefont {Gonz{\'a}lez-Herrero},
  \citenamefont {De~Laissardi{\`e}re}, \citenamefont {Ugeda}, \citenamefont
  {Magaud}, \citenamefont {G{\'o}mez-Rodr{\'\i}guez}, \citenamefont
  {Yndur{\'a}in},\ and\ \citenamefont {Veuillen}}]{brihuega2012unraveling}%
  \BibitemOpen
  \bibfield  {author} {\bibinfo {author} {\bibfnamefont {I.}~\bibnamefont
  {Brihuega}}, \bibinfo {author} {\bibfnamefont {P.}~\bibnamefont {Mallet}},
  \bibinfo {author} {\bibfnamefont {H.}~\bibnamefont {Gonz{\'a}lez-Herrero}},
  \bibinfo {author} {\bibfnamefont {G.~T.}\ \bibnamefont
  {De~Laissardi{\`e}re}}, \bibinfo {author} {\bibfnamefont {M.}~\bibnamefont
  {Ugeda}}, \bibinfo {author} {\bibfnamefont {L.}~\bibnamefont {Magaud}},
  \bibinfo {author} {\bibfnamefont {J.}~\bibnamefont
  {G{\'o}mez-Rodr{\'\i}guez}}, \bibinfo {author} {\bibfnamefont
  {F.}~\bibnamefont {Yndur{\'a}in}},\ and\ \bibinfo {author} {\bibfnamefont
  {J.-Y.}\ \bibnamefont {Veuillen}},\ }\bibfield  {title} {\bibinfo {title}
  {Unraveling the intrinsic and robust nature of van hove singularities in
  twisted bilayer graphene by scanning tunneling microscopy and theoretical
  analysis},\ }\href@noop {} {\bibfield  {journal} {\bibinfo  {journal}
  {Physical review letters}\ }\textbf {\bibinfo {volume} {109}},\ \bibinfo
  {pages} {196802} (\bibinfo {year} {2012})}\BibitemShut {NoStop}%
\bibitem [{\citenamefont {De~Laissardiere}\ \emph {et~al.}(2012)\citenamefont
  {De~Laissardiere}, \citenamefont {Mayou},\ and\ \citenamefont
  {Magaud}}]{trambly2012numerical}%
  \BibitemOpen
  \bibfield  {author} {\bibinfo {author} {\bibfnamefont {G.~T.}\ \bibnamefont
  {De~Laissardiere}}, \bibinfo {author} {\bibfnamefont {D.}~\bibnamefont
  {Mayou}},\ and\ \bibinfo {author} {\bibfnamefont {L.}~\bibnamefont
  {Magaud}},\ }\bibfield  {title} {\bibinfo {title} {Numerical studies of
  confined states in rotated bilayers of graphene},\ }\href@noop {} {\bibfield
  {journal} {\bibinfo  {journal} {Physical Review B}\ }\textbf {\bibinfo
  {volume} {86}},\ \bibinfo {pages} {125413} (\bibinfo {year}
  {2012})}\BibitemShut {NoStop}%
\bibitem [{\citenamefont {Faleev}\ \emph {et~al.}(2004)\citenamefont {Faleev},
  \citenamefont {Van~Schilfgaarde},\ and\ \citenamefont
  {Kotani}}]{faleev2004all}%
  \BibitemOpen
  \bibfield  {author} {\bibinfo {author} {\bibfnamefont {S.~V.}\ \bibnamefont
  {Faleev}}, \bibinfo {author} {\bibfnamefont {M.}~\bibnamefont
  {Van~Schilfgaarde}},\ and\ \bibinfo {author} {\bibfnamefont {T.}~\bibnamefont
  {Kotani}},\ }\bibfield  {title} {\bibinfo {title} {All-electron
  self-consistent g w approximation: Application to si, mno, and nio},\
  }\href@noop {} {\bibfield  {journal} {\bibinfo  {journal} {Physical review
  letters}\ }\textbf {\bibinfo {volume} {93}},\ \bibinfo {pages} {126406}
  (\bibinfo {year} {2004})}\BibitemShut {NoStop}%
\bibitem [{\citenamefont {Bruneval}\ \emph {et~al.}(2006)\citenamefont
  {Bruneval}, \citenamefont {Vast},\ and\ \citenamefont
  {Reining}}]{bruneval2006effect}%
  \BibitemOpen
  \bibfield  {author} {\bibinfo {author} {\bibfnamefont {F.}~\bibnamefont
  {Bruneval}}, \bibinfo {author} {\bibfnamefont {N.}~\bibnamefont {Vast}},\
  and\ \bibinfo {author} {\bibfnamefont {L.}~\bibnamefont {Reining}},\
  }\bibfield  {title} {\bibinfo {title} {Effect of self-consistency on
  quasiparticles in solids},\ }\href@noop {} {\bibfield  {journal} {\bibinfo
  {journal} {Physical Review B}\ }\textbf {\bibinfo {volume} {74}},\ \bibinfo
  {pages} {045102} (\bibinfo {year} {2006})}\BibitemShut {NoStop}%
\bibitem [{\citenamefont {Goodwin}\ \emph
  {et~al.}(2019{\natexlab{b}})\citenamefont {Goodwin}, \citenamefont
  {Corsetti}, \citenamefont {Mostofi},\ and\ \citenamefont
  {Lischner}}]{goodwin2019twist}%
  \BibitemOpen
  \bibfield  {author} {\bibinfo {author} {\bibfnamefont {Z.~A.}\ \bibnamefont
  {Goodwin}}, \bibinfo {author} {\bibfnamefont {F.}~\bibnamefont {Corsetti}},
  \bibinfo {author} {\bibfnamefont {A.~A.}\ \bibnamefont {Mostofi}},\ and\
  \bibinfo {author} {\bibfnamefont {J.}~\bibnamefont {Lischner}},\ }\bibfield
  {title} {\bibinfo {title} {Twist-angle sensitivity of electron correlations
  in moir{\'e} graphene bilayers},\ }\href@noop {} {\bibfield  {journal}
  {\bibinfo  {journal} {Physical Review B}\ }\textbf {\bibinfo {volume}
  {100}},\ \bibinfo {pages} {121106} (\bibinfo {year}
  {2019}{\natexlab{b}})}\BibitemShut {NoStop}%
\bibitem [{\citenamefont {Xian}\ \emph {et~al.}(2019)\citenamefont {Xian},
  \citenamefont {Kennes}, \citenamefont {Tancogne-Dejean}, \citenamefont
  {Altarelli},\ and\ \citenamefont {Rubio}}]{xian2019multiflat}%
  \BibitemOpen
  \bibfield  {author} {\bibinfo {author} {\bibfnamefont {L.}~\bibnamefont
  {Xian}}, \bibinfo {author} {\bibfnamefont {D.~M.}\ \bibnamefont {Kennes}},
  \bibinfo {author} {\bibfnamefont {N.}~\bibnamefont {Tancogne-Dejean}},
  \bibinfo {author} {\bibfnamefont {M.}~\bibnamefont {Altarelli}},\ and\
  \bibinfo {author} {\bibfnamefont {A.}~\bibnamefont {Rubio}},\ }\bibfield
  {title} {\bibinfo {title} {Multiflat bands and strong correlations in twisted
  bilayer boron nitride: Doping-induced correlated insulator and
  superconductor},\ }\href@noop {} {\bibfield  {journal} {\bibinfo  {journal}
  {Nano letters}\ }\textbf {\bibinfo {volume} {19}},\ \bibinfo {pages} {4934}
  (\bibinfo {year} {2019})}\BibitemShut {NoStop}%
\bibitem [{\citenamefont {Guo}\ \emph {et~al.}(2018)\citenamefont {Guo},
  \citenamefont {Zhu}, \citenamefont {Feng},\ and\ \citenamefont
  {Scalettar}}]{guo2018pairing}%
  \BibitemOpen
  \bibfield  {author} {\bibinfo {author} {\bibfnamefont {H.}~\bibnamefont
  {Guo}}, \bibinfo {author} {\bibfnamefont {X.}~\bibnamefont {Zhu}}, \bibinfo
  {author} {\bibfnamefont {S.}~\bibnamefont {Feng}},\ and\ \bibinfo {author}
  {\bibfnamefont {R.~T.}\ \bibnamefont {Scalettar}},\ }\bibfield  {title}
  {\bibinfo {title} {Pairing symmetry of interacting fermions on a twisted
  bilayer graphene superlattice},\ }\href@noop {} {\bibfield  {journal}
  {\bibinfo  {journal} {Physical Review B}\ }\textbf {\bibinfo {volume} {97}},\
  \bibinfo {pages} {235453} (\bibinfo {year} {2018})}\BibitemShut {NoStop}%
\bibitem [{\citenamefont {Neto}\ \emph {et~al.}(2009)\citenamefont {Neto},
  \citenamefont {Guinea}, \citenamefont {Peres}, \citenamefont {Novoselov},\
  and\ \citenamefont {Geim}}]{neto2009electronic}%
  \BibitemOpen
  \bibfield  {author} {\bibinfo {author} {\bibfnamefont {A.~C.}\ \bibnamefont
  {Neto}}, \bibinfo {author} {\bibfnamefont {F.}~\bibnamefont {Guinea}},
  \bibinfo {author} {\bibfnamefont {N.~M.}\ \bibnamefont {Peres}}, \bibinfo
  {author} {\bibfnamefont {K.~S.}\ \bibnamefont {Novoselov}},\ and\ \bibinfo
  {author} {\bibfnamefont {A.~K.}\ \bibnamefont {Geim}},\ }\bibfield  {title}
  {\bibinfo {title} {The electronic properties of graphene},\ }\href@noop {}
  {\bibfield  {journal} {\bibinfo  {journal} {Reviews of modern physics}\
  }\textbf {\bibinfo {volume} {81}},\ \bibinfo {pages} {109} (\bibinfo {year}
  {2009})}\BibitemShut {NoStop}%
\bibitem [{Note1()}]{Note1}%
  \BibitemOpen
  \bibinfo {note} {The hopping term is $1/6$ of the bandwidth associated with
  the flat bands.}\BibitemShut {Stop}%
\bibitem [{\citenamefont {Miyake}\ \emph {et~al.}(2009)\citenamefont {Miyake},
  \citenamefont {Aryasetiawan},\ and\ \citenamefont {Imada}}]{miyake2009ab}%
  \BibitemOpen
  \bibfield  {author} {\bibinfo {author} {\bibfnamefont {T.}~\bibnamefont
  {Miyake}}, \bibinfo {author} {\bibfnamefont {F.}~\bibnamefont
  {Aryasetiawan}},\ and\ \bibinfo {author} {\bibfnamefont {M.}~\bibnamefont
  {Imada}},\ }\bibfield  {title} {\bibinfo {title} {Ab initio procedure for
  constructing effective models of correlated materials with entangled band
  structure},\ }\href@noop {} {\bibfield  {journal} {\bibinfo  {journal}
  {Physical Review B}\ }\textbf {\bibinfo {volume} {80}},\ \bibinfo {pages}
  {155134} (\bibinfo {year} {2009})}\BibitemShut {NoStop}%
\bibitem [{\citenamefont {Ma}\ \emph {et~al.}(2021)\citenamefont {Ma},
  \citenamefont {Sheng}, \citenamefont {Govoni},\ and\ \citenamefont
  {Galli}}]{ma2021quantum}%
  \BibitemOpen
  \bibfield  {author} {\bibinfo {author} {\bibfnamefont {H.}~\bibnamefont
  {Ma}}, \bibinfo {author} {\bibfnamefont {N.}~\bibnamefont {Sheng}}, \bibinfo
  {author} {\bibfnamefont {M.}~\bibnamefont {Govoni}},\ and\ \bibinfo {author}
  {\bibfnamefont {G.}~\bibnamefont {Galli}},\ }\bibfield  {title} {\bibinfo
  {title} {Quantum embedding theory for strongly correlated states in
  materials},\ }\href@noop {} {\bibfield  {journal} {\bibinfo  {journal}
  {Journal of Chemical Theory and Computation}\ }\textbf {\bibinfo {volume}
  {17}},\ \bibinfo {pages} {2116} (\bibinfo {year} {2021})}\BibitemShut
  {NoStop}%
\bibitem [{Note2()}]{Note2}%
  \BibitemOpen
  \bibinfo {note} {When computing the screening for the approximate Dyson
  orbitals, $ \psi $, the weakly correlated subspace remains identical, since $
  \psi $ is composed of the combination of the quasi-degenerate
  states}\BibitemShut {NoStop}%
\bibitem [{\citenamefont {Lu}\ \emph {et~al.}(2016)\citenamefont {Lu},
  \citenamefont {Rodriguez-Vega}, \citenamefont {Li}, \citenamefont
  {Luican-Mayer}, \citenamefont {Watanabe}, \citenamefont {Taniguchi},
  \citenamefont {Rossi},\ and\ \citenamefont {Andrei}}]{lu2016local}%
  \BibitemOpen
  \bibfield  {author} {\bibinfo {author} {\bibfnamefont {C.-P.}\ \bibnamefont
  {Lu}}, \bibinfo {author} {\bibfnamefont {M.}~\bibnamefont {Rodriguez-Vega}},
  \bibinfo {author} {\bibfnamefont {G.}~\bibnamefont {Li}}, \bibinfo {author}
  {\bibfnamefont {A.}~\bibnamefont {Luican-Mayer}}, \bibinfo {author}
  {\bibfnamefont {K.}~\bibnamefont {Watanabe}}, \bibinfo {author}
  {\bibfnamefont {T.}~\bibnamefont {Taniguchi}}, \bibinfo {author}
  {\bibfnamefont {E.}~\bibnamefont {Rossi}},\ and\ \bibinfo {author}
  {\bibfnamefont {E.~Y.}\ \bibnamefont {Andrei}},\ }\bibfield  {title}
  {\bibinfo {title} {Local, global, and nonlinear screening in twisted
  double-layer graphene},\ }\href@noop {} {\bibfield  {journal} {\bibinfo
  {journal} {Proceedings of the National Academy of Sciences}\ }\textbf
  {\bibinfo {volume} {113}},\ \bibinfo {pages} {6623} (\bibinfo {year}
  {2016})}\BibitemShut {NoStop}%
\bibitem [{\citenamefont {Stauber}\ and\ \citenamefont
  {Kohler}(2016)}]{stauber2016quasi}%
  \BibitemOpen
  \bibfield  {author} {\bibinfo {author} {\bibfnamefont {T.}~\bibnamefont
  {Stauber}}\ and\ \bibinfo {author} {\bibfnamefont {H.}~\bibnamefont
  {Kohler}},\ }\bibfield  {title} {\bibinfo {title} {Quasi-flat plasmonic bands
  in twisted bilayer graphene},\ }\href@noop {} {\bibfield  {journal} {\bibinfo
   {journal} {Nano letters}\ }\textbf {\bibinfo {volume} {16}},\ \bibinfo
  {pages} {6844} (\bibinfo {year} {2016})}\BibitemShut {NoStop}%
\bibitem [{\citenamefont {Liu}\ \emph {et~al.}(2021{\natexlab{b}})\citenamefont
  {Liu}, \citenamefont {Wang}, \citenamefont {Watanabe}, \citenamefont
  {Taniguchi}, \citenamefont {Vafek},\ and\ \citenamefont
  {Li}}]{liu2020tuning}%
  \BibitemOpen
  \bibfield  {author} {\bibinfo {author} {\bibfnamefont {X.}~\bibnamefont
  {Liu}}, \bibinfo {author} {\bibfnamefont {Z.}~\bibnamefont {Wang}}, \bibinfo
  {author} {\bibfnamefont {K.}~\bibnamefont {Watanabe}}, \bibinfo {author}
  {\bibfnamefont {T.}~\bibnamefont {Taniguchi}}, \bibinfo {author}
  {\bibfnamefont {O.}~\bibnamefont {Vafek}},\ and\ \bibinfo {author}
  {\bibfnamefont {J.}~\bibnamefont {Li}},\ }\bibfield  {title} {\bibinfo
  {title} {Tuning electron correlation in magic-angle twisted bilayer graphene
  using coulomb screening},\ }\href@noop {} {\bibfield  {journal} {\bibinfo
  {journal} {Science}\ }\textbf {\bibinfo {volume} {371}},\ \bibinfo {pages}
  {1261} (\bibinfo {year} {2021}{\natexlab{b}})}\BibitemShut {NoStop}%
\bibitem [{\citenamefont {Zhu}\ \emph {et~al.}(2021)\citenamefont {Zhu},
  \citenamefont {Antezza},\ and\ \citenamefont {Wang}}]{zhu2101dynamical}%
  \BibitemOpen
  \bibfield  {author} {\bibinfo {author} {\bibfnamefont {T.}~\bibnamefont
  {Zhu}}, \bibinfo {author} {\bibfnamefont {M.}~\bibnamefont {Antezza}},\ and\
  \bibinfo {author} {\bibfnamefont {J.-S.}\ \bibnamefont {Wang}},\ }\bibfield
  {title} {\bibinfo {title} {Dynamical polarizability of graphene with spatial
  dispersions},\ }\href@noop {} {\bibfield  {journal} {\bibinfo  {journal}
  {arXiv preprint arXiv:2101.03278}\ } (\bibinfo {year} {2021})}\BibitemShut
  {NoStop}%
\bibitem [{\citenamefont {Vanhala}\ and\ \citenamefont
  {Pollet}(2020)}]{vanhala2020constrained}%
  \BibitemOpen
  \bibfield  {author} {\bibinfo {author} {\bibfnamefont {T.~I.}\ \bibnamefont
  {Vanhala}}\ and\ \bibinfo {author} {\bibfnamefont {L.}~\bibnamefont
  {Pollet}},\ }\bibfield  {title} {\bibinfo {title} {Constrained random phase
  approximation of the effective coulomb interaction in lattice models of
  twisted bilayer graphene},\ }\href@noop {} {\bibfield  {journal} {\bibinfo
  {journal} {Physical Review B}\ }\textbf {\bibinfo {volume} {102}},\ \bibinfo
  {pages} {035154} (\bibinfo {year} {2020})}\BibitemShut {NoStop}%
\bibitem [{\citenamefont {Nam}\ and\ \citenamefont
  {Koshino}(2017)}]{nam2017lattice}%
  \BibitemOpen
  \bibfield  {author} {\bibinfo {author} {\bibfnamefont {N.~N.}\ \bibnamefont
  {Nam}}\ and\ \bibinfo {author} {\bibfnamefont {M.}~\bibnamefont {Koshino}},\
  }\bibfield  {title} {\bibinfo {title} {Lattice relaxation and energy band
  modulation in twisted bilayer graphene},\ }\href@noop {} {\bibfield
  {journal} {\bibinfo  {journal} {Physical Review B}\ }\textbf {\bibinfo
  {volume} {96}},\ \bibinfo {pages} {075311} (\bibinfo {year}
  {2017})}\BibitemShut {NoStop}%
\bibitem [{\citenamefont {Leconte}\ \emph {et~al.}(2019)\citenamefont
  {Leconte}, \citenamefont {Javvaji}, \citenamefont {An},\ and\ \citenamefont
  {Jung}}]{leconte2019relaxation}%
  \BibitemOpen
  \bibfield  {author} {\bibinfo {author} {\bibfnamefont {N.}~\bibnamefont
  {Leconte}}, \bibinfo {author} {\bibfnamefont {S.}~\bibnamefont {Javvaji}},
  \bibinfo {author} {\bibfnamefont {J.}~\bibnamefont {An}},\ and\ \bibinfo
  {author} {\bibfnamefont {J.}~\bibnamefont {Jung}},\ }\bibfield  {title}
  {\bibinfo {title} {Relaxation effects in twisted bilayer graphene: a
  multi-scale approach},\ }\href@noop {} {\bibfield  {journal} {\bibinfo
  {journal} {arXiv preprint arXiv:1910.12805}\ } (\bibinfo {year}
  {2019})}\BibitemShut {NoStop}%
\bibitem [{\citenamefont {Perdew}\ \emph {et~al.}(1996)\citenamefont {Perdew},
  \citenamefont {Burke},\ and\ \citenamefont
  {Ernzerhof}}]{perdew1996generalized}%
  \BibitemOpen
  \bibfield  {author} {\bibinfo {author} {\bibfnamefont {J.~P.}\ \bibnamefont
  {Perdew}}, \bibinfo {author} {\bibfnamefont {K.}~\bibnamefont {Burke}},\ and\
  \bibinfo {author} {\bibfnamefont {M.}~\bibnamefont {Ernzerhof}},\ }\bibfield
  {title} {\bibinfo {title} {Generalized gradient approximation made simple},\
  }\href@noop {} {\bibfield  {journal} {\bibinfo  {journal} {Physical review
  letters}\ }\textbf {\bibinfo {volume} {77}},\ \bibinfo {pages} {3865}
  (\bibinfo {year} {1996})}\BibitemShut {NoStop}%
\bibitem [{\citenamefont {Hedin}(1965)}]{Hedin1965}%
  \BibitemOpen
  \bibfield  {author} {\bibinfo {author} {\bibfnamefont {L.}~\bibnamefont
  {Hedin}},\ }\bibfield  {title} {\bibinfo {title} {{New Method for Calculating
  the One-Particle Green's Function with Application to the Electron-Gas
  Problem}},\ }\href {https://doi.org/10.1103/PhysRev.139.A796} {\bibfield
  {journal} {\bibinfo  {journal} {Phys. Rev.}\ }\textbf {\bibinfo {volume}
  {139}},\ \bibinfo {pages} {A796} (\bibinfo {year} {1965})}\BibitemShut
  {NoStop}%
\bibitem [{\citenamefont {Hybertsen}\ and\ \citenamefont
  {Louie}(1986)}]{hybertsen1986electron}%
  \BibitemOpen
  \bibfield  {author} {\bibinfo {author} {\bibfnamefont {M.~S.}\ \bibnamefont
  {Hybertsen}}\ and\ \bibinfo {author} {\bibfnamefont {S.~G.}\ \bibnamefont
  {Louie}},\ }\bibfield  {title} {\bibinfo {title} {Electron correlation in
  semiconductors and insulators: Band gaps and quasiparticle energies},\
  }\href@noop {} {\bibfield  {journal} {\bibinfo  {journal} {Physical Review
  B}\ }\textbf {\bibinfo {volume} {34}},\ \bibinfo {pages} {5390} (\bibinfo
  {year} {1986})}\BibitemShut {NoStop}%
\bibitem [{\citenamefont {Aryasetiawan}\ and\ \citenamefont
  {Gunnarsson}(1998)}]{Aryasetiawan1998}%
  \BibitemOpen
  \bibfield  {author} {\bibinfo {author} {\bibfnamefont {F.}~\bibnamefont
  {Aryasetiawan}}\ and\ \bibinfo {author} {\bibfnamefont {O.}~\bibnamefont
  {Gunnarsson}},\ }\bibfield  {title} {\bibinfo {title} {{The GW method}},\
  }\href {https://doi.org/10.1088/0034-4885/61/3/002} {\bibfield  {journal}
  {\bibinfo  {journal} {Reports Prog. Phys.}\ }\textbf {\bibinfo {volume}
  {61}},\ \bibinfo {pages} {237} (\bibinfo {year} {1998})}\BibitemShut
  {NoStop}%
\bibitem [{Note3()}]{Note3}%
  \BibitemOpen
  \bibinfo {note} {The Dyson orbital is defined by an overlap of a many-body
  wavefunction for the ground state of $N$ particles, $\Psi _0^{N}$, and $N\pm
  1$ particles state, $\Psi _j^{N}$ for the $j^{\protect \rm th}$ excited
  state. For a hole in the $j^{\protect \rm th}$ state, the Dyson orbital is
  $\psi _j(x)\equiv \protect \sqrt {N} \Psi _j^{N-1}(\protect \bar x_1,\protect
  \bar x_2,\protect \dots ,\protect \bar x_N)\Psi _0^{N}(\protect \bar
  x_1,\protect \bar x_2,\protect \dots ,\protect \bar x_N,x)$, where $x_k$ is
  the spin-space coordinate of electron $k$ and one integrates over all
  coordinates with bar on top}\BibitemShut {NoStop}%
\bibitem [{\citenamefont {Kaplan}\ \emph {et~al.}(2015)\citenamefont {Kaplan},
  \citenamefont {Weigend}, \citenamefont {Evers},\ and\ \citenamefont {van
  Setten}}]{kaplan2015off}%
  \BibitemOpen
  \bibfield  {author} {\bibinfo {author} {\bibfnamefont {F.}~\bibnamefont
  {Kaplan}}, \bibinfo {author} {\bibfnamefont {F.}~\bibnamefont {Weigend}},
  \bibinfo {author} {\bibfnamefont {F.}~\bibnamefont {Evers}},\ and\ \bibinfo
  {author} {\bibfnamefont {M.~J.}\ \bibnamefont {van Setten}},\ }\bibfield
  {title} {\bibinfo {title} {Off-diagonal self-energy terms and partially
  self-consistency in gw calculations for single molecules: efficient
  implementation and quantitative effects on ionization potentials},\
  }\href@noop {} {\bibfield  {journal} {\bibinfo  {journal} {Journal of
  chemical theory and computation}\ }\textbf {\bibinfo {volume} {11}},\
  \bibinfo {pages} {5152} (\bibinfo {year} {2015})}\BibitemShut {NoStop}%
\bibitem [{\citenamefont {Fetter}\ and\ \citenamefont
  {Walecka}(2003)}]{FetterWalecka}%
  \BibitemOpen
  \bibfield  {author} {\bibinfo {author} {\bibfnamefont {A.~L.}\ \bibnamefont
  {Fetter}}\ and\ \bibinfo {author} {\bibfnamefont {J.~D.}\ \bibnamefont
  {Walecka}},\ }\href@noop {} {\emph {\bibinfo {title} {Quantum Theory of
  Many-Particle Systems}}}\ (\bibinfo  {publisher} {Dover Publications},\
  \bibinfo {year} {2003})\BibitemShut {NoStop}%
\bibitem [{\citenamefont {Gao}\ \emph {et~al.}(2015)\citenamefont {Gao},
  \citenamefont {Neuhauser}, \citenamefont {Baer},\ and\ \citenamefont
  {Rabani}}]{gao2015sublinear}%
  \BibitemOpen
  \bibfield  {author} {\bibinfo {author} {\bibfnamefont {Y.}~\bibnamefont
  {Gao}}, \bibinfo {author} {\bibfnamefont {D.}~\bibnamefont {Neuhauser}},
  \bibinfo {author} {\bibfnamefont {R.}~\bibnamefont {Baer}},\ and\ \bibinfo
  {author} {\bibfnamefont {E.}~\bibnamefont {Rabani}},\ }\bibfield  {title}
  {\bibinfo {title} {Sublinear scaling for time-dependent stochastic density
  functional theory},\ }\href@noop {} {\bibfield  {journal} {\bibinfo
  {journal} {The Journal of chemical physics}\ }\textbf {\bibinfo {volume}
  {142}},\ \bibinfo {pages} {034106} (\bibinfo {year} {2015})}\BibitemShut
  {NoStop}%
\bibitem [{\citenamefont {Rabani}\ \emph {et~al.}(2015)\citenamefont {Rabani},
  \citenamefont {Baer},\ and\ \citenamefont {Neuhauser}}]{Rabani2015}%
  \BibitemOpen
  \bibfield  {author} {\bibinfo {author} {\bibfnamefont {E.}~\bibnamefont
  {Rabani}}, \bibinfo {author} {\bibfnamefont {R.}~\bibnamefont {Baer}},\ and\
  \bibinfo {author} {\bibfnamefont {D.}~\bibnamefont {Neuhauser}},\ }\bibfield
  {title} {\bibinfo {title} {{Time-dependent stochastic Bethe-Salpeter
  approach}},\ }\href {https://doi.org/10.1103/PhysRevB.91.235302} {\bibfield
  {journal} {\bibinfo  {journal} {Phys. Rev. B}\ }\textbf {\bibinfo {volume}
  {91}},\ \bibinfo {pages} {235302} (\bibinfo {year} {2015})}\BibitemShut
  {NoStop}%
\bibitem [{\citenamefont {Neuhauser}\ \emph {et~al.}(2016)\citenamefont
  {Neuhauser}, \citenamefont {Rabani}, \citenamefont {Cytter},\ and\
  \citenamefont {Baer}}]{Neuhauser_2016}%
  \BibitemOpen
  \bibfield  {author} {\bibinfo {author} {\bibfnamefont {D.}~\bibnamefont
  {Neuhauser}}, \bibinfo {author} {\bibfnamefont {E.}~\bibnamefont {Rabani}},
  \bibinfo {author} {\bibfnamefont {Y.}~\bibnamefont {Cytter}},\ and\ \bibinfo
  {author} {\bibfnamefont {R.}~\bibnamefont {Baer}},\ }\bibfield  {title}
  {\bibinfo {title} {Stochastic optimally tuned range-separated hybrid density
  functional theory},\ }\href {https://doi.org/10.1021/acs.jpca.5b10573}
  {\bibfield  {journal} {\bibinfo  {journal} {The Journal of Physical Chemistry
  A}\ }\textbf {\bibinfo {volume} {120}},\ \bibinfo {pages} {3071} (\bibinfo
  {year} {2016})}\BibitemShut {NoStop}%
\bibitem [{\citenamefont {Baroni}\ \emph {et~al.}(2001)\citenamefont {Baroni},
  \citenamefont {de~Gironcoli},\ and\ \citenamefont {{Dal
  Corso}}}]{Baroni2001}%
  \BibitemOpen
  \bibfield  {author} {\bibinfo {author} {\bibfnamefont {S.}~\bibnamefont
  {Baroni}}, \bibinfo {author} {\bibfnamefont {S.}~\bibnamefont
  {de~Gironcoli}},\ and\ \bibinfo {author} {\bibfnamefont {A.}~\bibnamefont
  {{Dal Corso}}},\ }\bibfield  {title} {\bibinfo {title} {{Phonons and related
  crystal properties from density-functional perturbation theory}},\ }\href
  {https://doi.org/10.1103/RevModPhys.73.515} {\bibfield  {journal} {\bibinfo
  {journal} {Rev. Mod. Phys.}\ }\textbf {\bibinfo {volume} {73}},\ \bibinfo
  {pages} {515} (\bibinfo {year} {2001})}\BibitemShut {NoStop}%
\bibitem [{\citenamefont {Baer}\ and\ \citenamefont
  {Neuhauser}(2004)}]{BaerNeuhauser2004}%
  \BibitemOpen
  \bibfield  {author} {\bibinfo {author} {\bibfnamefont {R.}~\bibnamefont
  {Baer}}\ and\ \bibinfo {author} {\bibfnamefont {D.}~\bibnamefont
  {Neuhauser}},\ }\bibfield  {title} {\bibinfo {title} {Real-time linear
  response for time-dependent density-functional theory},\ }\href@noop {}
  {\bibfield  {journal} {\bibinfo  {journal} {The Journal of chemical physics}\
  }\textbf {\bibinfo {volume} {121}},\ \bibinfo {pages} {9803} (\bibinfo {year}
  {2004})}\BibitemShut {NoStop}%
\bibitem [{\citenamefont {Neuhauser}\ and\ \citenamefont
  {Baer}(2005)}]{Neuhauser2005}%
  \BibitemOpen
  \bibfield  {author} {\bibinfo {author} {\bibfnamefont {D.}~\bibnamefont
  {Neuhauser}}\ and\ \bibinfo {author} {\bibfnamefont {R.}~\bibnamefont
  {Baer}},\ }\bibfield  {title} {\bibinfo {title} {{Efficient linear-response
  method circumventing the exchange-correlation kernel: theory for molecular
  conductance under finite bias.}},\ }\href {https://doi.org/10.1063/1.2121607}
  {\bibfield  {journal} {\bibinfo  {journal} {J. Chem. Phys.}\ }\textbf
  {\bibinfo {volume} {123}},\ \bibinfo {pages} {204105} (\bibinfo {year}
  {2005})}\BibitemShut {NoStop}%
\bibitem [{\citenamefont {Hubbard}(1963)}]{hubbard1963electron}%
  \BibitemOpen
  \bibfield  {author} {\bibinfo {author} {\bibfnamefont {J.}~\bibnamefont
  {Hubbard}},\ }\bibfield  {title} {\bibinfo {title} {Electron correlations in
  narrow energy bands},\ }\href@noop {} {\bibfield  {journal} {\bibinfo
  {journal} {Proceedings of the Royal Society of London. Series A. Mathematical
  and Physical Sciences}\ }\textbf {\bibinfo {volume} {276}},\ \bibinfo {pages}
  {238} (\bibinfo {year} {1963})}\BibitemShut {NoStop}%
\bibitem [{\citenamefont {Springer}\ and\ \citenamefont
  {Aryasetiawan}(1998)}]{springer1998frequency}%
  \BibitemOpen
  \bibfield  {author} {\bibinfo {author} {\bibfnamefont {M.}~\bibnamefont
  {Springer}}\ and\ \bibinfo {author} {\bibfnamefont {F.}~\bibnamefont
  {Aryasetiawan}},\ }\bibfield  {title} {\bibinfo {title} {Frequency-dependent
  screened interaction in ni within the random-phase approximation},\
  }\href@noop {} {\bibfield  {journal} {\bibinfo  {journal} {Physical Review
  B}\ }\textbf {\bibinfo {volume} {57}},\ \bibinfo {pages} {4364} (\bibinfo
  {year} {1998})}\BibitemShut {NoStop}%
\bibitem [{\citenamefont {Kotani}(2000)}]{kotani2000ab}%
  \BibitemOpen
  \bibfield  {author} {\bibinfo {author} {\bibfnamefont {T.}~\bibnamefont
  {Kotani}},\ }\bibfield  {title} {\bibinfo {title} {Ab initio
  random-phase-approximation calculation of the frequency-dependent effective
  interaction between 3d electrons: Ni, fe, and mno},\ }\href@noop {}
  {\bibfield  {journal} {\bibinfo  {journal} {Journal of Physics: Condensed
  Matter}\ }\textbf {\bibinfo {volume} {12}},\ \bibinfo {pages} {2413}
  (\bibinfo {year} {2000})}\BibitemShut {NoStop}%
\bibitem [{\citenamefont {Romanova}\ and\ \citenamefont
  {Vl{\v{c}}ek}(2020)}]{romanova2020decomposition}%
  \BibitemOpen
  \bibfield  {author} {\bibinfo {author} {\bibfnamefont {M.}~\bibnamefont
  {Romanova}}\ and\ \bibinfo {author} {\bibfnamefont {V.}~\bibnamefont
  {Vl{\v{c}}ek}},\ }\bibfield  {title} {\bibinfo {title} {Decomposition and
  embedding in the stochastic gw self-energy},\ }\href@noop {} {\bibfield
  {journal} {\bibinfo  {journal} {The Journal of Chemical Physics}\ }\textbf
  {\bibinfo {volume} {153}},\ \bibinfo {pages} {134103} (\bibinfo {year}
  {2020})}\BibitemShut {NoStop}%
\bibitem [{Note4()}]{Note4}%
  \BibitemOpen
  \bibinfo {note} {The testing configuration was AMD EPYC 7502 with 2.5~GHz
  frequency using 10 out of 32 physical cores. The total computational time was
  9.6~hrs.}\BibitemShut {Stop}%
\bibitem [{\citenamefont {Giannozzi}\ \emph {et~al.}(2017)\citenamefont
  {Giannozzi}, \citenamefont {Andreussi}, \citenamefont {Brumme}, \citenamefont
  {Bunau}, \citenamefont {Nardelli}, \citenamefont {Calandra}, \citenamefont
  {Car}, \citenamefont {Cavazzoni}, \citenamefont {Ceresoli}, \citenamefont
  {Cococcioni} \emph {et~al.}}]{QE2017}%
  \BibitemOpen
  \bibfield  {author} {\bibinfo {author} {\bibfnamefont {P.}~\bibnamefont
  {Giannozzi}}, \bibinfo {author} {\bibfnamefont {O.}~\bibnamefont
  {Andreussi}}, \bibinfo {author} {\bibfnamefont {T.}~\bibnamefont {Brumme}},
  \bibinfo {author} {\bibfnamefont {O.}~\bibnamefont {Bunau}}, \bibinfo
  {author} {\bibfnamefont {M.~B.}\ \bibnamefont {Nardelli}}, \bibinfo {author}
  {\bibfnamefont {M.}~\bibnamefont {Calandra}}, \bibinfo {author}
  {\bibfnamefont {R.}~\bibnamefont {Car}}, \bibinfo {author} {\bibfnamefont
  {C.}~\bibnamefont {Cavazzoni}}, \bibinfo {author} {\bibfnamefont
  {D.}~\bibnamefont {Ceresoli}}, \bibinfo {author} {\bibfnamefont
  {M.}~\bibnamefont {Cococcioni}}, \emph {et~al.},\ }\bibfield  {title}
  {\bibinfo {title} {Advanced capabilities for materials modelling with quantum
  espresso},\ }\href@noop {} {\bibfield  {journal} {\bibinfo  {journal}
  {Journal of Physics: Condensed Matter}\ }\textbf {\bibinfo {volume} {29}},\
  \bibinfo {pages} {465901} (\bibinfo {year} {2017})}\BibitemShut {NoStop}%
\bibitem [{\citenamefont {Tkatchenko}\ and\ \citenamefont
  {Scheffler}(2009)}]{TS_2009}%
  \BibitemOpen
  \bibfield  {author} {\bibinfo {author} {\bibfnamefont {A.}~\bibnamefont
  {Tkatchenko}}\ and\ \bibinfo {author} {\bibfnamefont {M.}~\bibnamefont
  {Scheffler}},\ }\bibfield  {title} {\bibinfo {title} {Accurate molecular van
  der waals interactions from ground-state electron density and free-atom
  reference data},\ }\href {https://doi.org/10.1103/PhysRevLett.102.073005}
  {\bibfield  {journal} {\bibinfo  {journal} {Phys. Rev. Lett.}\ }\textbf
  {\bibinfo {volume} {102}},\ \bibinfo {pages} {073005} (\bibinfo {year}
  {2009})}\BibitemShut {NoStop}%
\bibitem [{\citenamefont {Otani}\ and\ \citenamefont
  {Sugino}(2006)}]{Otani_2006}%
  \BibitemOpen
  \bibfield  {author} {\bibinfo {author} {\bibfnamefont {M.}~\bibnamefont
  {Otani}}\ and\ \bibinfo {author} {\bibfnamefont {O.}~\bibnamefont {Sugino}},\
  }\bibfield  {title} {\bibinfo {title} {First-principles calculations of
  charged surfaces and interfaces: A plane-wave nonrepeated slab approach},\
  }\href {https://doi.org/10.1103/PhysRevB.73.115407} {\bibfield  {journal}
  {\bibinfo  {journal} {Phys. Rev. B}\ }\textbf {\bibinfo {volume} {73}},\
  \bibinfo {pages} {115407} (\bibinfo {year} {2006})}\BibitemShut {NoStop}%
\bibitem [{\citenamefont {Troullier}\ and\ \citenamefont
  {Martins}(1991)}]{TroullierMartins1991}%
  \BibitemOpen
  \bibfield  {author} {\bibinfo {author} {\bibfnamefont {N.}~\bibnamefont
  {Troullier}}\ and\ \bibinfo {author} {\bibfnamefont {J.~L.}\ \bibnamefont
  {Martins}},\ }\bibfield  {title} {\bibinfo {title} {Efficient
  pseudopotentials for plane-wave calculations},\ }\href@noop {} {\bibfield
  {journal} {\bibinfo  {journal} {Phys. Rev. B}\ }\textbf {\bibinfo {volume}
  {43}},\ \bibinfo {pages} {1993} (\bibinfo {year} {1991})}\BibitemShut
  {NoStop}%
\bibitem [{\citenamefont {Perdew}\ and\ \citenamefont
  {Wang}(1992)}]{PerdewWang}%
  \BibitemOpen
  \bibfield  {author} {\bibinfo {author} {\bibfnamefont {J.~P.}\ \bibnamefont
  {Perdew}}\ and\ \bibinfo {author} {\bibfnamefont {Y.}~\bibnamefont {Wang}},\
  }\bibfield  {title} {\bibinfo {title} {{Accurate and simple analytic
  representation of the electron-gas correlation energy}},\ }\href
  {https://doi.org/10.1103/PhysRevB.45.13244} {\bibfield  {journal} {\bibinfo
  {journal} {Phys. Rev. B}\ }\textbf {\bibinfo {volume} {45}},\ \bibinfo
  {pages} {13244} (\bibinfo {year} {1992})}\BibitemShut {NoStop}%
\bibitem [{\citenamefont {Murnaghan}(1944)}]{murnaghan1944}%
  \BibitemOpen
  \bibfield  {author} {\bibinfo {author} {\bibfnamefont {F.}~\bibnamefont
  {Murnaghan}},\ }\bibfield  {title} {\bibinfo {title} {The compressibility of
  media under extreme pressures},\ }\href@noop {} {\bibfield  {journal}
  {\bibinfo  {journal} {Proceedings of the national academy of sciences of the
  United States of America}\ }\textbf {\bibinfo {volume} {30}},\ \bibinfo
  {pages} {244} (\bibinfo {year} {1944})}\BibitemShut {NoStop}%
\bibitem [{\citenamefont {Rozzi}\ \emph {et~al.}(2006)\citenamefont {Rozzi},
  \citenamefont {Varsano}, \citenamefont {Marini}, \citenamefont {Gross},\ and\
  \citenamefont {Rubio}}]{Rozzi_2006}%
  \BibitemOpen
  \bibfield  {author} {\bibinfo {author} {\bibfnamefont {C.~A.}\ \bibnamefont
  {Rozzi}}, \bibinfo {author} {\bibfnamefont {D.}~\bibnamefont {Varsano}},
  \bibinfo {author} {\bibfnamefont {A.}~\bibnamefont {Marini}}, \bibinfo
  {author} {\bibfnamefont {E.~K.~U.}\ \bibnamefont {Gross}},\ and\ \bibinfo
  {author} {\bibfnamefont {A.}~\bibnamefont {Rubio}},\ }\bibfield  {title}
  {\bibinfo {title} {Exact coulomb cutoff technique for supercell
  calculations},\ }\href {https://doi.org/10.1103/PhysRevB.73.205119}
  {\bibfield  {journal} {\bibinfo  {journal} {Phys. Rev. B}\ }\textbf {\bibinfo
  {volume} {73}},\ \bibinfo {pages} {205119} (\bibinfo {year}
  {2006})}\BibitemShut {NoStop}%
\bibitem [{\citenamefont {Towns}\ \emph {et~al.}(2014)\citenamefont {Towns},
  \citenamefont {Cockerill}, \citenamefont {Dahan}, \citenamefont {Foster},
  \citenamefont {Gaither}, \citenamefont {Grimshaw}, \citenamefont {Hazlewood},
  \citenamefont {Lathrop}, \citenamefont {Lifka}, \citenamefont {Peterson},
  \citenamefont {Roskies}, \citenamefont {Scott},\ and\ \citenamefont
  {Wilkins-Diehr}}]{Towns_2014}%
  \BibitemOpen
  \bibfield  {author} {\bibinfo {author} {\bibfnamefont {J.}~\bibnamefont
  {Towns}}, \bibinfo {author} {\bibfnamefont {T.}~\bibnamefont {Cockerill}},
  \bibinfo {author} {\bibfnamefont {M.}~\bibnamefont {Dahan}}, \bibinfo
  {author} {\bibfnamefont {I.}~\bibnamefont {Foster}}, \bibinfo {author}
  {\bibfnamefont {K.}~\bibnamefont {Gaither}}, \bibinfo {author} {\bibfnamefont
  {A.}~\bibnamefont {Grimshaw}}, \bibinfo {author} {\bibfnamefont
  {V.}~\bibnamefont {Hazlewood}}, \bibinfo {author} {\bibfnamefont
  {S.}~\bibnamefont {Lathrop}}, \bibinfo {author} {\bibfnamefont
  {D.}~\bibnamefont {Lifka}}, \bibinfo {author} {\bibfnamefont {G.~D.}\
  \bibnamefont {Peterson}}, \bibinfo {author} {\bibfnamefont {R.}~\bibnamefont
  {Roskies}}, \bibinfo {author} {\bibfnamefont {J.}~\bibnamefont {Scott}},\
  and\ \bibinfo {author} {\bibfnamefont {N.}~\bibnamefont {Wilkins-Diehr}},\
  }\bibfield  {title} {\bibinfo {title} {Xsede: Accelerating scientific
  discovery},\ }\href {https://doi.org/10.1109/MCSE.2014.80} {\bibfield
  {journal} {\bibinfo  {journal} {Computing in Science \& Engineering}\
  }\textbf {\bibinfo {volume} {16}},\ \bibinfo {pages} {62} (\bibinfo {year}
  {2014})}\BibitemShut {NoStop}%
\end{thebibliography}
\end{document}